\documentclass[twocolumn]{aastex63}
\usepackage{graphicx}
\usepackage{comment}
\usepackage{natbib}
\usepackage{amsmath}
\usepackage{graphics}
\usepackage{ulem}

\def\etal{{et~al.\null}}
\newcommand{\noop}[1]{}

\def\bv{\hbox{$B\!-\!V$}}

\def\gp{\hbox{$g^{\prime}$}}

\newcommand{\elixer}{\texttt{ELiXer}}
\newcommand{\OII}{[\ion{O}{2}]}
\newcommand{\OIII}{[\ion{O}{3}]}
\newcommand{\hbeta}{H$\beta$}
\newcommand{\hgamma}{H$\gamma$}
\newcommand{\hdelta}{H$\delta$}
\newcommand{\HeII}{\ion{He}{2}}
\newcommand{\CIV}{\ion{C}{4}}
\newcommand{\CIII}{\ion{C}{3}}
\newcommand{\lya}{Ly$\alpha$}
\newcommand{\nsource}{223,641}
\newcommand{\nsourceobs}{232,650}
\newcommand{\nstar}{37,916}
\newcommand{\nstarobs}{40,088}
\newcommand{\nlae}{51,863}
\newcommand{\nlaeobs}{52,681}
\newcommand{\noii}{123,891}
\newcommand{\noiiobs}{129,511}
\newcommand{\nagn}{4976}
\newcommand{\nagnobs}{4971}
\newcommand{\nlzg}{5274}
\newcommand{\nlzgobs}{5399}
\newcommand{\nfall}{56,251}
\newcommand{\nspring}{172,831}
\newcommand{\ncosmos}{2447}
\newcommand{\ngoods}{1121}
\newcommand{\zhet}{\texttt{z\_hetdex}}
\newcommand{\zdiagnose}{\texttt{z\_diagnose}}
\newcommand{\linedetcount}{236,354}
\newcommand{\contdetcount}{60,907}
\newcommand{\sdss}{SDSS}
\newcommand{\hetg}{$g_\mathrm{HETDEX}$}
\newcommand{\fluxden}{$10^{-17}$ erg/s/cm$^{2}$/\AA}

\begin{document}

\title{HETDEX Public Source Catalog 1: 220\,K Sources Including Over 50\,K Lyman Alpha Emitters from an Untargeted Wide-area Spectroscopic Survey
\footnote{Based on observations obtained with the Hobby-Eberly Telescope, which is a joint project of the University of Texas at Austin, the Pennsylvania State University, Ludwig-Maximilians-Universit\"at M\"unchen, and Georg-August-Universit\"at G\"ottingen.  The HET is named in honor of its principal benefactors, William P.~Hobby and Robert E.~Eberly.}
}

\author[0000-0002-2307-0146]{Erin Mentuch Cooper}
\affiliation{Department of Astronomy, The University of Texas at Austin, 2515 Speedway Boulevard, Austin, TX 78712, USA}
\affiliation{McDonald Observatory, The University of Texas at Austin, 2515 Speedway Boulevard, Austin, TX 78712, USA}
\email{erin@astro.as.utexas.edu }

\author[0000-0002-8433-8185]{Karl Gebhardt}
\affiliation{Department of Astronomy, The University of Texas at Austin, 2515 Speedway Boulevard, Austin, TX 78712, USA}

\author[0000-0002-8925-9769]{Dustin Davis}
\affiliation{Department of Astronomy, The University of Texas at Austin, 2515 Speedway Boulevard, Austin, TX 78712, USA}

\author[0000-0003-2575-0652]{Daniel J. Farrow}
\affiliation{University Observatory, Fakult\"at f\"ur Physik, Ludwig-Maximilians University Munich, Scheinerstrasse 1, 81679 Munich, Germany}
\affiliation{Max-Planck Institut f\"ur extraterrestrische Physik, Giessenbachstrasse 1, 85748 Garching, Germany}

\author[0000-0001-5561-2010]{Chenxu Liu}
\affiliation{Department of Astronomy, The University of Texas at Austin, 2515 Speedway Boulevard, Austin, TX 78712, USA}

\author[0000-0003-2307-0629]{Gregory Zeimann}
\affiliation{Hobby Eberly Telescope, University of Texas, Austin, Austin, TX, 78712}

\author[0000-0002-1328-0211]{Robin Ciardullo}
\affiliation{Department of Astronomy \& Astrophysics, The Pennsylvania State University, University Park, PA 16802, USA}
\affiliation{Institute for Gravitation and the Cosmos, The Pennsylvania State University, University Park, PA 16802, USA}

\author[0000-0003-2908-2620] {John J. Feldmeier}
\affiliation{Department of Physics, Astronomy, Geology, and Environmental Sciences, Youngstown State University
Youngstown, OH 44555, USA}

\author{Niv Drory}
\affiliation{McDonald Observatory, The University of Texas at Austin, 2515 Speedway Boulevard, Austin, TX 78712, USA}

\author[0000-0002-8434-979X]{Donghui Jeong}
\affiliation{Department of Astronomy \& Astrophysics, The Pennsylvania State University, University Park, PA 16802, USA}
\affiliation{Institute for Gravitation and the Cosmos, The Pennsylvania State University, University Park, PA 16802, USA}

\author{Barbara Benda}
\affiliation{Physics and Astronomy Department, Rutgers, The State University, Piscataway, NJ 08854-8019}

\author[0000-0003-4381-5245]{William P. Bowman}
\affiliation{Department of Astronomy \& Astrophysics, The Pennsylvania State University, University Park, PA 16802, USA}
\affiliation{Institute for Gravitation and the Cosmos, The Pennsylvania State University, University Park, PA 16802, USA}

\author[0000-0002-9604-343X]{Michael Boylan-Kolchin}
\affiliation{Department of Astronomy, The University of Texas at Austin, 2515 Speedway Boulevard, Austin, TX 78712, USA}

\author[0000-0003-2332-5505]{\'Oscar A. Ch\'avez Ortiz}
\affiliation{Department of Astronomy, The University of Texas at Austin, 2515 Speedway Boulevard, Austin, TX 78712, USA}

\author[0000-0002-1998-5677]{Maya H. Debski}
\affiliation{Department of Astronomy \& Astrophysics, The Pennsylvania
State University, University Park, PA 16802, USA}

\author[0000-0002-5149-2282]{Mona Dentler}
\affiliation{Institut f\"{u}r Astrophysik, Universit\"{a}t G\"{o}ttingen, Friedrich-Hund-Platz 1, 37077 G\"{o}ttingen, Germany}

\author{Maximilian Fabricius}
\affiliation{Max-Planck Institut f\"ur extraterrestrische Physik, Giessenbachstrasse 1, 85748 Garching, Germany}
\affiliation{University Observatory, Fakult\"at f\"ur Physik, Ludwig-Maximilians University Munich, Scheinerstrasse 1, 81679 Munich, Germany}

\author{Rameen Farooq}
\affiliation{Physics and Astronomy Department, Rutgers, The State University, Piscataway, NJ 08854}

\author[0000-0001-8519-1130]{Steven L. Finkelstein}
\affiliation{Department of Astronomy, The University of Texas at Austin, 2515 Speedway Boulevard, Austin, TX 78712, USA}

\author[0000-0003-1530-8713]{Eric Gawiser}
\affiliation{Physics and Astronomy Department, Rutgers, The State University, Piscataway, NJ 08854}

\author[0000-0001-6842-2371]{Caryl Gronwall}
\affiliation{Department of Astronomy \& Astrophysics, The Pennsylvania
State University, University Park, PA 16802, USA}
\affiliation{Institute for Gravitation and the Cosmos, The Pennsylvania State University, University Park, PA 16802, USA}

\author[0000-0001-6717-7685]{Gary J. Hill}
\affiliation{McDonald Observatory, The University of Texas at Austin, 2515 Speedway Boulevard, Austin, TX 78712, USA}
\affiliation{Department of Astronomy, The University of Texas at Austin, 2515 Speedway Boulevard, Austin, TX 78712, USA}

\author[0000-0003-1008-225X]{Ulrich Hopp}
\affiliation{University Observatory, Fakult\"at f\"ur Physik, Ludwig-Maximilians University Munich, Scheinerstrasse 1, 81679 Munich, Germany}
\affiliation{Max-Planck Institut f\"ur extraterrestrische Physik, Giessenbachstrasse 1, 85748 Garching, Germany}

\author[0000-0002-1496-6514]{Lindsay R. House}
\affiliation{Department of Astronomy, The University of Texas at Austin, 2515 Speedway Boulevard, Austin, TX 78712, USA}

\author[0000-0001-9165-8905]{Steven Janowiecki}
\affiliation{University of Texas, Hobby-Eberly Telescope, McDonald Observatory, TX 79734, USA}

\author[0000-0001-5610-4405]{Hasti Khoraminezhad}
\affiliation{Institute for Multi-messenger Astrophysics and Cosmology, Department of Physics, Missouri University of Science and Technology, 1315 N Pine St, Rolla, MO 65409}

\author[0000-0002-0417-1494]{Wolfram Kollatschny}
\affiliation{Institut f\"{u}r Astrophysik, Universit\"{a}t G\"{o}ttingen, Friedrich-Hund-Platz 1, 37077 G\"{o}ttingen, Germany}

\author[0000-0002-0136-2404]{Eiichiro Komatsu}
\affiliation{Max-Planck-Institut f\"{u}r Astrophysik, Karl-Schwarzschild-Str. 1, 85741 Garching, Germany}
\affiliation{Kavli Institute for the Physics and Mathematics of the Universe (WPI), Todai Institutes for Advanced Study, the University of Tokyo, Kashiwanoha, Kashiwa, Chiba 277-8583, Japan}

\author[0000-0003-1838-8528]{Martin Landriau}
\affiliation{Lawrence Berkeley National Laboratory, 1 Cyclotron Road, Berkeley, CA 94720, USA}

\author[0000-0002-6907-8370]{Maja Lujan Niemeyer}
\affiliation{Max-Planck-Institut f\"{u}r Astrophysik, Karl-Schwarzschild-Str. 1, 85741 Garching, Germany}

\author[0000-0002-3559-5310]{Hanshin Lee}
\affiliation{McDonald Observatory, The University of Texas at Austin, 2515 Speedway Boulevard, Austin, TX 78712, USA}

\author{Phillip MacQueen}
\affiliation{McDonald Observatory, The University of Texas at Austin, 2515 Speedway Boulevard, Austin, TX 78712, USA}

\author[0000-0003-4985-0201]{Ken Mawatari}
\affiliation{National Astronomical Observatory of Japan, Osawa 2-21-1, Mitaka, Tokyo 181-8588, Japan}
\affiliation{Institute for Cosmic Ray Research, The University of Tokyo, 5-1-5 Kashiwanoha, Kashiwa, Chiba 277-8582, Japan}

\author{Brianna McKay}
\affiliation{Department of Astronomy, University of Washington, Seattle, 3910 15th Ave NE, Room C319, Seattle WA 98195-0002}

\author[0000-0002-1049-6658]{Masami Ouchi}
\affiliation{National Astronomical Observatory of Japan, 2-21-1 Osawa, Mitaka, Tokyo 181-8588, Japan}
\affiliation{Institute for Cosmic Ray Research, The University of Tokyo, 5-1-5 Kashiwanoha, Kashiwa, Chiba 277-8582, Japan}
\affiliation{Kavli Institute for the Physics and Mathematics of the Universe (WPI), Todai Institutes for Advanced Study, the University of Tokyo, Kashiwanoha, Kashiwa, Chiba 277-8583, Japan}

\author{Jennifer Poppe}
\affiliation{Department of Astronomy, The University of Texas at Austin, 2515 Speedway Boulevard, Austin, TX 78712, USA}

\author[0000-0002-6186-5476]{Shun Saito}
\affiliation{Institute for Multi-messenger Astrophysics and Cosmology, Department of Physics, Missouri University of Science and Technology, 1315 N Pine St, Rolla, MO 65409}
\affiliation{Kavli Institute for the Physics and Mathematics of the Universe (WPI), Todai Institutes for Advanced Study, the University of Tokyo, Kashiwanoha, Kashiwa, Chiba 277-8583, Japan}

\author[0000-0001-7240-7449]{Donald P. Schneider}
\affiliation{Department of Astronomy \& Astrophysics, The Pennsylvania State University, University Park, PA 16802, USA}
\affiliation{Institute for Gravitation and the Cosmos, The Pennsylvania State University, University Park, PA 16802, USA}


\author[0000-0003-4044-5357]{Jan Snigula}
\affiliation{Max-Planck Institut f\"ur extraterrestrische Physik, Giessenbachstrasse 1, 85748 Garching, Germany}

\author[0000-0002-0977-1974]{Benjamin P. Thomas}
\affiliation{Department of Astronomy, The University of Texas at Austin, 2515 Speedway Boulevard, Austin, TX 78712, USA}

\author[0000-0002-7327-565X]{Sarah Tuttle}
\affiliation{Department of Astronomy, University of Washington, Seattle, 3910 15th Ave NE, Room C319, Seattle WA 98195-0002}

\author[0000-0001-6746-9936]{Tanya Urrutia}
\affiliation{Leibniz-Institut f\"ur Astrophysik Potsdam (AIP), An der Sternwarte 16, 14482 Potsdam, Germany}

\author[0000-0002-4974-1243]{Laurel Weiss}
\affiliation{Department of Astronomy, The University of Texas at Austin, 2515 Speedway Boulevard, Austin, TX 78712, USA}

\author{Lutz Wisotzki}
\affiliation{Leibniz-Institut f\"ur Astrophysik Potsdam (AIP), An der Sternwarte 16, 14482 Potsdam, Germany}

\author[0000-0003-3817-8739]{Yechi Zhang}
\affiliation{Institute for Cosmic Ray Research, The University of Tokyo, 5-1-5 Kashiwanoha, Kashiwa, Chiba 277-8582, Japan}
\affiliation{Department of Astronomy, Graduate School of Science, the University of Tokyo, 7-3-1 Hongo, Bunkyo, Tokyo 113-0033, Japan}

\collaboration{50}{The HETDEX collaboration}


\begin{abstract}

We present the first publicly released catalog of sources obtained from the Hobby-Eberly Telescope Dark Energy Experiment (HETDEX)\null. HETDEX is an integral field spectroscopic survey designed to measure the Hubble expansion parameter and angular diameter distance at $1.88<z<3.52$ by using the spatial distribution of more than a million Ly$\alpha$-emitting galaxies over a total target area of 540\,deg$^2$. The catalog comes from contiguous fiber spectra coverage of 25\,deg$^2$ of sky from January 2017 through June 2020, where object detection is performed through two complementary detection methods: one designed to search for line emission and the other a search for continuum emission. The HETDEX public release catalog is dominated by emission-line galaxies and includes \nlae\ Ly$\alpha$-emitting galaxy (LAE) identifications and {\noii}  \OII-emitting galaxies at $z<0.5$. Also included in the catalog are \nstar\ stars, \nlzg\ low-redshift ($z<0.5$) galaxies without emission lines, and \nagn\ active galactic nuclei. The catalog provides sky coordinates, redshifts, line identifications, classification information, line fluxes, \OII~and Ly$\alpha$ line luminosities where applicable, and spectra for all identified sources processed by the HETDEX detection pipeline. Extensive testing demonstrates that HETDEX redshifts agree to within $\Delta z < 0.02$, 96.1\% of the time to those in external spectroscopic catalogs.  We measure the photometric counterpart fraction in deep ancillary Hyper Suprime-Cam imaging and find that only 55.5\% of the LAE sample has an $r$-band continuum counterpart down to a limiting magnitude of  $r\sim26.2$\,mag (AB) indicating that an LAE search of similar sensitivity to HETDEX with photometric preselection would miss nearly half of the HETDEX LAE catalog sample.

Data access and details about the catalog can be found online at \url{http://hetdex.org/}. A copy of the catalogs presented in this work (Version 3.2) is available to download at Zenodo \dataset[doi:10.5281/zenodo.7448504]{https://doi.org/10.5281/zenodo.7448504}.


\end{abstract}

\keywords{Catalogs (205) -- Emission line galaxies(459) -- Lyman-alpha galaxies(978) -- Redshift surveys(1378)}

\section{Introduction} 
\label{sec:intro}

Systematic wide-area spectroscopic surveys undertaken in the past two decades, such as the Sloan Digital Sky Survey \citep[\sdss;][]{sdss2000}, BOSS \citep{boss2013}, eBOSS \citep{eboss2016} and DESI \citep{desi2022}, have resulted in orders of magnitude increase in the number of moderate-resolution spectra available for study.  These investigations, thus far, select their spectroscopic targets based upon multiwavelength photometric imaging. Targets are chosen based on continuum brightness, color, morphology, determined stellar mass, and determined star formation rates. These surveys, with their well-defined observing limits and well-characterized systematic uncertainties, have greatly advanced our understanding of the universe.

The above surveys have compiled extensive galaxy samples out to $z \sim 1$.  At higher redshifts, spectroscopic surveys of galaxies are limited to relatively small solid angle regions, where deep imaging aids in the construction of magnitude-limited samples that are sufficiently bright to yield spectroscopic redshifts. Examples of these surveys include the Cosmic Evolution Survey (\citealp[COSMOS;][]{cosmos2007}), and the Great Observatories Origins Deep Survey (\citealp[GOODS;][]{Giavalisco2004}), which both provide unprecedented views of our universe with \textit{HST} and complementary ground-based imaging data. Spectroscopic redshifts in both of these fields number in the tens of thousands (e.g., \citealt{Deimos10K, Reddy2006, Barger2008, Wirth2004, Wirth2015, Ferreras2009, Cooper2011, Kriek2015, Momcheva2016}) and provide important benchmarks for photometric redshifts, as well as numerous targeted investigations in these legacy fields. 

At redshifts larger than two, galaxy samples are often targeted based upon color and magnitude, depending on the science goals. In most cases, these datasets will be biased toward bright, high stellar-mass objects \citep[e.g.,][]{kriek2008,marsan2017} and come from a variety of observatories and heterogeneous sensitivity limits. However, at high redshift, the higher spatial densities of low-mass galaxies provide a stronger tracer of the galaxy distribution  \citep{muzzin2013,Finkelstein2015, song2016}.  For these faint galaxies, spectroscopic redshifts are difficult to obtain from absorption features, and it is most efficient to rely on emission lines for redshifts.  The strong line emission from Lyman-$\alpha$-emitting galaxies (LAEs) allows detection over a wide range of stellar mass (e.g., \citealt{Shapley2003, HuCowie2006}) and redshifts for objects generally too faint for detection in broadband images \citep{Hagen2016, Oyarzun2017, Santos2020}. See \citealt{Ouchi2020} and references therein for a thorough review.

LAE surveys are traditionally conducted by comparing an object's flux through a narrowband filter with that seen in broadband imaging (e.g., \citealt{cowie1998,rhoads2000, Gronwall2007a, Ouchi2008, Konno2016, Sobral2018, Spinoso+2020, ono2021}).  Such searches can be quite successful, but cover relatively small slices in redshift space, as only those objects that have \lya\ redshifted into the bandpass of the narrowband filter are detected.  Recent searches \citep{benitez+2014,eriksen2019,Bonoli+2021} are optimizing the technique by utilizing a high number of narrowband filters, providing for higher redshift coverage, improved source identification and efficient, homogeneous sky coverage.

Alternatively, an efficient method to survey large volumes of high redshift (high-$z$) space is through Integral Field Unit (IFU) observations \citep{vanB2005, Bacon2015, Adams2011a, Urrutia2019}.  IFU observations provide simultaneous redshift coverage along with spatial information in the field-of-view (FOV), limiting the need for follow-up spectroscopy and providing spectral information for neighbouring sources which can aid in identifying contaminants due to spatially extended line emission from low-$z$ galaxies. Though IFU surveys can still be subject to occasional contamination by lower-redshift galaxies and active galactic nuclei (AGN), especially when the wavelength coverage of the spectrograph is limited, such instruments are more efficient at detecting high-$z$ LAEs than narrowband imaging.

One such instrument is the Visible IFU Replicable Unit Spectrograph  \citep[VIRUS;][]{Hill2021}, 
on the 10\,m Hobby Eberly Telescope (HET; \citealt{Ramsey1998,Hill2021},
which can obtain $\approx$35,000 spectra simultaneously, each covering the wavelength range $3500 \mathrm{\AA} \lesssim \lambda \lesssim 5500 \mathrm{\AA}$ with spectral resolving power $750<R<950$. VIRUS is the primary instrument of the Hobby-Eberly Telescope Dark Energy Experiment \citep[HETDEX;][]{Gebhardt2021}, whose goal is to measure the Hubble parameter, $H(z)$, and the angular diameter distance, $D_{A}(z)$, to better than 1\% accuracy in the redshift range $1.9<z<3.5$. HETDEX uses LAEs as a (biased) tracer of dark matter density; by measuring their clustering, HETDEX characterizes the evolution in the universe's dark energy density and tests for potential evolution \citep{Shoji2009}.  To achieve the desired accuracy, HETDEX needs to measure at least one million LAEs over 540 deg$^2$ of sky, or $10.9$~Gpc$^3$ in the targeted redshift range.  The project does not need complete coverage within this sky area to accomplish its scientific goals; as discussed by \citet{Chiang2013}, a fill factor of 0.22 (1/4.6), which optimizes the number of IFUs given the area of the focal plane of the HET, is sufficient.  For the target number of LAEs, HETDEX needs an exposure time sufficient to reach about 2.5 LAEs per IFU\null. The typical total exposure time is 20~minutes per field. With 468,000 IFU observations, at 2.5 LAEs per IFU, we reach the goal of one million LAEs. If the experiment falls short of this goal, the sky area can be adjusted if needed to reach the target number of objects.


\begin{figure*}[t]
    \centering
    \includegraphics[width=\linewidth,trim=1cm 3cm 1cm 3cm,clip]{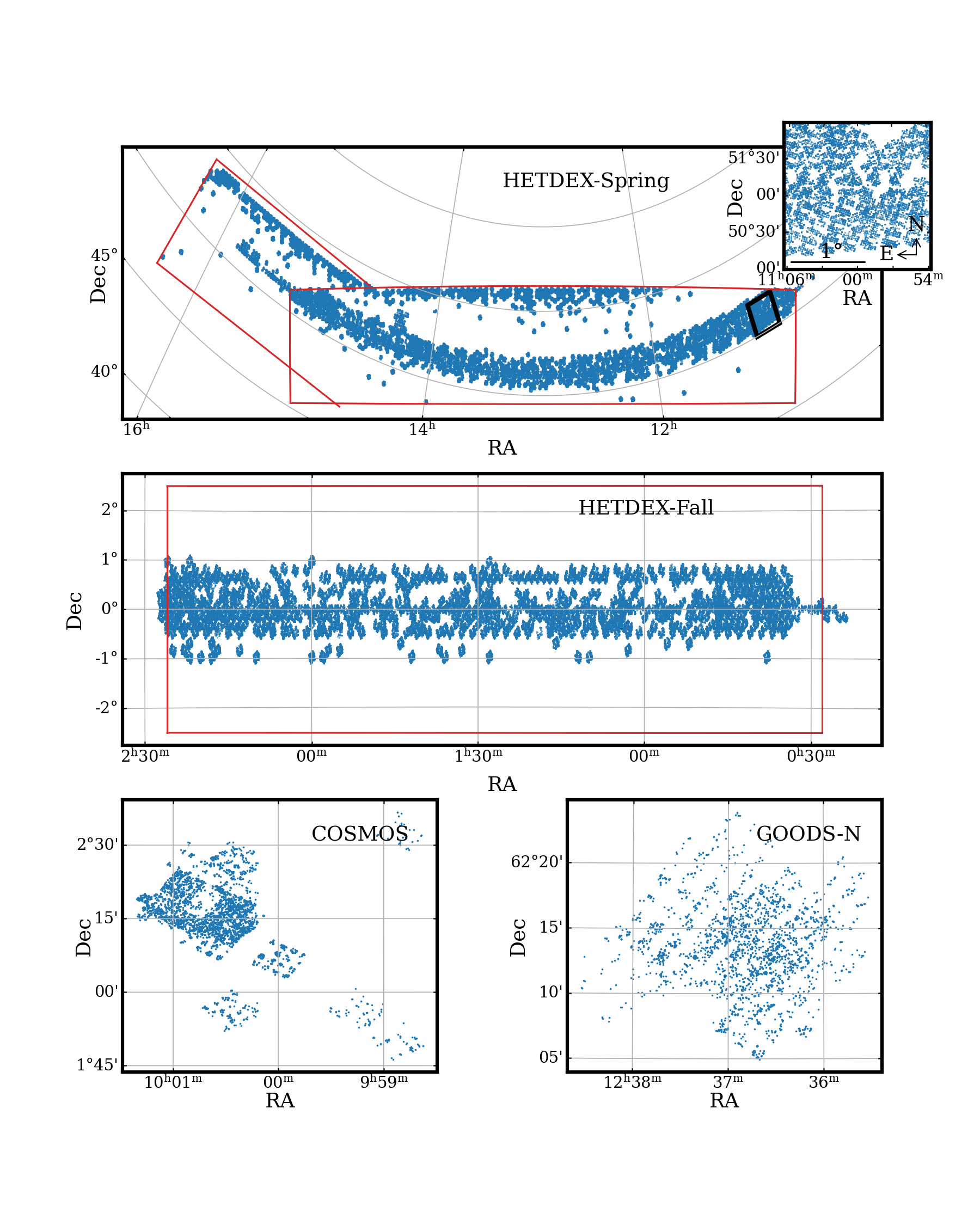}
    \caption{Sky coverage of the planned HETDEX science fields (in red) and the footprint of this catalog release (in blue): (1)~the high decl. Spring Field (top), which is centered at~13.5$^{\rm h}$, +51$^{\circ}$ and covers $\sim390$~deg$^2$ of the sky, and (2)~the equatorial Fall Field (middle), which is centered at 1.5$^{\rm h}$, 0$^{\circ}$, covers $\sim150$~deg$^2$ of the sky. Also highlighted are the two legacy fields, COSMOS (bottom-left) and GOODS-N (bottom-right), where some coverage is included in this catalog. The blue points represent catalog sources, which effectively trace the Integral Field Unit (IFU) array footprint of this release. Each IFU has a field-of-view of $51\arcsec \times 51\arcsec$, which means that the full VIRUS IFU array covers a 0.22 fill factor in the HET's~22$'$-diameter FOV. The expanded inset on the top-right presents a typical 2~deg$^2$ region in the HETDEX-Spring field (representing the rectangular region in the top panel).}
    \label{fig:coverage}
\end{figure*}

The first observations of HETDEX were obtained in January 2017, with VIRUS in commissioning mode at a 
fraction of its current capability. In 2017 the project started with 11 working IFUs; by August 2021  the maximum number of 78 IFUs were operational in the focal plane. This paper presents the first general public catalog of HETDEX sources acquired over the first three years of the survey. These sources come from HETDEX's dual object detection method, described in \citet{Gebhardt2021}, that searches for line emission sources and continuum emission sources independently within the HETDEX IFU dataset. Although designed to find LAEs, the untargeted IFU data also observes a wide range of astronomical sources. This catalog provides coordinates, redshifts, spectra and measured properties of \nsource\ objects which we organize into five source types that are referenced throughout this paper: \lya-emitter as \texttt{lae}, \OII-emitting galaxy as \texttt{oii}, active galactic nuclei as \texttt{agn}, low-$z$ galaxy (with no measured \OII\ line emission) as \texttt{lzg} and $z$=0 sources as \texttt{star}. Transient objects such as meteors and satellites are not included, nor are large nearby galaxies: these objects will be published at a later time.

The outline of this paper is as follows. Section~\ref{sec:observations} describes the observations obtained for the HETDEX survey and details concerning the quality assessment of the observations. Section~\ref{sec:catalog} describes the process of going from raw object  detection to an astronomical source. Section \,\ref{sec:classification} describes source classification and redshift assignment. In Section~\ref{sec:format}, we provide the data format of the catalog, and Section~\ref{sec:sampleproperties} presents properties of the catalog samples. 

\begin{figure*}[t]
    \centering
    \includegraphics[width=3.5in]{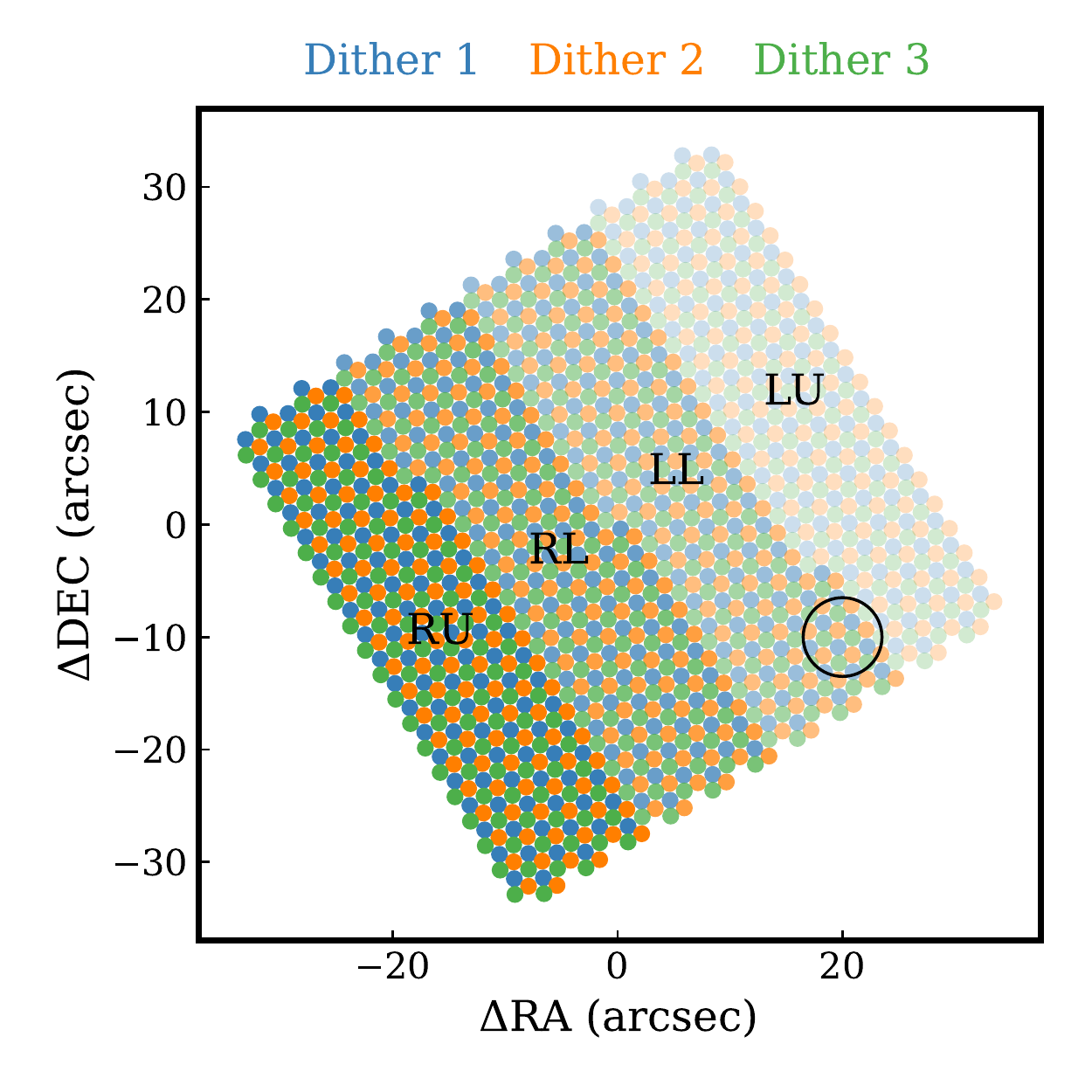}
    \includegraphics[width=3.in,trim=0 -0.5in 0 0]{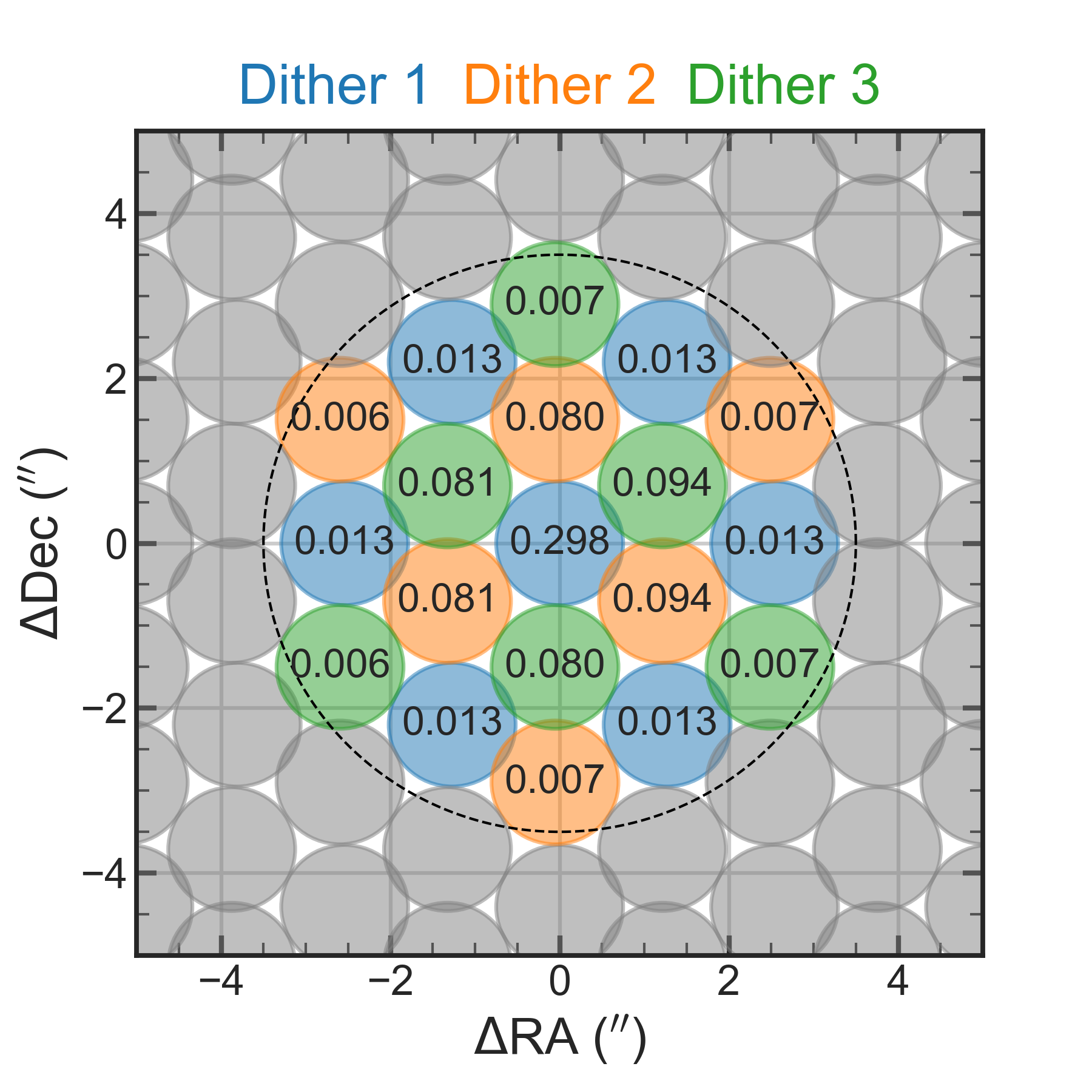}
    \caption{Fiber layout for a typical three-dither HETDEX/VIRUS observation for a single 51\arcsec\ $\times$ 51\arcsec\ IFU is shown on the left panel. The filled circles represent the fiber footprint on the sky illustrating the $1 \farcs 5$ diameter fiber locations for three separate dithered exposures as colored. The three dithers provide complete coverage of the IFU. The different shades indicate the 4 amplifier channels of the IFU. The black outlined circle indicates an example aperture used for a HETDEX PSF-weighted spectral extraction (r=3\farcs 5). The right panel provides a close-up example of a spectral extraction with the axis centered relative to aperture center. The values in each circle represent the fractional contribution of the specific fiber to the extracted summed spectrum at~4500\,\AA\ for a point-source model centered at 0,0 with $1 \farcs 8$ FWHM\null.  These fractions change with wavelength due to differential atmospheric refraction resulting in asymmetrical fiber weights, depending on the zenith direction. The dashed line is a 3\farcs 5 aperture and is the radius at which we collect fibers for extraction.}
    \label{fig:extract}
\end{figure*}
Accompanying this paper are two separate catalogs. The first we call the \textit{Source Observation Table} (columns described in Table~\ref{tab:column_info}) which is a summary of information for each HETDEX observation of a single astronomical source. Its position, classification, redshift, as well as line flux and luminosity measurements are provided for each observation. Here, the group of detections that comprise the source are reduced to one representative detection per source observation and we provide the spectrum for that detection in a separate FITS file. In addition to this aggregate table, we provide a table, called the \textit{Detection Info Table} (columns described in Table~\ref{tab:det_col_info}), which provides information for every HETDEX detection that has passed a series of quality checks and object detection criteria. Line emission detection information, such as observed wavelength and line fluxes are provided for every HETDEX detection in this table and can include a variety of line species, unlike the \textit{Source Observation Table} which is limited specifically to \lya\ and \OII\ line flux and luminosity measurements for simplicity if they are relevant for a source (e.g. a star or low-$z$ galaxy will not have an accompanying \lya\ or \OII\ measurement).

All positions reported in this paper are in the International Celestial Reference System (ICRS)\null.  We adopt the flat $\Lambda$-cold-dark-matter cosmology with $H_0=67.7\,\mathrm{km}\,\mathrm{s}^{-1}\,\mathrm{Mpc}^{-1}$ and $\Omega_{\mathrm{m},0}=0.31$ measured by \citet{Planck2018}. All magnitudes are expressed in the AB system \citep{oke1983}. We assume a rest-frame vacuum wavelength of $\lambda=1215.67$~\AA\ for Ly$\alpha$ and rest-frame air wavelength of $\lambda=3727.8$~\AA\ for the \OII\ doublet, integrated to our instrumental resolution. Observed wavelengths expressed in this paper and associated data products are as measured in air. All redshifts are appropriately calculated for any differences between air and vacuum wavelengths using the standard in \cite{Morton1991}.

\section{Observations}
\label{sec:observations}

\begin{figure*}[t]
    \centering
    \includegraphics[width=\linewidth,trim=0cm 0cm 0cm 0cm,clip]{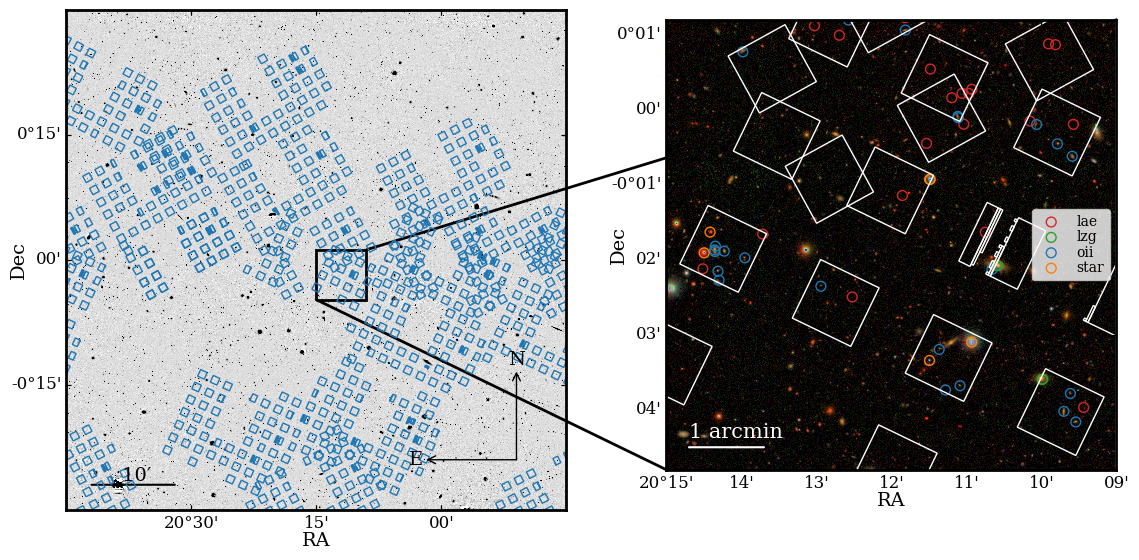}
    \caption{Left panel: Observational footprint of a $1\,\deg^2$ area of the HETDEX Fall field. The blue outlines show 51\arcsec\ $\times$ 51\arcsec\ IFU footprints. The IFU array design results in gaps in observations with a 0.22 fill factor.  Since the IFUs are oriented in the direction of the parallactic angle at the time of observation, they are oriented in differing directions. Right panel: An expanded view of a $6\arcmin \times 6\arcmin$ region of the field, superposed on a color image from the DESI Legacy Imaging Surveys \citep{Dey2019}. White squares indicate the boundaries of HETDEX IFU observations. Catalog sources, identified by our classification scheme (Ly$\alpha$ emitter, low-$z$ galaxy, \OII\  emitter, star), are coded according to the legend.  Note that one of the IFU observations was affected by a bad amplifier so that some coverage is missing in the IFU located at $20^\circ$~11\arcmin, $00^\circ$~02\arcmin.}
    \label{fig:coverage2}
\end{figure*}

The data on which these catalogs are based were all obtained in HETDEX survey observations \citep{Gebhardt2021} using the IFUs of VIRUS, the fiber-fed, multi-spectrograph instrument of the Hobby Eberly Telescope \citep{Hill2021}.  Each IFU feeds a pair of VIRUS spectrographs with 448~1\farcs5 diameter fibers positioned on a rectangular array with the fiber center separations of~2\farcs5. At a given pointing, three exposures, each typically 6-7 minutes in duration (the exposure times range from 3.6 to 12 minutes, depending upon observing conditions), are obtained; the telescope is dithered in a triangular pattern to obtain a complete fill factor for each of the \hbox{$51'' \times 51''$} IFU fields (see \citealt{Hill2021, Gebhardt2021}).  Figure\,\ref{fig:extract} provides an example IFU fiber layout for this three-dither pattern. In a single IFU observation, 1344 fiber spectra are obtained providing full sky coverage of the IFU. Also shown in varying shades of color are the four amplifiers that compose the IFU. Because each amplifier channel is fed to its own detector channel, we consider these components individually in the quality assessment of HETDEX observations. At full completion of the VIRUS instrument, its 78 IFUs cover approximately~21.7\% (a factor of 4.6) of the HET's 22$'$-diameter field~of-view.

Survey data for this catalog release come from the internal HETDEX Data Release 2 (HDR2). This release consists of $2797$ observations obtained starting in January 2017, when the VIRUS IFU assembly contained just 16 operational IFUs, and ending on 26~June~2020 when 71 IFUs were installed within the VIRUS array.  The full complement of 78 IFU units became operational in August 2021. This catalog is generated from $134,831$ IFU observations of which $124,472$ (92.3\%) pass our quality control pipeline described in detail in the following sections. A tally of IFU observations in each field is given in Table~\ref{tab:n_ifu}. The total sky coverage of the catalog is 25.0~deg$^2$.

The HETDEX footprint consists of two primary fields that allow for full-year surveying as shown in Figure~\ref{fig:coverage}. The Spring Field, labeled as the \texttt{dex-spring} field throughout this paper and in the associated catalog, covers 390\,deg$^2$ of high-declination ($\delta \sim 51^{\circ}$) sky while the Fall Field, labeled as \texttt{dex-fall} in the catalog, covers 150\,deg$^2$ along the celestial equator (see \citealt{Gebhardt2021} for full details on field selection). To reach the survey science requirements, 468,000~IFU observations are needed at the current technical specifications, resulting in~94\,deg$^2$ of complete sky coverage. In addition to the two primary fields, HETDEX obtained a number of science verification observations of COSMOS ($\sim$2.0~deg$^2$) and GOODS-N ($\sim$0.09~deg$^2$).  While most of these data were acquired using the exposure times and dithering pattern described above, several of these fields were taken with longer exposures or were visited multiple times during the survey. 

\begin{table*}[t]
    \centering
    \caption{Catalog Release Survey Statistics}
    \begin{tabular}{lllllll}
    \hline
    Field  &  field id & Center & N(IFU)   & N(IFU) & Area       & N(source) \\
           &   & (R.A., Decl.) &  Observed & Included   & {(deg$^2$)} & \\
    \hline
    \hline
  HETDEX-Spring  & dex-spring & $13^{\rm h} \ +51^{\circ}$ & 96,955 & 89,603 & 17.97 & \nspring \\
  HETDEX-Fall   & dex-fall & $1.5^{\rm h} \ 0^{\circ}$  & 35,741  & 33,001 &  6.62 & \nfall \\
  COSMOS         & cosmos & $10^{\rm h} \  +2^{\circ}$ & 1506   &  1340 &  0.27 & \ncosmos \\
  GOODS-N        & goods-n & $12.5^{\rm h} \ +62.2^{\circ}$  & 638 & 528 &  0.11 & \ngoods \\
  \hline
    \end{tabular}
    
    \raggedright Note: Listed is the number count of IFU observations observed and the count included after observation quality inclusion criteria. The field ID is the string match to find each field in the catalog.
    \label{tab:n_ifu}
\end{table*}

HETDEX observations are expected to be completed in 2024, and eventually cover 540 deg$^2$ with partial fill. The final effective sky coverage is expected to be about 94 deg$^2$ with non-contiguous tiling over the two main HETDEX fields in combination with the 21.7\% fill factor of the VIRUS IFU array. Figure~\ref{fig:coverage} shows the survey boundaries in red and the source positions in blue, which effectively map out the HETDEX IFU field boundaries in this release. Examples of IFU boundaries are overlaid in blue over DESI Legacy Imaging Data \citep{Dey2019} in the left panel of Figure~\ref{fig:coverage2} for a $1\deg^2$ region in the HETDEX-Fall field. The right panel zooms into a much smaller $6\arcmin \times 6\arcmin$ region, indicating the IFU boundaries in white and source positions of HETDEX sources as described in the legend. This cropped region covers only a quarter of the HET FOV.  Overlapping observations are seen from two independent observations (taken at different HET track angles). Overlapping IFUs such as this provide valuable repeated observations for validation tests discussed later in Sections\,\ref{sec:counterpartfraction} and \ref{sec:sample_validation}.

The data processing of HETDEX frames is detailed in \citet{Gebhardt2021}. Briefly, bias frames, pixel flats, twilight sky flats, and the background on the science frames themselves are used to produce a wavelength calibrated, sky-subtracted spectrum for each fiber in the array.  Astrometric calibrations are achieved by measuring the centroid of each field star from fiber counts between 4400\,\AA\ and 5200\,\AA\ and comparing their IFU positions to the stars’ equatorial coordinates in the \sdss\ \citep{sdss2000} and GAIA \citep{Gaia2018} catalogs.  This process typically results in global solutions that are accurate
to $\sim 0\farcs 2$.  The absolute flux calibrations are produced by using $g < 24$ \sdss\ field stars as \textit{in situ} standards and using their $ugriz$ colors \citep{Padmanabhan2008}, \textit{Gaia} parallaxes \citep{Gaia2018}, and foreground reddenings \citep{Schlafly2011} to determine their most likely spectral energy distribution in a grid of model spectra \citep{Cenarro2007, Falcon-Barroso2011}.  The final system throughput curve is derived from the most likely flux distribution of $\sim 20$ stars, and is generally accurate to $\sim 5\%$ \citep{Gebhardt2021}.

\subsection{Data Quality Control}

\label{sec:quality}

An accurate description of sky sensitivity and coverage is essential for HETDEX. Each IFU consists of 448 fibers which are divided into two spectrograph channels.  Each channel has a \hbox{$2064 \times 2064$} detector.  We use two amplifiers per spectrograph channel and bin $2\times$ in the spectral direction during readout. Thus, each IFU consists of four amplifier channels, labeled "RU", "RL", "LL", "LU" as demonstrated in the left panel of Figure\,\ref{fig:extract}. Each amplifier generates a FITS image that is \hbox{$1032 \times 1032$}, each with 112 fiber spectra. With a full 78 IFU installation, each single exposure consists of  data from 312 CCD amplifiers, which corresponds to about 35,000 fiber spectra. Our standard three-dithered observation set generates 936 FITS files, each an image of a single amplifier, and 104,000 fiber spectra. Although the IFU spectrographs are designed to be identical, in practice, there are important variations from amplifier to amplifier that we track (see, for example, Figure 6 from \citealt{Gebhardt2021}). Over its lifetime, including calibration frames, HETDEX will generate about 20 million FITS files. Each one of the amplifier images consisting of 112 fiber images is considered individually for quality assessment.

Over the first three years of the HETDEX survey, we have seen a variety of detector and calibration issues. These include dead amplifiers, variable electronic noise, low count rates, scrambled pixels, among others. Calibration issues include vignetting of some IFUs, saturation problems from bright objects, astrometric uncertainties from fields with low number of stars, large variation in throughput over a dither set, large variations in the wavelength solution for some spectrographs, among others. We robustly track these issues and find that for a given exposure set, about 92\% of the FITS files are useful and make it into the catalog. This percentage has increased over the years as we have fixed various detector issues, and we expect an even higher rate of return in the future.


\subsubsection{Detector Issues}

Instrument deficiencies can result in a number of failures. Specific detectors may exhibit low response or spatially distorted features that cannot be removed by flat field corrections.  These failures can vary with time and significantly impact our detection methods. Certain failures result in the creation of many false detections and can dramatically affect our detected sample. Building from an initial sample that was visually flagged, we have developed a set of criteria from statistics generated by our image calibration that automatically identifies substandard amplifier readouts.  These criteria are summarized in Table~\ref{tab:amp_quality}. We provide the quality inclusion criterion limits for each statistic and a short description. We also indicate the fraction of amplifiers that pass each criteria. Ultimately, an amplifier must satisfy all of the criteria to be included.  Unfortunately not all issues can be caught automatically, and an extensive list of detector issues for each detector are maintained\footnote{\url{https://github.com/HETDEX/hetdex_api/tree/master/known_issues}} so that both the catalog and the survey selection function are consistently masked. For this catalog release, about 92\% of the FITS files pass our quality control, although we note that the first year of observations were particularly poor with the fraction of usable data below 90\%. The quality fraction generally averages about 94\% in recent years. An example where a specific amplifier is removed from the survey is shown in the right panel of Figure~\ref{fig:coverage2}. The IFU at roughly $20^\circ~11\arcmin$, $00^\circ~02\arcmin$ has a single amplifier masked out of the catalog.

\subsubsection{Calibration Failures}

Science frames that cannot be calibrated to the HETDEX specification \citep{Gebhardt2021} are also removed from our catalog. These data are usually produced by the presence of bright stars or large galaxies on or near an IFU\null. While the HETDEX tiling attempts to avoid the very brightest stars, the spectra of objects brighter than $g\sim14$ will typically saturate a detector, and flood nearby fibers (on the detector) with excess signal. The counts in these fibers cannot be flux calibrated.   The criteria set out in the previous section and summarized in Table~\ref{tab:amp_quality} also help to automatically remove any frames which have calibration issues. 

\begin{table*}[t]
    \centering
    \caption{Statistics in Image Processing Used for Amplifier Quality Assessment} 
    \begin{tabular}{p{0.2\linewidth}  p{0.1\linewidth}  p{0.6\linewidth} }
    \hline
    Quality Criteria & Quality Fraction & Description \\
    \hline
    \hline
    im\_median $> 0.05$  & 98.7\%  & Median counts in unprocessed amplifier image frame \\ 
    $-10<$ background $<100$ & 98.0\% & Median counts value in sky-subtracted image \\
    $0.2 <$ sky\_sub\_rms $< 1.5$ & 99.3\% & RMS in sky-subtracted image counts \\
    sky\_sub\_rms\_rel $<1.5$ & 97.7\% & Ratio of sky RMS in individual amplifier relative to all other amplifiers in all IFUs within the same exposure\\
    n\_cont $<35$ & 98.7\% & Number of fibers above a certain counts threshold. A good indication of an amplifier saturated by a bright star or nearby galaxy. \\ 
    norm $<0.5$ & 99.3\% & Relative normalization for a dithered exposure. \\
    maskfraction $<0.2$ & 98.3\% & Rejected if more than 20\% of the frame is masked \\   
  \hline
  \end{tabular}
   \raggedright Note: Inclusion quality criteria for each image statistic, the fraction of amplifiers that pass the criteria and a short description are listed. Each HETDEX IFU is fed to four amplifiers, each containing spectral information for 112 fibers.  For HETDEX survey data, we consider each amplifier to be an independent observation with its own quality criteria.
    \label{tab:amp_quality}
\end{table*}

\subsubsection{Observation Quality Criteria}

Each dither in a HETDEX observation is individually flux calibrated, as there may be small differences in their relative throughputs due to variations in the  observing conditions. For the HETDEX catalog we require that a nominal throughput, assuming a 360\,s exposure time, must be greater than~0.08, and that the relative throughputs of each dithered exposure cannot differ by more than a factor of three. The most common reason for rejection by this criterion is a significant drop in transparency during the third dithered exposure when clouds drifted into the FOV.

\subsubsection{Pixel Masks}

As described in \citet{Gebhardt2021}, several detectors have significant features, including large dust spots, many charge traps, and a “pox” contamination where the quantum efficiency of individual pixels can be suppressed by 10\%-40\%. While the flat-field calibrations can identify many of the worst features automatically, many low-count defects remain in the data and can produce false-positive line detections. 

For each pixel, we track the sky-subtracted residuals divided by the sky at that location. We then average over all observed fields (a few thousand in this case), and generate the scatter of the residuals for each pixel. We use these “residual maps” to highlight regions that have poor or variable sky subtraction. In this way, flat-field defects, charge traps, and pixel defects show up more easily. Pixel masks are then created from the visual inspection in these residual science frames. Additionally, charge traps related to a deficiency of counts are identified as vertical features in the detectors with a width of one pixel.  (They can start at any $y$-position on the detector and either continue to the top or bottom of the frame depending on the readout direction.)  A mask three pixels wide, centered on the affected line and covering the length of the defect, is then applied to the two-dimension fiber data frame, and propagated in the one-dimensional flux-calibrated fiber spectra.

\subsection{Large Galaxy Masks}

Galaxies larger than roughly 1$'$ are excluded from our catalog using the positions and optically-defined elliptical apertures provided by the Third Reference Catalog of Bright Galaxies \citep[RC3;][]{rc3} and the Uppsala General Catalogue of Galaxies \citep[UGC;][]{1973ugcg.book.....N}.  The Spring field contains 644 such galaxies; the fall field, 447.  For each galaxy, we use the catalogs' basic parameters for position, position angle, ellipticity, and D25 semi-major axis (i.e., the size of the galaxy defined by its $B$-band isophote at 25.0 mag~arcsec$^{-2}$). Each galaxy is visually inspected through photometric imaging to confirm that these default values are reasonable. Where the parameters are uncertain, they are corrected to values listed in the NASA/IPAC Extragalactic Database\footnote{\url{http://ned.ipac.caltech.edu}} or through visual inspection of the galaxy in SDSS $g$-band images. All detections that fall within $1.5\times$ the D25 scale of a bright galaxy's elliptical aperture are removed. This factor was determined by examining the HETDEX spectra at different scalings and ensuring that all detections related to the bright galaxy were encompassed in the aperture mask. These galaxy masks are consistently applied to the HETDEX catalog and accompanying survey area mask through the HETDEX python-based, software repository \texttt{hetdex-api}\footnote{\url{https://github.com/HETDEX/hetdex_api}} to provide proper survey volume accounting. 

\subsection{Satellites and Meteors}

Both satellites and meteors generate signals that produce detections in both our emission-line and continuum emission catalogs. Meteors largely appear as line emission at multiple wavelengths and therefore contaminate our LAE samples because of their lack of strong continuum underlying the emission. Fiber spectra from HDR2 contain at least 31 meteors resulting in thousands of spatially extended emission-line detections, as the meteor travels across the HET focal plane. We use a systematic search method for these objects as part of our Emission Line Explorer software tool (\elixer; \citealt{Davis2022}). Strong line emission appearing in just a single dithered HETDEX observation is flagged as a meteor candidate. For any observation with over ten associated meteor candidate emission-line detections, we visually inspect the detections to confirm the presence of the meteor. We create a simple linear mask by fitting to the positions of the flagged meteor detections. This mask extends 12\,\arcsec\ above and below the linear fit to the meteor positions; in many cases, a smaller mask could be used, but this width is needed for the brightest events.  We therefore chose to be conservative with this mask. This linear mask is consistently applied to both the line emission and continuum emission raw catalogs as well as to our survey area mask. 

Satellites are identified when the continuum flux density measured in the HETDEX spectral data differs from that estimated within photometric imaging data (see Figure~3 in \citealt{Gebhardt2021} for an example of a satellite trail across the HETDEX FOV). Each HETDEX detection is processed with the \elixer\ software tool, in which forced aperture photometry is performed on all available imaging. If any reported photometric measurement is more than two magnitudes fainter than the measured HETDEX value, the source is rejected as it indicates the signal is from a transient source, such as a satellite, only briefly observed at that point in the sky. Visual inspection confirmed that the majority of these detections are indeed satellites, scattered light from bright star or artifacts caused by improper flux calibration. For more discussion about finding transient sources in HETDEX see \citet{Vinko2022}.

\section{Catalog Generation}
\label{sec:catalog}

\begin{figure}[t]
    \centering
    \includegraphics[width=3.5in]{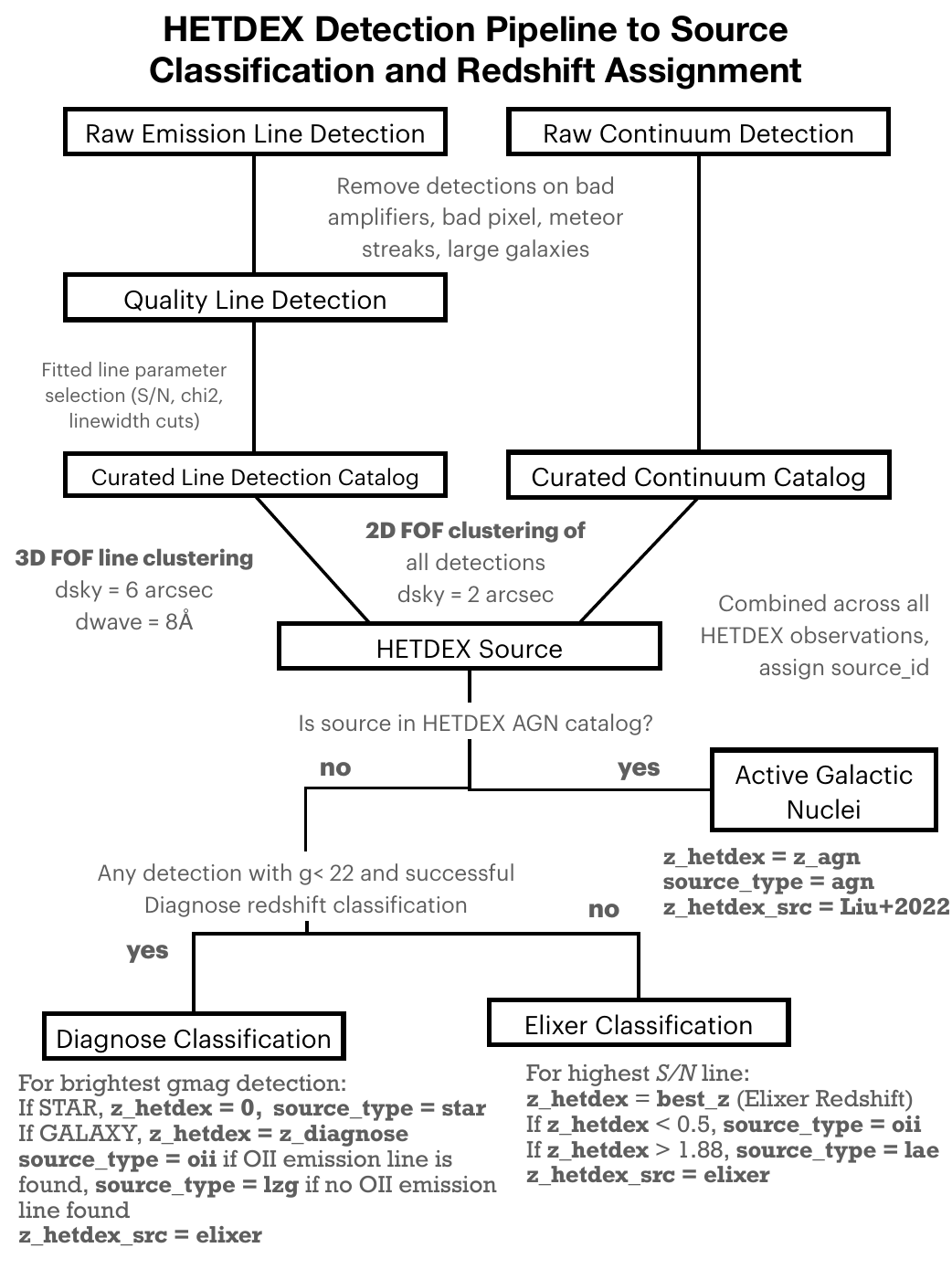}
    \caption{The steps and decisions made to generate the HETDEX source catalog from the raw detection pipelines. A HETDEX source is created from detection grouping in both 3D and 2D space through friends-of-friends (FOF) clustering to create unique sources on the sky. Multiple observations of the same source are assigned just one identifier (\texttt{source\_id}).  Described in detail in Section\,\ref{sec:classify}, each source is assigned a redshift and classification. If any detection has been identified as an AGN from \citet{liu2022}, it is assigned the redshift from the AGN catalog and is classified as an AGN\null. If a source has \hetg\  brighter than 22 mags, the \texttt{Diagnose} classification and redshift are assigned to the source; otherwise the \elixer\ redshift and classification is used. }
    \label{fig:flowchart}
\end{figure}

With the HETDEX data frames reduced and verified, the data is organized into a database of flux-calibrated, one-dimensional fiber spectra each with their own corresponding sky coordinate. In this section, we describe the steps taken to create a catalog of astronomical sources from initial object detection to final source identification. This process involves assessing data quality, as outlined in the previous sections, performing a grid search for potential object detection, then reducing the initial raw databases of potential line and continuum emission detections into high-quality detections. These two independent catalogs of high-quality detections are combined to create a single list of astronomical sources through detection grouping.

The flowchart in Figure~\ref{fig:flowchart} illustrates the process of producing a source classification from the raw line and continuum emission detection pipeline. This section describes the steps from detection to source object, including spectral extractions from the IFU data, detection search methods and line parameter measurements. We also detail our method of deriving spatially resolved line fluxes for resolved sources which are applied to the low-$z$ galaxy sample exclusively. Following this section, we describe source type identification and redshift assignment in detail in Section\,\ref{sec:classification}.


\subsection{Object Detection}
\label{sec:detection}
Two independent, but complementary, object detection search techniques are performed as part of the main HETDEX reduction pipeline: one to identify emission lines, the other to detect continuum sources.  In the second internal data release for HETDEX (HDR2) a search was performed across 210 million flux-calibrated fibers as described in detail in Section 7 of \citet{Gebhardt2021}. We briefly summarize the procedures here. During this process, no imaging pre-selection is used and the HETDEX data itself provides object detection. Both the emission line and continuum detection algorithms are designed to identify point-sources and account for the variable image quality, or point-spread-function (PSF) of each independent HETDEX three-dither observation. To move from object detection to source classification, the output from both object detection methods are combined as described in Section~\ref{sec:det_group} below.

\subsubsection{Spectral Extraction}
\label{sec:extract}
Each $51'' \times 51''$ IFU observation consists of 448 fibers $\times$ 1036 spectral elements $\times$ three dithers. A demonstrative example of the fiber layout is found in the left panel of Figure\,\ref{fig:extract}. Each IFU is searched individually for line emission in a grid of one-dimensional (1D), PSF-weighted spectral extractions. A single fiber alone does not provide evidence for line emission; instead we assume that the signal-to-noise ($S/N$) ratio of an object can best be measured in a collection of nearby fibers rather than individual fibers.
We therefore use the collection of all fibers within a $3\farcs 5$ radius aperture about a candidate line. The image quality of the observation, assumed to be described by a symmetric two-dimensional Moffat function ($\beta = 3.0$, \citealt{Moffat1969}), assigns the weights to each fiber in an aperture according to the optimal extraction algorithm of \citet{Horne1986}.

An example $3\farcs 5$ radius aperture is displayed in both panels in  Figure~\ref{fig:extract}. The text in each circle in the right panel displays the fractional flux contribution from each individual fiber for the case of average HETDEX image quality ($1\farcs 8$) with a detection centered on the central fiber. The fraction that each fiber contributes depends on the location of each fiber with respect to the aperture center, the wavelength (due to the effects of atmospheric diffraction), and the measured image quality PSF\null. In this extraction example, the central seven fibers contribute~80\% of the extracted flux at~4500\,\AA. 

The fiber weights are dependent on image quality. For the best HETDEX image quality ($\sim 1\farcs2$), a fiber centered on the source contains more than 50\% of the flux; the poorest-quality HETDEX observation ($\sim 2\farcs 5$) has 10\% of the flux in the central fiber. As the HET does not have an atmospheric dispersion corrector, the weighted signal contribution from each fiber varies as a function of wavelength due to differential atmospheric refraction (DAR)\null. Since the HET is a fixed-altitude telescope, the magnitude of the differential refraction is essentially constant for all observations. As described in \citet{Gebhardt2021}, our data demonstrate that from 3500 to 5500 Å, a source position moves by $0\farcs 95$.

\subsubsection{Line Emission Search}
\label{sec:line}

\begin{figure}[t]
    \centering
    \includegraphics[width=3.5in]{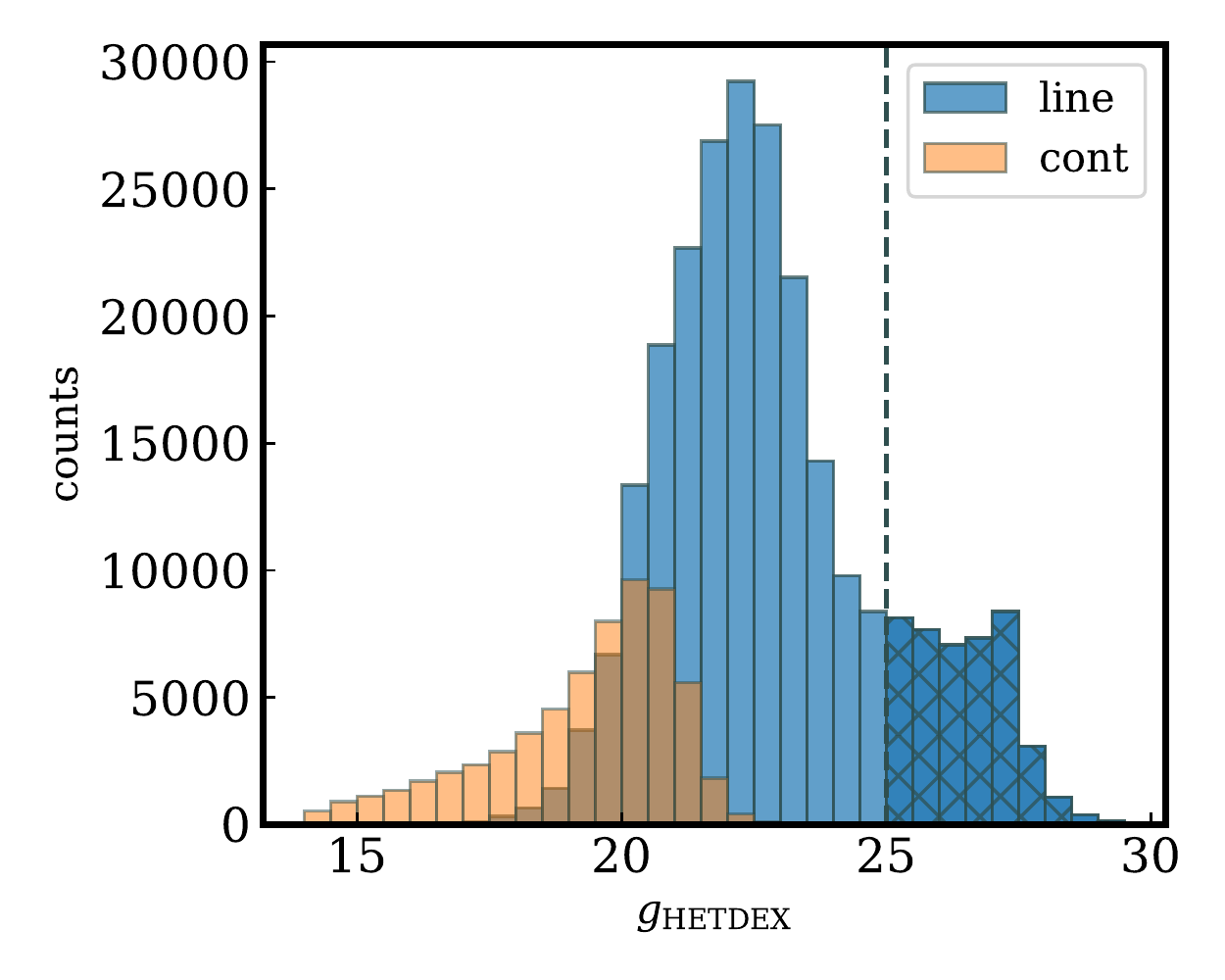}
    \caption{Distribution of emission line detections and continuum detections as a function of \hetg, a pseudo magnitude calculated by integrating the extracted spectrum at the detection location, weighted by the \sdss\ \gp~filter transmission curve.  At magnitudes brighter than 22.5 mag continuum and line emission detections overlap. The bimodality in the line emission detections is due to  \OII-emitting galaxies and LAEs being the dominant samples HETDEX can detect. The vertical dashed line and hatched region at \hetg=25 represents the average HETDEX continuum sensitivity limit.}
    \label{fig:dettype}
\end{figure}

The initial grid search for an emission line is performed in steps of 0$\farcs$5 in the spatial direction and steps of 8\,\AA\ in the spectral dimension, guided by the simulations described in \citet{Gebhardt2021}. At each grid step, a Gaussian line profile with the instrumental line width ($\sigma$=1.7\,\AA) is used for the initial fit. Continuum emission is subtracted by fitting a constant intensity value to the spectrum $\pm 50$\,\AA\ around the Gaussian's central wavelength. The signal-to-noise of the line fit is measured by integrating the flux in the Gaussian model out to $\pm 3.5$\,\AA\ around the central wavelength, then dividing by the noise, which is the quadratic sum of the uncertainties in the same wavelength range. All line fits with $S/N>4.0$ and $\chi^2<3.0$ are submitted to the next stage of line fitting to better constrain the line parameter measurements.

For the emission-line candidates identified in this first search, an optimized grid of spectral extractions is performed at a higher ($0\farcs 15$) rastering resolution, with a Gaussian line width, $\sigma$, that is now allowed to be a free parameter. The location within the raster that provides the highest $S/N$ of the emission line is assumed to be the true source position. The amplitude of the Gaussian fit then yields the measured continuum-subtracted line flux. In the case of duplicate detections (defined as emission lines lying within 3\arcsec\ and 3\,\AA\ of each other), only the line-fit with the highest $S/N$ detection is accepted. The resulting line-fit parameters, including the measured observed line flux (\texttt{flux\_obs}), the continuum measurement (\texttt{continuum\_obs}) line width, (Gaussian $\sigma$ -- listed as \texttt{sigma}), and quality of fit ($\chi^2$ -- listed as \texttt{chi2}), are included in the \textit{Detection Info Table} described in the Appendix.

\begin{figure*}[ht]
    \centering
    \includegraphics[trim=2 8 2 2,clip, width=6.7in]{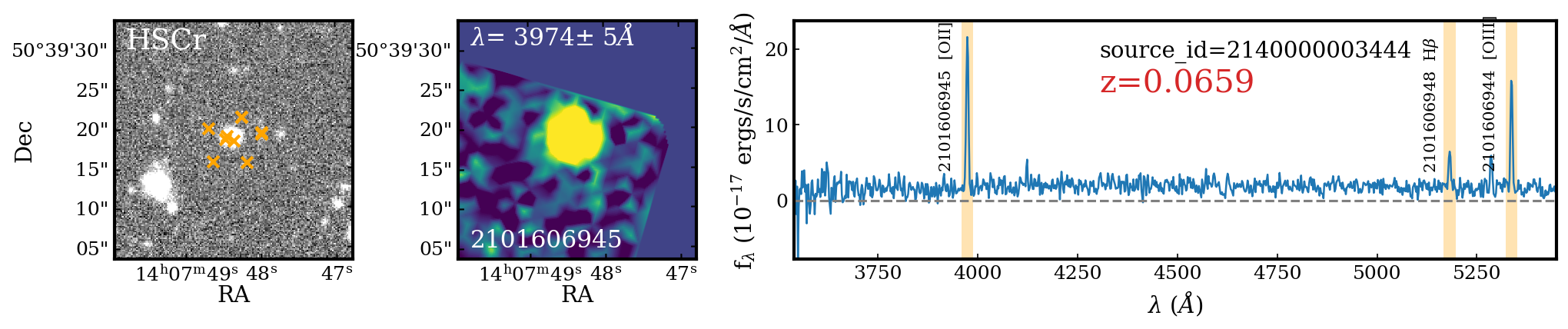}
    \includegraphics[trim=2 8 2 2,clip, width=6.7in]{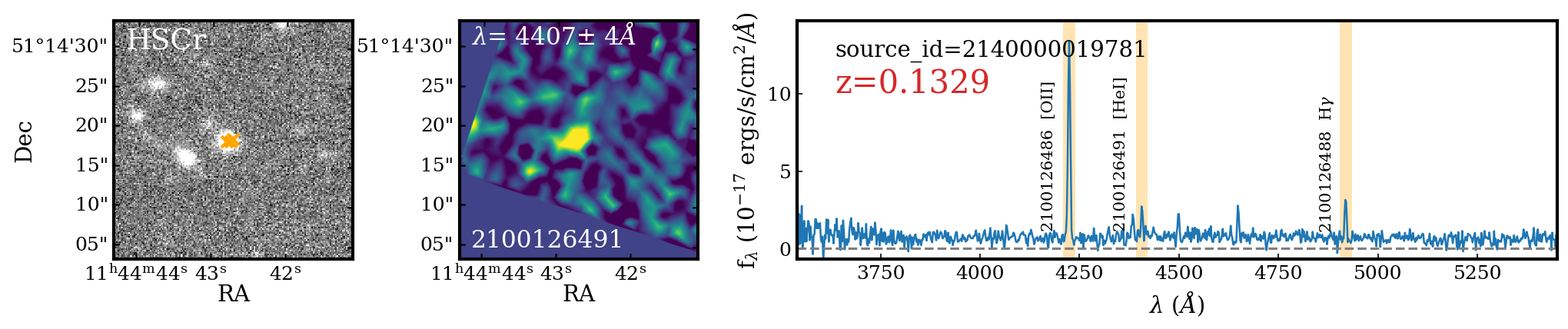}
    \includegraphics[trim=2 8 2 2,clip, width=6.7in]{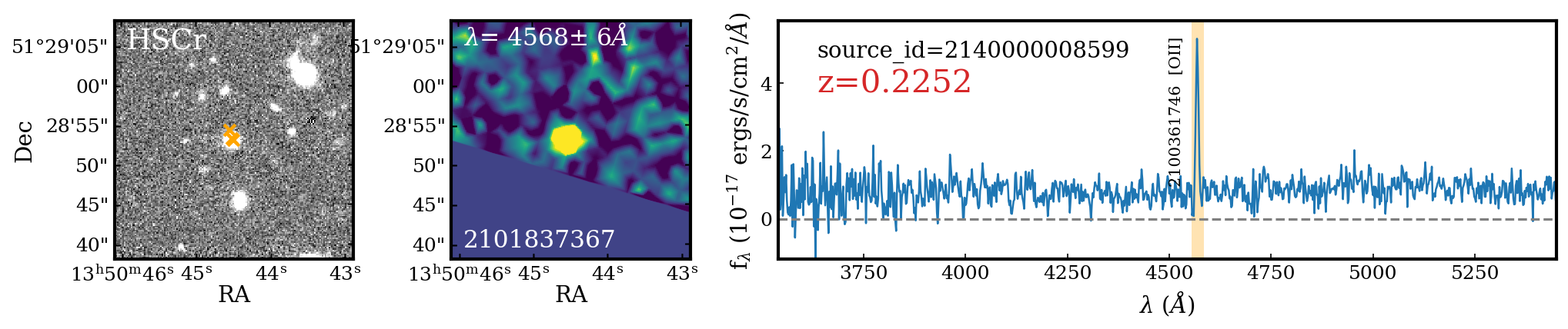}
    \includegraphics[trim=2 8 2 2,clip, width=6.7in]{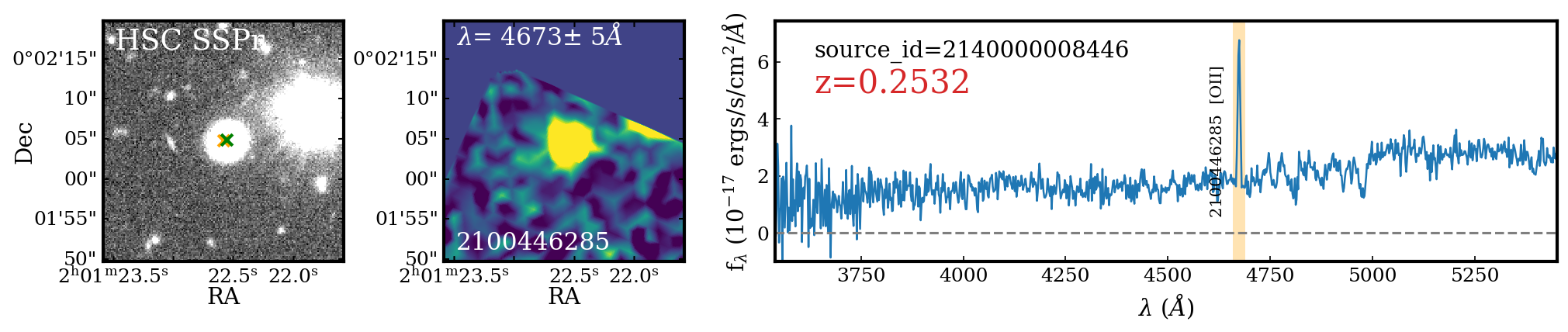}
    \includegraphics[trim=2 8 2 2,clip, width=6.7in]{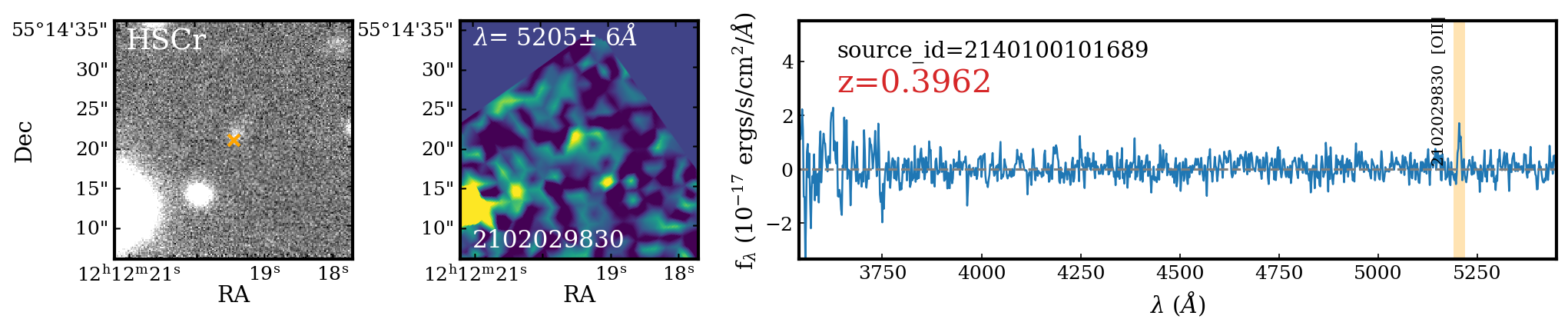}
    \includegraphics[trim=2 8 2 2,clip, width=6.7in]{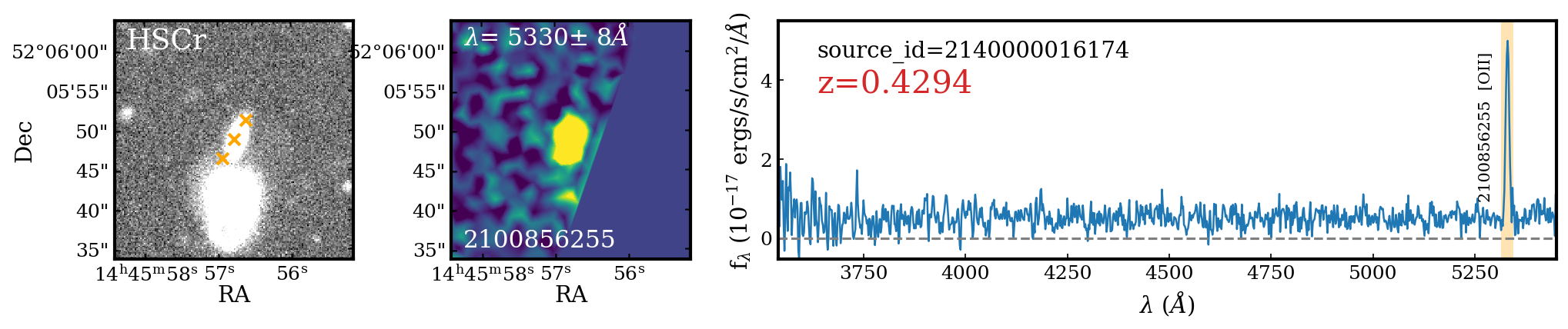}
    \caption{Example low-$z$ sources in the catalog in order of increasing redshift.  For each object, the left panel displays a~$30'' \times 30''$ HSC-$r$ band image; the middle panel is a wavelength-collapsed, continuum-subtracted, flux map of line emission (same area and orientation as left panel), at the HETDEX detected line indicated in the text. Solid blue in the line flux map indicates area that is not covered by an IFU (i.e.,  no HETDEX data exists).
    The right panel presents the HETDEX spectrum for the best detection for the source (indicated by \texttt{selected\_det==True} in the \textit{Detection Info Table} described in Appendix\,\ref{appendix:1}).
    Text on this panel indicates the \texttt{source\_id}, the individual emission-line detections (as indicated by the vertical \texttt{detectid} text), and the HETDEX redshift, \zhet. Multiple detections can often arise from a single source if they possess both continuum and line emission. Extended emitters can appear multiple times in the detection catalog as seen in the top two examples. Note that some emission lines are missed from the catalog in the second spectrum from the top where a single Gaussian model cannot sufficiently identify H$\delta$. The position of each detection is indicated in the left-hand column images as orange and green crosses, for line and continuum detections, respectively.}
    \label{fig:spec_exampleslowz}
\end{figure*}

\begin{figure*}[ht]
    \centering
    \includegraphics[trim=2 8 2 2,clip, width=7in]{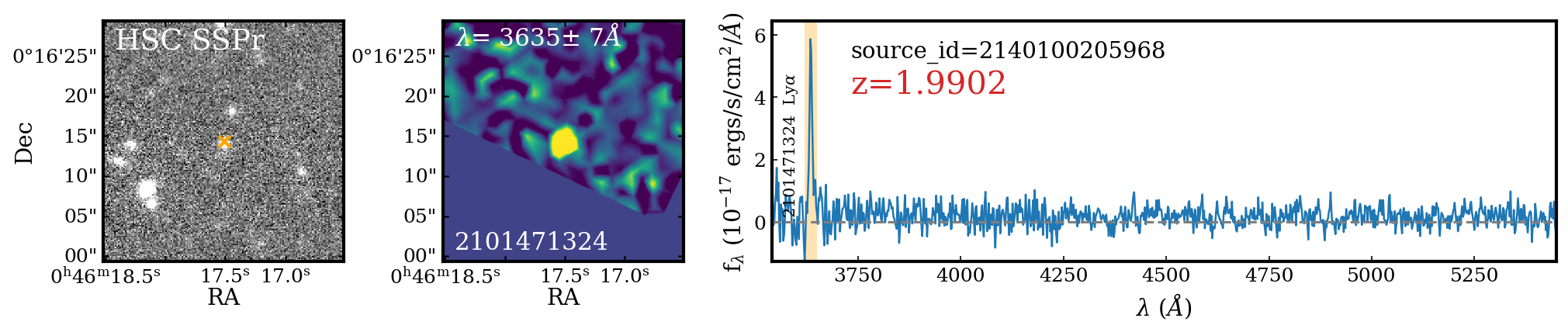}
    \includegraphics[trim=2 8 2 2,clip, width=7in]{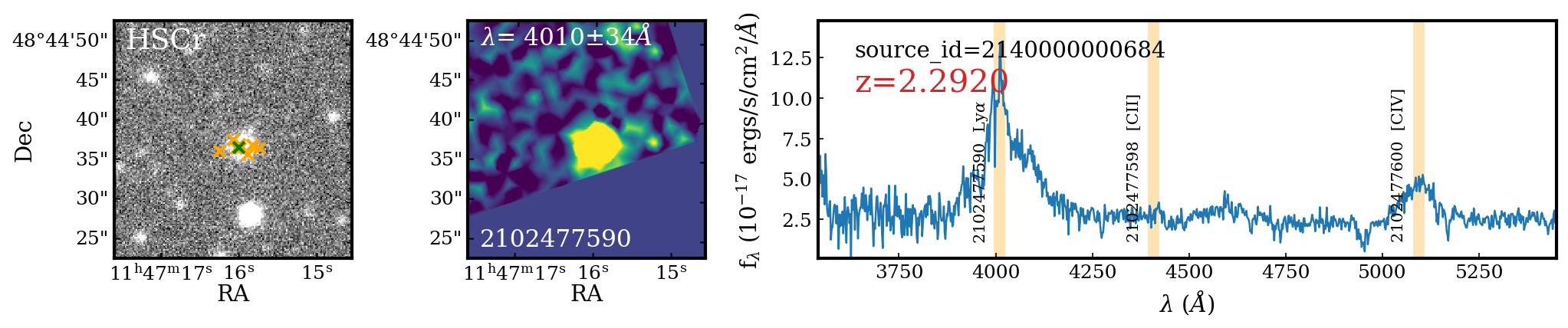}
    \includegraphics[trim=2 8 2 2,clip, width=7in]{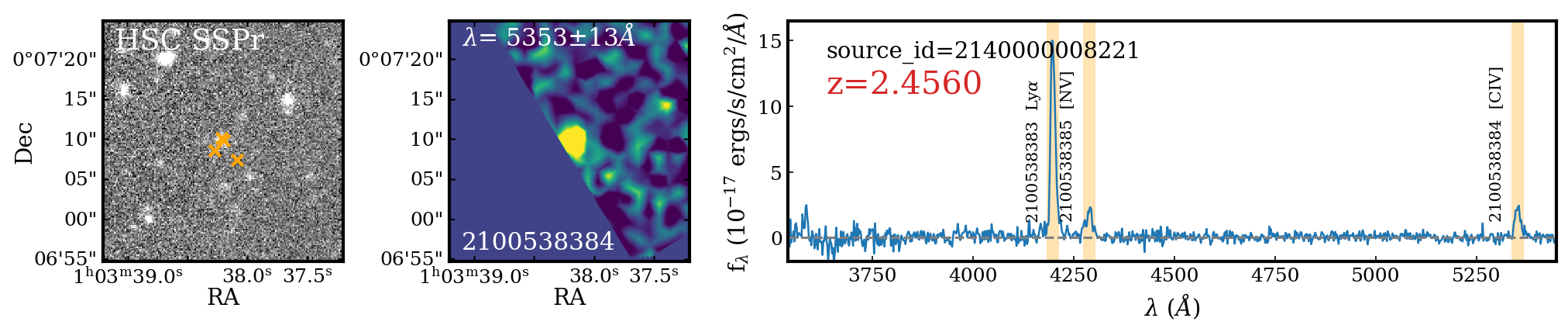}
    \includegraphics[trim=2 8 2 2,clip, width=7in]{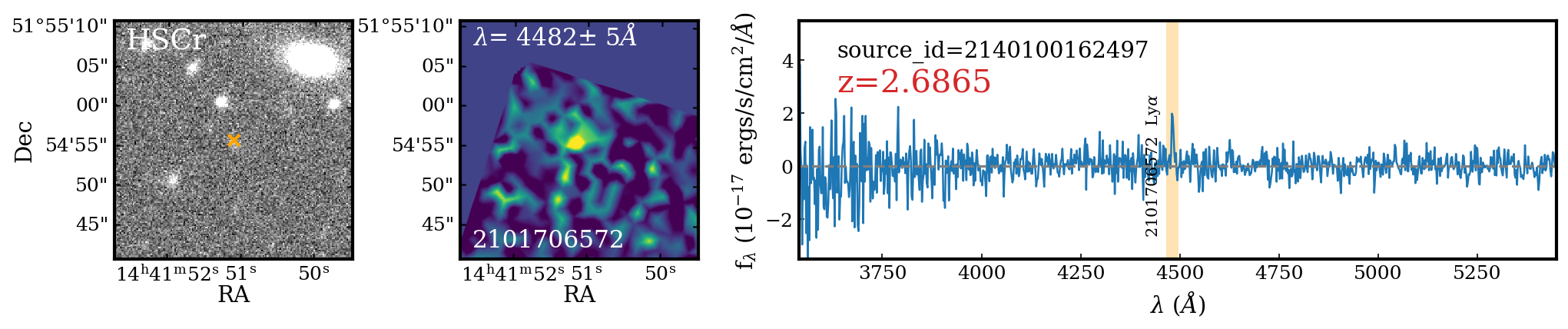}
    \includegraphics[trim=2 8 2 2,clip, width=7in]{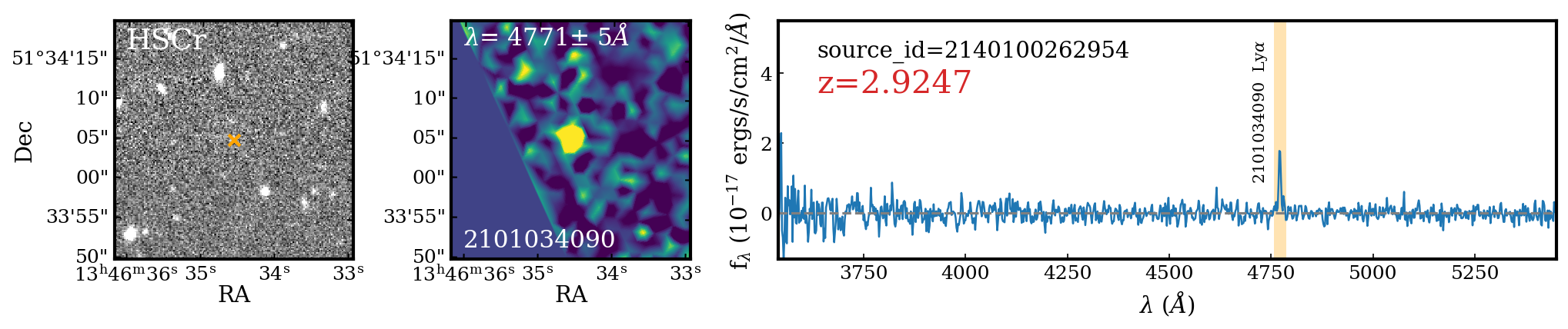}
    \includegraphics[trim=2 8 2 2,clip, width=7in]{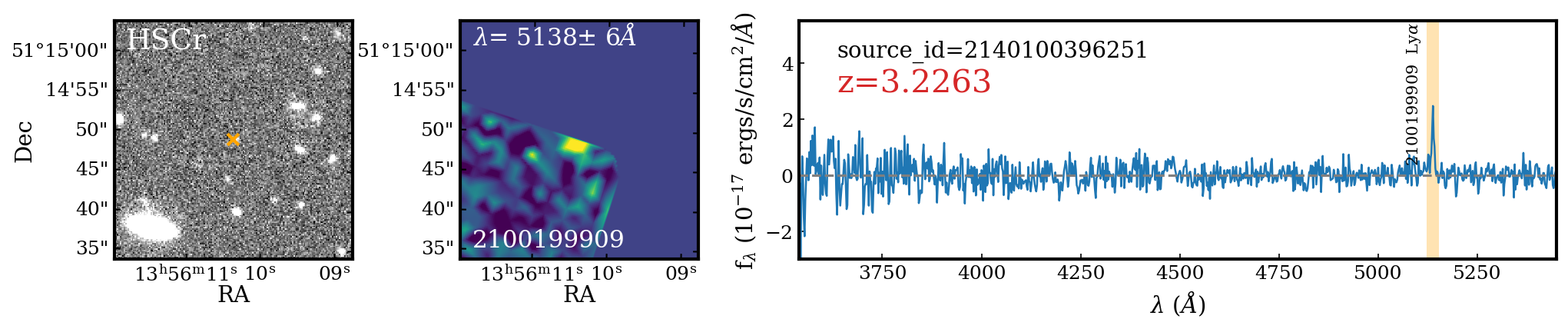}
    \caption{Examples of high-$z$ HETDEX sources in increasing redshift order. The panels are described as in Figure~\ref{fig:spec_exampleslowz}. The second row is an AGN exhibiting broad-line emission;  the third row is an AGN with narrow-line emission. Both of these objects exhibit multiple line emission detections as shown by the orange crosses in the left image.  The broad AGN is also a continuum emission detection source (indicated as a green cross in the left image). The remaining objects in the figure are typical LAEs in the catalog, i.e., pure line emission sources, and are frequently found without an imaging counterpart.}
    \label{fig:spec_exampleshighz}
\end{figure*}

\subsubsection{Emission-Line Fits and Criteria}
\label{sec:linecriteria}
The raw emission-line database is produced from all available HETDEX observations and consists of both real sources and artifacts that can arise from the data quality issues described in \ref{sec:quality}. Each raw line detection is subjected to a series of tests, which check whether the candidate line is close to a large galaxy, a meteor trail, a known detector feature, or subject to a poorly performing amplifier.  A criterion is then applied, based on the $S/N$ measured in the continuum-subtracted emission line, the Gaussian width of the fitted line in Angstroms, $\sigma$, and the quality of line fit, $\chi^2$, measured in a $\pm2\times\sigma$ wide spectral window.  Specifically, to be classified as a detection, a line must satisfy either
\begin{equation}
    5.5 < S/N < 6.5~~\&~~1.7\,\text{\AA}<\sigma<6\,\text{\AA}~~\&~~\chi^2<1.2
    \label{eq:criterion-1}
\end{equation}
or 
\begin{equation}
    S/N > 6.5~~\&~~1.7\,\text{\AA} <\sigma <14\,\text{\AA}~~\&~~\chi^2 <2.5~~\&~~g>19
    \label{eq:criterion-2}
\end{equation}
where $g$, hereafter labelled \hetg, is an equivalent broad-band photometric measurement obtained by summing up the flux densities in the HETDEX spectrum, weighted by the \sdss\ $g$-band filter curve.

This combination of constraints means that high line width sources can have poorer fits (i.e., a higher $\chi^2$) if they also have a relatively high $S/N$ and faint \hetg. Sources in the high line width regime tend to suffer from a higher contamination rate, due to calibration issues or the existence of broad continuum emission from nearby galaxies and late-type stars.  However, the broad-line identifications do contain interesting high-$z$ sources, including AGN \citep[see the HETDEX AGN Catalog;][]{liu2022} and extended Ly$\alpha$ emitters (Mentuch Cooper \etal, in prep). We therefore allow a more liberal $\chi^2$ quality of fit for these objects, as their lines are not typically well-described by a single Gaussian line model (especially in the case of AGN\null).  Additional to the above criteria, narrow emission line detections are cataloged in the wavelength range between 3510 and 5490\,\AA\, while broad-line features are only catalogued between 3550 and 5460~\AA, as many spurious high line width sources were identified by the detection software on the spectral edges.

Detector artifacts are a major issue with some HETDEX spectra.  In some cases, the fiber spectrum is poorly calibrated, resulting in measured continuum flux
that is negative, leading to a false positive detection. To mitigate this issue, we apply a lower cut of $-3\times$\fluxden\ to the local continuum measured in the Gaussian line fit. Additionally, the fiber profile quality of fit can also help identify detector artifacts. If the quality fit of the fiber profile solution, $\chi^2_\mathrm{fiber}$, is high, we remove the detection. In practice this value is measured for each of the five highest weight fibers in an aperture extraction, 5\,\AA~above and below the central wavelength of the detected emission line. If any of the fibers in this spectral window have a $\chi^2_\mathrm{fiber}>4.5$ or $\chi^2_\mathrm{fiber}>3$ and \hbox{$\mathrm{continuum}<0.5\times10^{-17}$ ergs s$^{-1}$ cm$^{-2}$ \AA$^{-1}$} the detection is removed from further consideration. We opted for this dual criterion because fibers with a significant continuum signal can produce higher reduced $\chi^2_\mathrm{fiber}$ values;  we are particularly concerned with finding artifacts in the low continuum regime, where their presence can lead to false-positive LAE candidates.

Our final curated line emission detection catalog consists of \linedetcount\ line emission detections. They can be identified in the \textit{Detection Info Table} (see Table~\ref{tab:det_col_info}) in the column \texttt{det\_type==line}. The line flux sensitivity limit for a HETDEX line emission source depends on observed wavelength, image quality, exposure time, other observing conditions, and instrument component inconsistencies, but, on average, 50\% completeness is reached at roughly $\sim 7 \times 10^{-17}$ erg/s/cm$^{2}$. The reader is referred to Section\,8.2 of \citet{Gebhardt2021} for a detailed discussion on HETDEX emission line sensitivity and completeness with an updated discussion on these topics and the HETDEX selection function to be presented in \citet{Farrow2023}.

\subsubsection{Continuum Emission Search}
\label{sec:cont}

For each of the 448 fibers in an IFU, the detector counts are measured in two 200\,\AA\ regions: one in a blue region of the spectrum (from 3700 to 3900\,\AA) and one in a red region of the spectrum (from 5100 to 5300\,\AA\null). If either region contains more than 50 counts per 2\,\AA\ pixel on average (corresponding to  $g \sim 22.5$), it is collected as a possible continuum source. The 50-count limit is arbitrary and designed to be conservative (future HETDEX catalogs reach significantly lower
fluxes, as objects can be detected more than two magnitudes fainter than this limit). Once we detect a possible source, we search about the fiber position, using a $15\times15$ element raster with 0\farcs1 spatial bins. The spatial location that achieves the lowest $\chi^2$ fit to our PSF model defines the center of source (as opposed to the line emission search, which peaked up on the $S/N$ of the Gaussian fit), and a point-source extraction at that position provides the detection spectrum.

Each continuum source undergoes a series of quality checks to ensure it lies on a high quality detector and is not flagged as an artifact or satellite. The final curated continuum emission detection catalog consists of \contdetcount\ detections. These detections can be identified in the \textit{Detection Info Table} (see Table~\ref{tab:det_col_info}) in the column \texttt{det\_type==cont}. The sensitivity of the continuum catalog is based on a photon count threshold and depends on observing conditions and exposure time. On average the HDR2 continuum detection sensitivity is equal to \hetg~$\approx$~22.5\,mag. But we note that the counts threshold can be adjusted to reach fainter sensitivities down to \hetg~$\approx$~24 - 25.

\subsection{AGN Catalog}
\label{sec:agn}

The HETDEX AGN catalog from the same internal data release, HDR2, is presented in \citet{liu2022}, and contains the same base sample that is included in the catalog presented in this paper.  There are, however, several selection differences between \citet{liu2022} and the current work; some of these add candidates that are not in the \citet{liu2022} HETDEX AGN catalog, while others reject \citet{liu2022} objects.

The catalog presented here includes additional data quality criteria that limit the sample relative to \citet{liu2022}. For example, some frames with poor observational conditions or sources on amplifiers that failed our quality assessment remain in the HETDEX AGN catalog, which mitigated these issues through visual inspection. Our sample is also limited to the HETDEX Fall and Spring fields, as well as the COSMOS and GOODS-N legacy fields, whereas the AGN catalog includes additional data from a North Ecliptic Pole survey (NEP; Chavez Ortiz\etal, submitted).

Roughly a quarter of the AGN sample overlap with the curated line emission catalog and the continuum detection catalog; the main divergence between the catalogs arises from AGN that exhibit broad line emission that is not well fit by the Gaussian model implemented in the HETDEX line detection algorithm. These detections occupy both high line width and high $\chi^2$ parameter space and do not meet the line parameter criteria for line-emission in our curated line detection catalog. As described in \citet{liu2022}, visual inspection of this broad line emission sample is essential for classifying a source as an AGN rather than a calibration artifact.

We include the AGN catalog in our combined source catalog and allow individual detections to be grouped according to the process described in Section~\ref{sec:det_group}. This approach allows for additional line and continuum emission detections to be associated with the AGN source and assigns an AGN classification and its associated redshift.

\subsection{Detection Grouping}
\label{sec:det_group}

Both the line emission and continuum emission pipelines are designed to identify point-source emission.  For LAEs, the primary target of interest for HETDEX, and many \OII\ emitting galaxies, the point-source approximation is a good one. However, $\sim 40\%$ of the high quality detections identified in the emission-line and continuum source pipeline have multiple identifications. This situation can arise in extended objects, where emission is found at more than one spatial location, or with sources where more than a single emission-line has been detected.  Similarly, bright astronomical sources can have both line emission and continuum emission, leading to entries in both the continuum and line curated catalogs.

The overlap in point-source brightness between the detection samples is shown in Figure~\ref{fig:dettype}.  As expected, the continuum sources have much brighter \hetg\ values than the objects found by the line detection algorithm. Starting near \hetg~$\approx 20$ there is considerable overlap in the catalogs, while at magnitudes fainter than \hetg~$\approx$~22, the line emission sources dominate. A particularly important challenge to creating a robust LAE sample is extended \OII\ line emission surrounding low-$z$ galaxies; these features can often be confused for Ly$\alpha$ emission due to the lack of detectable continuum emission at large galactocentric radii.  To mitigate the impact of these contaminants and to properly associate extended line emission to a single emission-line source, we apply a three-dimensional friend-of-friends (FOF) clustering algorithm to our list of line detections.   Our code\footnote{\url{https://github.com/HETDEX/hetdex_api/blob/hdr2.1.3/hetdex_tools/fof_kdtree.py}} uses \texttt{cKDtree} from \texttt{SciPy} \citep{scipy} but is modified to use normalized coordinates in a pseudo-spherical space consisting of projected separation on sky and normalized wavelength difference.  Specifically, we adopt a spatial linking length of 6\arcsec\ and a linking length in the spectral direction of 8\,\AA.  Information for the 3D clustering of the emission-line detections is provided in the columns  \texttt{wave\_group\_XX} of the \textit{Detection Info Table} (see Table~\ref{tab:det_col_info}). These columns contain the identification for the wave group, its mean equatorial coordinate, and mean central wavelength. Also included is the group's semi-major \texttt{wave\_group\_a} and semi-minor \texttt{wave\_group\_b} axes, as determined from the line flux-weighted first order moment of the line detection group.  These values can be considered as a rough approximation to the extended group's emission-line size, but we caution that many sources that consist of just two matched line detections will be elongated in shape, while sources that have incomplete IFU coverage (as in the cases with extended \OII-emitting galaxies) will be limited by the IFU edge of fiber coverage.

In addition to 3D clustering in wavelength and position, we also link all detections on sky together with a spatial linking length of $2\arcsec$. This will ensure that sources with emission-lines at multiple wavelengths will be grouped as one source. If those lines are themselves extended then this step will group extended emission at multiple wavelengths together as is the case for nearby galaxies that might, for example, exhibit extended \OII\, \hbeta\ and \OIII\ emission at multiple wavelengths. For blended objects or cases where a background object lies behind an extended foreground group, we accept that this linking may cause background detections to be lost and ultimately merged into the foreground object. Some of these sources may be quite interesting, particularly those with the potential to be gravitationally lensed, as demonstrated for the sample in \citet{Laseter2022}. For fainter source groups that are ultimately classified as LAE after detection grouping, we separate these sources spatially and assign a redshift according to each detection's observed wavelength assuming it to be Ly$\alpha$. Although we note a possible exception is if the line emission wavelengths contain a pair of emission lines that can be associated with a common redshift, such as Ly$\alpha$, \HeII, and \CIV. Redshift assignment and source classification is described further in Section\,\ref{sec:classification}.

For each source, we select a single representative detection for each source observation. This is listed as \texttt{selected\_det==True} in the Table\,\ref{tab:det_col_info} and \texttt{detectid} in Table\,\ref{tab:column_info}. This detectid corresponds to the detection member with the brightest (i.e., smallest) \hetg\ value for all sources that are not LAEs. For LAEs, we use the highest $S/N$ \lya\ line detection as the selected representative \texttt{detectid}. To remove detections that are identified by the HETDEX detection pipeline due to sharp discontinuities in the detection spectrum that do not correspond to true line emission we opt to remove all emission-lines with a line width greater than 6\,\AA\ that are not identified as a representative source (with \texttt{selected\_det==True}) or are not included in the AGN catalog.

Examples of the data (broad-band images, reconstructed IFU images, and spectra) for a range of objects in the catalog are presented in Figures~\ref{fig:spec_exampleslowz} ($z<0.5$) and~\ref{fig:spec_exampleshighz} \hbox{($1.9 < z < 3.3$).} Individual detections are overplot on the imaging data in the left-panel in each row. Line emission detections are marked by orange crosses and continuum detections are marked by green crosses. The spectrum for a single representative detection for the source group (identified by \texttt{selected\_det==True}) is shown on right. Yellow bars indicate line emission detections. In the first example in Figure~\ref{fig:spec_exampleslowz}, multiple emission lines are found but notably a line near 5250\,\AA\ is missing from the curated catalog. This is because the other line emission causes the single Gaussian fit of that detection to have a poor quality of fit and does not make the curated catalog. Continuum-subtracted emission, line flux maps, shown in the middle column, at the observed wavelength indicated in text at the top, demonstrate differences between the line emission distribution and continuum emission morphology as shown in the imaging data on the left. 

\subsection{Spatially Resolved Line Fluxes}
\label{sec:aperflux}

\begin{figure}[t]

    \centering
    \includegraphics[width=3.2in]{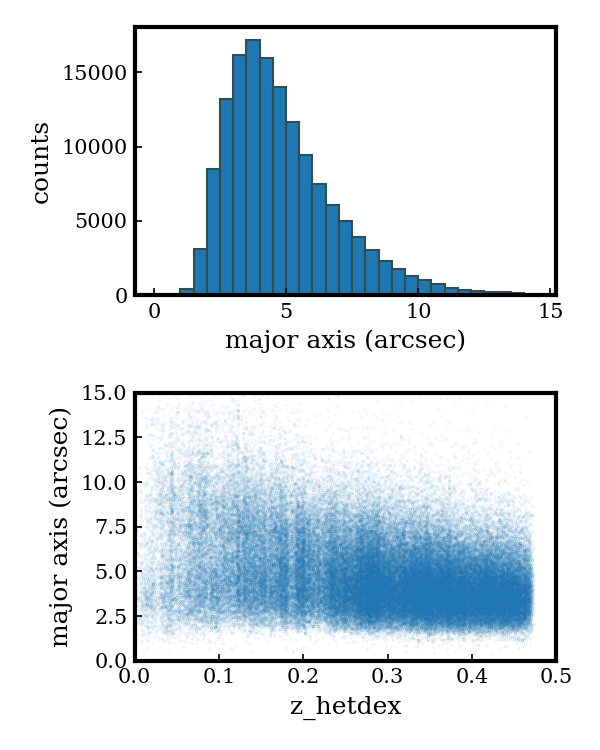}
    \caption{Distribution of sizes for the $z<0.5$ galaxy sample as derived from object detection using Source Extraction and Photometry (SEP) on ancillary $r$-band or $g$-band photometric imaging. For the low-$z$ galaxy sample, 78.5\% of the galaxies have sizes greater than 3\arcsec, stressing the need for aperture flux values that encompass a galaxy's full extended emission. }
    \label{fig:oii_size_dist}
\end{figure}

For every galaxy with \zhet $<0.5$, we measure spatially resolved \OII\ line fluxes at the galaxy's redshift in addition to those provided by the HETDEX line detection algorithm which are point-source, PSF-weighted line flux values. We note that while a resolved line emitter can appear as multiple detections in the line database, some flux will inevitably be missed even if each detection is summed. In addition, the line detection pipeline used for this catalog release contained an upper limit on the continuum value, so some very bright line emitting galaxies are completely missing from our line emission database, even though they are found in the HETDEX continuum catalog.

A major strength of the wide-IFU (dithered) coverage with HETDEX is that the observations automatically produce an emission-line map of resolved galaxies.  However, due to the IFU layout in the HET's focal plane, many of these systems have incomplete coverage, as their light extends off the edge of their IFU\null. The angular resolution of our imaging observations is substantially better than the IFU fibers.  As a result, object shapes and sizes are better measured from direct imaging.

 Object shapes are automatically included in HETDEX's \elixer\ classification tool \citep{Davis2022}, as it applies \texttt{Source Extraction and Photometry} \citep[\texttt{SEP};][]{sep_pkg} to all available broadband imaging at the location of each HETDEX detection. This step provides the major and minor axes of an ellipse fit to the second-order moment of each object's surface brightness distribution \citep{source_extractor}. We use the \elixer\ catalog \texttt{selected=True} option and preferentially choose $r$-band over $g$-band measurements to define each galaxy's elliptical aperture. In general, the $r$-band imaging we have obtained has a fainter limiting magnitude, and significantly better image quality.
The image selection can be found in the columns \texttt{catalog\_name\_aper} and \texttt{filter\_name\_aper} in the \textit{Detection Info Table} (see Table~\ref{tab:det_col_info}). Elliptical parameters for each low-$z$ galaxy can be found in both the \textit{Source Observation Table} (Table~\ref{tab:column_info}) and the \textit{Detection Info Table} (Table~\ref{tab:det_col_info}) under the columns \texttt{major}, \texttt{minor}, and \texttt{theta}. Additionally the aperture center and the measured continuum aperture magnitude are in the \textit{Detection Info Table} (see Table~\ref{tab:det_col_info}) in the columns \texttt{ra\_aper}, \texttt{dec\_aper}, \texttt{mag\_aper}, and \texttt{mag\_aper\_err}.


Figure~\ref{fig:oii_size_dist} presents the major axis distribution of the low-$z$ sample. More than three-quarters of the sample has a major axis larger than 3\arcsec\ and is thus spatially resolved by VIRUS\null. We create continuum-subtracted line flux and flux uncertainty maps for each source's \OII\ emission by summing the fiber data in a $\pm$15\,\AA\ window around the wavelength of observed \OII, redshifted from $\lambda\,3727.8$\,\AA\ according to \zhet. We then subtract local spectral continuum by making two narrowband-like images, each 50\,\AA\ wide, shifted by an additional 10\,\AA\ blue and red of the line emission. We subtract the average of these two images from the line flux map to produce a continuum-subtracted line-flux map.

The flux and associated error in the galaxy's elliptical aperture is summed using the \texttt{photutils} software package \citep{photutils_1.3.0}. The resulting aperture and dust-corrected aperture fluxes are found in columns \texttt{flux\_aper} and \texttt{flux\_aper\_err}. The photometric information has an aperture correction applied, \texttt{im\_apcor}, for sources that lack full IFU coverage. For each source, we opt to use the \texttt{flux\_aper} value for \texttt{flux\_oii} if the value is positive and the major-axis of the low-$z$ galaxy is greater than 2\,\arcsec.

\subsubsection{Comparison to SDSS \OII\ Line Fluxes}
\label{sec:sdss_comp}
\begin{figure}[t]
    \centering
    \includegraphics[width=3.5in]{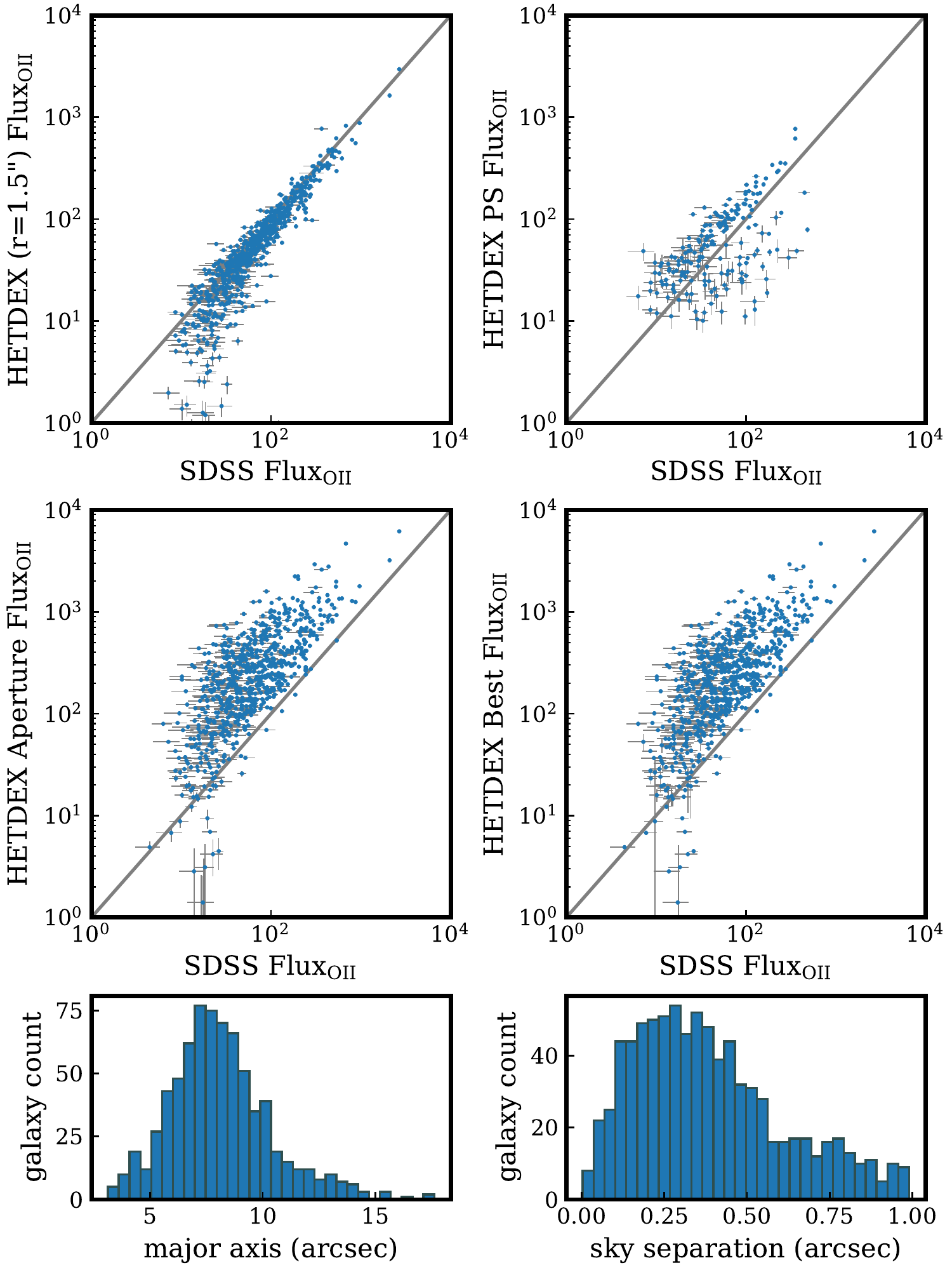}
    \caption{HETDEX \OII\ line fluxes compared to \sdss\ \OII\ line fluxes. All units in  $10^{-17} \mathrm{ergs/s/cm^{2}}$. The top-left panel is a comparison of $3\arcsec$ diameter aperture \OII\ line flux measurements from HETDEX in which the continuum-subtracted \OII\ line fluxes within a 3\arcsec\ diameter aperture are measured at the exact location of the \sdss\ fibers.  The top-right panel shows the HETDEX pipeline continuum-subtracted \OII\ line fluxes, which are measured using PSF-weighted extracted spectra at the location of the object's peak \OII. The middle-left panel compares HETDEX aperture line fluxes with the \sdss. Here, the HETDEX values lie significantly above those of \sdss\ due to the larger aperture area. The middle-right shows the optimal HETDEX \OII\ line flux, which can be either from the HETDEX pipeline or the aperture flux measurement, depending on conditions.  The range of major axis sizes for the HETDEX measurements is shown in the bottom-left panel and separations between HETDEX catalog detections and \sdss\ fiber positions is shown in bottom-right.}
    \label{fig:sdss_comps}
\end{figure}

Since the HETDEX survey fields lie completely within the SDSS footprint, we can compare spectra that are in common between the two surveys. Figure~\ref{fig:sdss_comps} presents HETDEX measurements of the continuum-subtracted emission-line fluxes for \OII\ emitting galaxies found in the MPA-JHU value-added catalogs from \sdss\,DR8\footnote{https://www.sdss.org/dr16/spectro/galaxy\_mpajhu/} \citep[based on the methods described in ][]{Tremonti2004, Brinchmann2004}.  In the top-left, for optimal comparison, the HETDEX line fluxes are measured in a circular $3\arcsec$ diameter aperture to match the $3\arcsec$ diameter fibers of \sdss. The figure demonstrates that for forced aperture line fluxes, the HETDEX measurements are well matched to \sdss\ $3\arcsec$ values to an RMS of~26\% for objects with line fluxes above \hbox{of $\sim 3 \times 10^{-16}$ ergs s$^{-1}$ cm$^{-2}$}. Differences in positioning are mitigated by placing apertures exactly at the location quoted by \sdss; however, the derived line fluxes also depend on how the continuum is measured, and our IFU data have a different spatial profile than that of the single \sdss\ fiber.

Comparisons between the pipeline point-source fluxes show much greater scatter in the top-right panel in Figure~\ref{fig:sdss_comps}. As with the forced line fluxes, differences in the measurement method can create scatter, but here the bigger culprit is positioning. The HETDEX detection pipeline is designed to peak on the highest $S/N$ line detection in a spatial grid of 1D extracted spectral data. This peak can vary from the location of the \sdss\ fiber by up to 1\,\arcsec. The bottom-right panel shows the distribution of sky separations between the \sdss\ fiber and the position of best HETDEX peak detection. In cases where the \sdss\ fiber is on the edge of the IFU, some flux is lost and underestimated in the pipeline point-source fluxes. In addition, fewer data points are shown here because the continuum detection search was performed with an upper limit threshold on counts, thus excluding the brightest \OII-emitting galaxies.

The middle-left panel compares HETDEX aperture fluxes, \texttt{flux\_aper}, with \sdss\ measurements. Here, the HETDEX values lie significantly above those of \sdss\ due to the larger aperture area. The middle-right shows the optimal HETDEX \OII\ line flux, \texttt{flux\_oii} in the catalog, which can be either from the HETDEX pipeline or the aperture flux measurement. It is assigned \texttt{flux\_aper} if it is a positive value and the major-axis of the galaxy based on broadband imaging is greater than 2\arcsec. Otherwise the line flux measured comes from the \texttt{flux} measured from the HETDEX pipeline. The flag listed as \texttt{flag\_aper} is 1 if \texttt{flux\_aper} is used, 0 if \texttt{flux} is used and -1 if it is not relevant, as is the case for LAEs, AGN, stars and low-$z$ galaxies (LZGs). As shown in the bottom-left of Figure~\ref{fig:sdss_comps}, the galaxies in the comparison have a wide range of sizes, so it is not surprising that the HETDEX spatially resolved fluxes are much larger than the fiber fluxes from \sdss\null.

\subsection{Dust Correction}
\label{sec:dustcorrection}

Reported fluxes are provided both as measured at the top of the atmosphere, and corrected for Galactic extinction. The python software package \texttt{dustmaps} \citep{dustmaps} is employed to access the local Milky Way Dust reddening values for each source's coordinates as measured by \citet{Schlegel1998}. The software returns the locally measured color excess value, $E$(\bv), based on a source coordinate. We assume the ratio of $V$-band extinction, $A_V$, to color excess, $E$(\bv), to be $R_V = 3.1$ and apply a factor of 2.742 to measure the local $V$-band extinction as $A_V = 2.742\times E(\bv)$ according to the re-calibration using \sdss\ stars of the \citet{Schlegel1998} maps by \citet{Schlafly2011}. $A_V$ values range from 0.01 to 1.44 with a median value of 0.04.  Dust correction of line fluxes are applied at the central wavelength of the line emission according to the $R_V=3.1$ extinction curve of \citet{Fitzpatrick1999}, implemented using the open source python software \texttt{extinction}\footnote{https://github.com/kbarbary/extinction}.
The measured values are designated with the notation \texttt{XXX\_obs}, where \texttt{XXX\_obs} can be \texttt{flux\_obs}, \texttt{flux\_obs\_err}, \texttt{continuum\_obs}, or \texttt{continuum\_obs\_err} without any dust correction, while those without the \texttt{obs} suffix have the local dust correction applied. Included source spectra are also offered with and without an applied dust correction. Their format is described further in Section~\ref{sec:format}.

\section{Source classification and Redshift Determination}
\label{sec:classification}
Once emission-line and continuum detections are placed into common source groups, we assign a source classification and redshift to each group. As the two HETDEX detection methods probe different astronomical sources, we must take a multi-pronged approach to classify our sample. In this section, we first outline the three methods of source classification and redshift assignment (Section~\ref{sec:zmethods}) and then present the decision logic to assign a classification to a source (Section~\ref{sec:classify}). In Section~\ref{sec:redshifts} we compare our measured redshifts to those available in the literature and quantify the accuracy of our redshift measurements.

\subsection{Methods}
\label{sec:zmethods}
The continuum sample is comprised of  objects in the magnitude range of $14 \lesssim g \lesssim 21.5$.  In contrast,  the line-emission sample probes a broad range of continuum levels, with a quarter of the curated, high-quality sample having \hetg$>25$, which we consider to be the approximate sensitivity limit of HETDEX 1D spectra. 

The line emission algorithms probe a wide range of sources, which include galaxies with no detectable continua and bright objects with multiple emission-line entries in our detection catalog. Depending on the choice of line-fit parameters, the detection algorithms also include high line-width detections that are are actually sharp discontinuities in the spectra, caused by absorption features in late-type stars, others are due to the broad, complex emission lines of AGN.

For the brighter continuum spectra, we employ the software package \texttt{Diagnose} to determine a source's classification and redshift (see Section~\ref{sec:diagnose}). For fainter objects, we rely on the properties of the line emission and assumptions about the expected luminosity function and equivalent width distribution of line emitting sources to assign a redshift; this process is described in detail in \citet{Davis2022} and briefly summarized in Section~\ref{sec:elixer}. If any source contains a detection that is found within the HETDEX AGN Catalog, then the redshift and classification from \citet{liu2022} is applied as detailed in Section~\ref{sec:agnredshift}.

\subsubsection{Diagnose}
\label{sec:diagnose}

\texttt{Diagnose}, a software package developed for the Hobby Eberly Telescope VIRUS Parallel Survey (HETVIPS; \citealt{Zeimann2022}), uses a principal component analysis (PCA) algorithm to classify sources as stars, galaxies, quasars, or unknown. The \texttt{redrock}\footnote{https://github.com/desihub/redrock-templates} spectral templates used by \texttt{Diagnose} are the same as those employed by \sdss -IV \citep{Ross2020} for their classification/redshift measurements.  The templates include spectra from ten galaxies, four quasars, and three cataclysmic variables. Stars are classified by spectral type, and are assigned to the subclasses B, A, F, G, K, M, and white dwarf.  \citet{Zeimann2022} report that for objects with both SDSS and HETVIPS classifications, the \texttt{Diagnose}  values match those of \sdss\ for 96.9\%, 94.7\%, and 92.3\% of stars, galaxies, and quasars, respectively. 

Unsurprisingly, the fraction of sources that achieve a successful \texttt{Diagnose} classification and redshift assignment decreases as a function of VIRUS spectral signal-to-noise, and is correlated with a source's $g$-magnitude. \citet{Zeimann2022} demonstrate they reach $\sim$90\% recovery of classifications at a spectral \hbox{$\langle {\rm S/N} \rangle = 8$}, where $\langle \textrm{S/N} \rangle$ is the mean $S/N$ measured per 2\,\AA\ spectral resolution element, and a value of 8 corresponds roughly to $g=20$. For our sample of sources with \hetg $<22$, \texttt{Diagnose} reports a confident classification for 98.5\% of the detections; At brightnesses in the range of $22<g<23$, classifications are reported for 86.4\% of the detections. However, we do not rely on \texttt{Diagnose} at these fainter magnitudes because of possible confusion between \OII\ and Ly$\alpha$ line emission. Often these sources have little detected continuum signal, causing \texttt{Diagnose} to automatically default to a low-$z$ star-forming template with many \lya-emission being falsely identified as \OII, \hbeta, or \OIII\ emission. Faint line emission classification is better assessed by \elixer.

\subsubsection{Emission Line eXplorer (ELiXer)}
\label{sec:elixer}

The majority ($\gtrsim60\%$, although this number is much higher when considering lower signal-to-noise detections) of HETDEX line emission detections consist of just a single emission-line and line identification cannot be trivially deduced from the spectrum itself. For LAEs, the largest contaminant is $z<0.5$ \OII\ emitting galaxies. Historically, a 20\,\AA\ equivalent width cut (in the rest-frame of \lya) has been used to segregate \OII\ from \lya\ \citep[][]{Gronwall2007a,Adams2011a} where the continuum is measured from either the spectrum itself (if sensitive enough) or in accompanying deep photometric imaging. In practice, this criterion typically results in more than~4\% contamination, and excludes all lower equivalent width \lya\ lines \citep{Acquaviva2014}. For HETDEX, this can be a problem as the $H(z)$ and $D_A(z)$ measures are sensitive to interloper clustering\citep{Leung2017, grasshorn2019,Farrow2021} and HETDEX requires contamination in the LAE sample to be $\lesssim$ 2\% \citep{Gebhardt2021}. \cite{Leung2017} improved on the 20\,\AA\ cut by adopting a Bayesian approach and including additional information about the equivalent width distributions of \lya\ and \OII\ using $g$ and $r$-band photometric info, and the systems' emission-line luminosity functions. From their modeled data, \citet{Leung2017} reported an expected contamination rate of \lya\ by \OII\ of between approximately 0.5\% and 3.0\% at cost of $\sim$ 6.0\% to 2.4\% lost LAEs. HETDEX implements this line discrimination approach and builds upon it in its line emission classifying software \texttt{Emission Line eXplorer} (\elixer, \citealt{Davis2022}). It adds in a suite of additional information such as multiple line emission considerations, photometric imaging counterpart information (galaxy size and magnitude for example) and additional data quality checks that assign a probability likelihood, \texttt{P(\lya)}, that a HETDEX emission-line detection is due to \lya. This value, \texttt{plya\_classification} and a number of other measurements from \elixer\ related to the detection's imaging counterpart is presented for each detection in the \textit{Detection Info Table} (described in Appendix\,\ref{appendix:1}). 

\citet{Davis2022} report a projected HETDEX LAE contamination rate from \OII\ of 1.3\% ($\pm$0.1\%) and an additional 0.8\% ($\pm$0.1\%) from all other sources, along with an LAE recovery rate of 95.7\% ($\pm$3.4\%) with \elixer\ version 1.16.5 and the current internal HETDEX catalog (based on its third internal data release). For the work in this paper, we use an earlier \elixer\ version (1.9.1) and find an LAE contamination rate from \OII\ of 2.4\% with an LAE recovery rate of 95.2\% for galaxies with  \hetg$>22$ mags. These rates do not include the bias in the spectroscopic sample used to measure the contamination rate and recovery rates; this sample tends to be brighter than the main HETDEX LAE sample. For details on projecting these values to unbiased rates please see \citet{Davis2022}.

\subsubsection{AGN Catalog Redshifts}
\label{sec:agnredshift}

As discussed in Section~\ref{sec:agn}, a systematic search for AGN within HDR2 is performed to identify both broad-lined emitting AGN and narrow-line AGN with two confirmed emission lines. This AGN catalog (\citealt{liu2022}) consists of~5,322 AGN, of which~3,733 have spectroscopic redshifts secured by either (1) two emission line confirmations and/or (2) a positional match to AGN within the \sdss\,DR14 Quasar Catalog \citep{Paris2018}. These sources are identified with \texttt{zflag=1} in \citet{liu2022} and are identified in this catalog release by \texttt{agn\_flag=1}. The remaining single broad-lined sources are assumed to be due to \lya\ emission from AGN and are identified in the catalog by \texttt{agn\_flag=0}. Sources that are not AGN are given \texttt{agn\_flag=-1}. 

\subsection{Assignment}
\label{sec:classify}

A sequence of logic is implemented to assign a redshift and classification to each source.  We highlight the general logic in the bottom part of the flowchart presented in Figure~\ref{fig:flowchart} and describe the details here. 

The method of the assigned redshift is found in the column \texttt{z\_src\_redshift}. If any detection in the group is found in the HETDEX AGN catalog, the redshift from \citet{liu2022} is assigned to the source group, the source type is labeled \texttt{agn}, and the source's redshift confidence is taken from the \texttt{z\_flag} column from the HETDEX AGN Catalog.  This value is 1 if either the redshift is derived from multiple emission lines or the object has an \sdss\ counterpart with a measured redshift consistent with the HETDEX observations.  A small (2.2\%) fraction of our catalog (\nagn/\nsource) is assigned its redshift from the HETDEX AGN Catalog and can be found in the catalog under \texttt{z\_src\_redshift==`Liu+2022'}

For source groups that contain one or more detections with \hetg $<22$ mags and a  \texttt{Diagnose} classification,  we adopt the \texttt{Diagnose} redshift, \zdiagnose. We assign a confidence to this redshift of \texttt{z\_conf=0.9} (an arbitrary high-confident number here, but we aim to provide better calibration of our redshift assignment in the future) as redshifts assigned from \texttt{Diagnose} are highly reliable, with a 97.1\% accuracy for our sample (described further in Section~\ref{sec:redshifts}). If the detection has a \texttt{STELLAR} classification, we assign \texttt{source\_type=`star'} and \texttt{\zhet=0}. We note that additional classification information can be found from the \texttt{Diagnose} spectral fits in columns (\zdiagnose, \texttt{cls\_diagnose}, \texttt{stellartype}) in the \textit{Detection Info Table} (described in Appendix~\ref{appendix:1} and Table~\ref{tab:det_col_info}). If the \texttt{Diagnose} classification is \texttt{GALAXY}, we label the source type as \texttt{oii} if an \OII\ emission-line is present in the spectrum with a line flux value, \texttt{flux\_oii}, reported, and \texttt{lzg} (low-$z$ galaxy) if no emission-line has been detected. If the \texttt{Diagnose} classification is \texttt{QSO}, we assign a source type of \texttt{agn}. Less than half (41.1\%,  91,885/\nsource) of the catalog redshifts are assigned using \texttt{Diagnose} and can be isolated in the catalog under \texttt{z\_src\_redshift==`Diagnose'}.

For all other source groups, we rely on \elixer\ to assign source redshifts. For the public HDR2 catalog that is limited to higher $S/N>5.5$ detections, 60.7\% (135,789/\nsource) sources are classified by \elixer. A few steps of logic are involved when making the final selection which we briefly outline here. For a single emission-line source group, we simply assign the \elixer\ redshift, \texttt{best\_z}, to \zhet. We also transfer the \elixer\ redshift confidence, \texttt{best\_pz} (described in detail in \citealt{Davis2022}) to \texttt{z\_conf}. The redshift confidence should not be used for selection criteria, however, as we have not calibrated it. This will be applied in later HETDEX catalogs.

If multiple line detections are found, we first check to see if any of the detections are part of a common wavelength linked group. We then use the redshift for the detection closest to the center of the wave group  (listed as the minimum value of \texttt{src\_separation} for the detection group).  Next we check to see if any of the detections are confidently at low-$z$, with $\texttt{plya\_classification}<0.4$. This value is empirically chosen to maximize the LAE recovery fraction (96\%), while minimizing the \OII\ contamination fraction (at 3\%). It also differs from the built-in threshold ($\texttt{plya\_classification}=0.5$) that \elixer\ users for its redshift-assignment as this earlier software version was found to put low quality LAE candidates at $0.4<\texttt{plya\_classification}<0.5$. If this is the case, we assign the \elixer\ \texttt{best\_z} to the detection closest to the source group center.  This can result in background line emitters getting blended with the foreground source and ultimately not classified as an LAE or more distant galaxy. If neither of these cases are found, we go through the source group assigned redshifts and make a choice of which is the best redshift to use. 

If the collection of redshifts has a standard deviation less than 0.02, then we can simply assign the redshift of the detection closest to the source center. This will happen if the source is an extended \lya-emitter or if observed emission lines are a pair match such as \lya\ and \CIV. Extended \lya-emitters will be analyzed in a future HETDEX paper but can be found in this catalog by searching for sources with a defined \texttt{wave\_group\_id} at \zhet$>1.88$. Both AGN and LAE source types (e.g. via a logical search of \texttt{source\_type} == \texttt{lae} or \texttt{agn}) can exhibit extended emission. If the standard deviation of \zhet\ in the detection grouping is larger than 0.02 and all detections are classified as high-$z$ according to \elixer's \texttt{best\_z}, then we assume the detections to be independent from each other and be line of sight interlopers. We disassociate the group of detections and assign each detection a unique \texttt{source\_id}, and classify each detection as an LAE at redshift corresponding to \lya\ the observed line wavelength. Sources that have been assigned their redshift from \elixer\ are found in the \textit{Detection Info Table} (see Table~\ref{tab:det_col_info}) under \texttt{z\_src\_redshift==elixer}.

Examples of the source classification and redshift assignment for a range of objects in the catalog are presented in Figures~\ref{fig:spec_exampleslowz} ($z<0.5$) and~\ref{fig:spec_exampleshighz} \hbox{$1.9 < z < 3.3$).}

\subsection{Accuracy}
\label{sec:redshifts}
\begin{figure*}[t]
    \centering
    \includegraphics[width=7in]{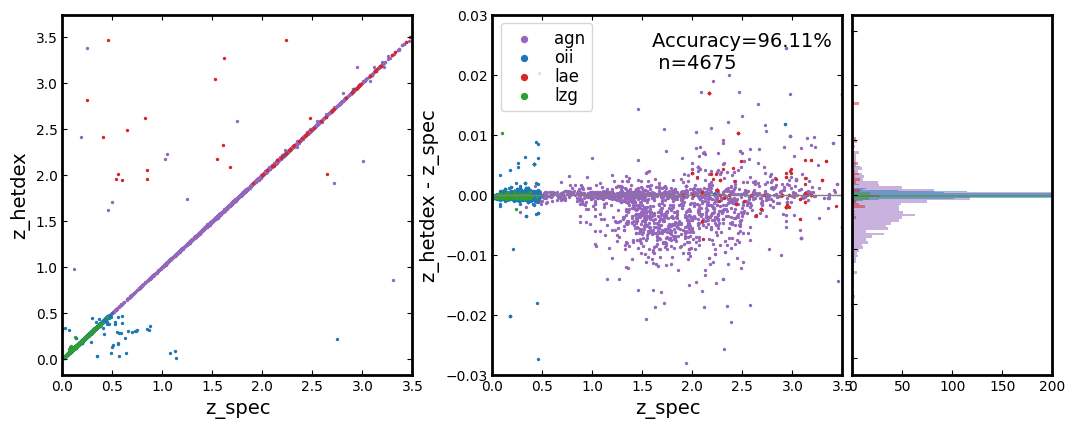}
    \caption{In the left panel, a comparison plot of HETDEX spectroscopic redshifts, \zhet, and spectroscopic redshifts, $\mathrm{z\_spec}$, from multiple publicly available redshift catalogs is shown. In the right panel, a zoom-in on redshift differences for well-matched sources. The histogram on far-right panel shows the distribution of differences. Our net accuracy of spectroscopic redshift agreement to within \hbox{$\Delta z < 0.02$} is 96.11\%.  }
    \label{fig:redshift}
\end{figure*} 

To assess the accuracy of our redshift assignments, we can compare our cataloged values, \zhet, to spectroscopically determined redshifts from other surveys.  These literature redshifts are generally quite reliable, as they tend to be derived from spectra with higher spectral resolution, and/or broader wavelength coverage than that from HETDEX.  However, because these targets were often pre-selected from broadband imaging data, they also tend to have continuum magnitudes significantly brighter than the bulk of the HETDEX detections. 

In the COSMOS legacy field, we use redshifts from zCOSMOS DR3 \citep{Lilly2009} and the DEIMOS 10k Sample \citep{Deimos10K}; in the GOODS-N legacy field, the redshifts come from \citet{Reddy2006}, \citet{Barger2008}, \citet{Wirth2004}, \citet{Wirth2015}, PEARS \citep{Ferreras2009} and DEEP3 \citep{Cooper2011}. Redshifts are also used from MOSDEF \citep{Kriek2015} and 3D-HST \citep{Momcheva2016}, which cover multiple deep legacy fields.  Finally, we also use measurements from the \sdss\ DR16 Redshift Catalog \citep{SDSS_DR16}, which covers brighter objects over all our fields. For all these data, we apply quality criteria to select the surveys' most confident redshifts as described in their corresponding papers. 

Shown in Figure~\ref{fig:redshift}, 4675 of our catalog sources have a spectroscopic redshift match within 1\farcs2. The source type breakdown of the catalog-matched sources is 134~LAEs, 1592 AGNs, 2560~\OII-emitting low-$z$ galaxies, 311 other low-$z$ galaxies (LZGs) without \OII\ line emission, and 78 stars. The accuracy of our combined redshift assignment method is~96.1\% (88.4\%), where  accuracy is defined as agreement with an external spectroscopic redshift value to within
\hbox{$\Delta z < 0.02$} (\hbox{$\Delta z < 0.005$.}). Restricting the matches to those sources with redshifts assigned from the HETDEX AGN Catalog, 98.2\% (1564/1592) are in agreement. The redshift assignment for AGN, however, included cross-matches to AGN spectroscopic redshifts from \sdss\,DR14 \citep{Paris2018} which were adopted if they agreed with line emission measured in the HETDEX data. Removing the AGN population from the catalog and considering just \texttt{Diagnose} and \elixer\ redshift assignments, the net redshift accuracy to within
\hbox{$\Delta z < 0.02$} is 95.0\% (2929/3083).  The sample assigned redshifts by \texttt{Diagnose} are 95.7\% (1994/2084) accurate; this result is slightly higher than that reported by \citet{Zeimann2022} because we have excluded the poorer performing AGN/Quasar assignments which are affected by narrower wavelength coverage of VIRUS\null.  The remaining redshifts assigned by \elixer\ have a relatively poorer redshift accuracy of 93.6\% (935/999) primarily because \elixer\ is the catchall for all the remaining sources that are not bright enough for \texttt{Diagnose} nor are they visually classified as an AGN\null. False-positive line-emission detections, which generally do not have associated continuum emission, fall into this category, as do transient detections (such as missed meteor and satellite features), and local line emitters with little continuum (e.g., young stellar objects, active late-type stars, and planetary nebulae).  Finally, the overall density of faint line emitters is considerably larger than that for bright continuum sources, so some line-of-sight mismatches with external redshift catalogs are likely.

One approach to measure the true success rate of identifying \OII\ and \lya\ sources and mitigate possible contamination from false positives and other interlopers is to only consider line detections at observed wavelengths that match the spectroscopic redshift from the external catalogs at either observed frame of \OII\ or \lya.  This requirement increases the accuracy of \elixer\ to 95.24\% and the accuracy of HETDEX redshift assignments to~96.8\%. For greater, in-depth discussion on assigning classifications with \elixer\ and its success rates, see \citet{Davis2022}.

\section{Catalog Format}
\label{sec:format}

\begin{table*}
\begin{center}
\caption{\textit{Source Observation Table} Column Descriptions\label{tab:column_info}}
\begin{tabular}{ll}
\hline \hline
Name & Description \\
\hline
source\_name & HETDEX IAU designation (ie. \texttt{HETDEX~J123449.19+511733.7}) \\
source\_id & HETDEX Source Identifier \\
shotid & integer representing observation ID: int( date+obsid) \\
RA & source\_id R.A. (ICRS deg) \\
DEC & source\_id decl. (ICRS deg) \\
gmag & (\hetg) SDSS $g$-magnitude measured in HETDEX spectrum \\
Av & applied dust correction in V band \\
z\_hetdex & HETDEX spectroscopic redshift \\
z\_hetdex\_src & HETDEX spectroscopic redshift source \\
z\_hetdex\_conf & 0 to 1 confidence HETDEX spectroscopic redshift source \\
source\_type & options are \texttt{star}, \texttt{lae}, \texttt{agn}, \texttt{lzg}, \texttt{oii}, \texttt{none}  \\
n\_members & number of detections in the source group \\
detectid & detection ID of representative detection for the source (selected\_det == True in \textit{Detection Info Table}) \\
field & field ID: cosmos, goods-n, dex-fall, dex-spring \\
flux\_aper & Dust corrected, OII line flux measured in elliptical galaxy aperture in $10^{-17}\mathrm{erg/s/cm^2}$  \\
flux\_aper\_err & error in flux\_aper \\
flag\_aper & 1 = aperture line flux used for lum\_oii, 0= PSF-line flux used from `flux' column \\
major & major axis in arcsec of aperture ellipse of resolved OII galaxy defined by imaging \\
minor & minor axis in arcsec of aperture ellipse of resolved OII galaxy defined by imaging \\
theta & angle in aperture ellipse \\
lum\_lya & Ly$\alpha$ luminosity and error calculated from "flux" column (ie. dust corrected Lya line flux) in ergs/s \\
lum\_lya\_err & error in lum\_lya \\
lum\_oii & OII line luminosity calculated from `flux' column if flag\_aper=0 or 'flux\_aper' column if flag\_aper=1 in ergs/s \\
lum\_oii\_err & error in flux\_oii \\
flux\_lya & Ly$\alpha$ flux calculated from `flux' column for (ie. dust corrected Lya line flux) in ergs/s \\
flux\_lya\_err & error in flux\_lya \\
flux\_oii &  \OII\ flux in ergs/s calculated from `flux' if flag\_aper=0 or 'flux\_aper' if flag\_aper=1 (ie. dust corrected)' \\
flux\_oii\_err &   error in flux\_oii\\
sn & signal-to-noise for line emission \\
apcor & aperture correction applied to spectrum at 4500\AA \\
\hline
\end{tabular}
\end{center}
\end{table*}

The information in this release is presented in two separate catalog formats: the curated \textit{Detection Info Table}, with columns described in Table\,\ref{tab:det_col_info}, which contains information about every HETDEX line emission and continuum detection that has passed the quality checks and line parameter criteria described in \S\,\ref{sec:quality} and \ref{sec:linecriteria} respectively; and the \textit{Source Observation Table}, which contains aggregate information from the more detailed \textit{Detection Info Table} for each source observation. It contains fundamental information on a source (position, redshift, physical size if relevant, \OII\ or \lya\ flux and luminosity where appropriate) and is repeated for each separate HETDEX observation of the source. For most users, the \textit{Source Observation Table} will be the sufficient and it is a limited, easier-to-parse summary of the \textit{Detection Info Table} (which is provided in more detail in Appendix~\ref{appendix:1}).

A HETDEX source, identified by \texttt{source\_id}, is a collection of all detections at the same on-sky position combined through the detection grouping method described in Section~\ref{sec:det_group}. If the source is observed more than once, its \texttt{source\_id} and \texttt{source\_name} remain the same but the observation id (\texttt{shotid}) will be different as will the reported catalog measurements. We report a single representative detection identifier, \texttt{detectid}, which may be matched to the \textit{Detection Info Table} for each source observation in the \texttt{detectid} column; this column corresponds to the detection member with the brightest (i.e., smallest) \hetg\ value for all sources that are not LAEs. For LAEs, we use the highest $S/N$ \lya\ line detection as the selected representative \texttt{detectid}. A user may search the \textit{Detection Info Table} for this representative detectid by the selecting the column \texttt{selected\_det==True}.

For the \OII\ and \lya\ line fluxes, we provide the columns \texttt{flux\_oii} and \texttt{flux\_lya} and corresponding error columns for sources identified as \texttt{lae} and \texttt{oii}. As discussed in Section~\ref{sec:aperflux}, for each low-$z$ galaxy, an aperture \OII\ line flux is measured: \texttt{flux\_aper} at \zhet. This flux is assigned as the source's \texttt{flux\_oii} if it is a positive value and the major-axis of the galaxy based on broadband imaging is greater than 2\arcsec. Otherwise the line flux measured comes from the \texttt{flux} measured from the line fit to the extracted spectrum of the brightest \texttt{detectid} in the source group.  Line fluxes and associated errors are converted to intrinsic \OII\ and \lya\ line luminosities using our best measured redshift, \zhet, and the cosmology defined by \citet{Planck2018}.

The following files are included in this release:
\begin{itemize}
    
    \item \textit{Source Observation Table} (columns described in Table~\ref{tab:column_info}):
\begin{verbatim}
hetdex_sc1_vX.X.dat/.ecsv
\end{verbatim}
    This table consists of one row per source observation. For each source observation, it provides the source's J2000 equatorial coordinates, and redshift (\zhet). Every source is classified into one of the following source\_type options: \texttt{lae}, \texttt{oii}, \texttt{agn}, \texttt{lzg}, or \texttt{star} as described in Section~\ref{sec:classify}. For sources with either \lya\ or \OII\ line emission, the table provides the optimal measurement for the dust-corrected, aperture-corrected flux and luminosity in \texttt{flux\_lya}, \texttt{flux\_oii}, \texttt{lum\_lya} and \texttt{lum\_oii}. 
    
    \item \textit{Source Observation Table Spectra} in FITS format:
\begin{verbatim}
hetdex_sc1_spec_vX.X.fits
\end{verbatim}
    For each row in the \textit{Source Observation Table}, we provide a corresponding 1D extracted spectra in a FITS file format consisting of 7 Header Data Units (HDUs). Multiple HDUs are included as listed in Table\,\ref{tab:spec_format}. The primary HDU is empty. HDU1:INFO contains a copy of the \textit{Source Observation Table}. At the same row index for each source in this table, HDU2:SPEC and HDU3:SPEC\_ERR contain the aperture corrected, 1D PSF-weighted dust-corrected spectra and their associated uncertainties in \fluxden, computed according to the procedure outlined in Section~\ref{sec:dustcorrection}. HDU4:SPEC\_OBS and HDU5:SPEC\_OBS\_ERR are the aperture corrected, observed spectrum and associated uncertainty in \fluxden. HDU6:APCOR contains the applied aperture correction for each spectral element in the spectrum. The correction varies by wavelength due to the atmospheric diffraction correction. The final HDU7:WAVELENGTH is a 1036 array corresponding to the spectral dimension in \AA\. All spectra have the same spectral range from 3470\,\AA\ to 5540\,\AA\ in steps of 2\,\AA.
    
    \item \textit{Detection Info Table} (columns described in Table~\ref{tab:det_col_info}): 
\begin{verbatim}
hetdex_sc1_detinfo_vX.X.dat/.ecsv/.fits 
\end{verbatim}
    This table contains specific information for every curated line emission and continuum detection. Every emission-line detection row contains all parameter information, including the emission-line's observed wavelength, its fitted parameters and measured flux. If the observed wavelength corresponds commonly found spectral species\footnote{http://classic.sdss.org/dr6/algorithms/linestable.html} at redshift \zhet, the species is indicated in the column \texttt{line\_id}. There are also several columns related to imaging counterpart matches, redshift assignments, and emission-line classification as found by \elixer\ \citep{Davis2022}. Also included are a number of columns containing details about the specific observation, the instrument, and the detection grouping parameters. A full column description is provided in Table~\ref{tab:det_col_info}. Detailed information concerning this catalog is provided in Appendix\,\ref{appendix:1}.

\end{itemize}

\begin{table*}[t]
    \centering
    \caption{Format of the \textit{Source Observation Table} (columns described in Table\,\ref{tab:column_info}) Spectra FITS file}
    \begin{tabular}{lllcp{0.45\linewidth}}
    \hline
    HDU No. \& Name &   Type  &   Dimensions &  Format & Description \\ 
    \hline
    0:PRIMARY       & PrimaryHDU       &    &  & Empty\\     
    1:INFO       &    BinTableHDU     &   7367R x 27C   & & Source information for each catalog source, one row per source observation \\
    2:SPEC        &   ImageHDU         &   (1036, \nsourceobs) &  float32 & Dust- and aperture-corrected, PSF-weighted 1D spectrum at \texttt{detectid} in units of \fluxden. \\
    3:SPEC\_ERR     & ImageHDU         &  (1036, \nsourceobs)   & float32 & Uncertainty in SPEC \fluxden.   \\
    4:SPEC\_OBS      & ImageHDU         &   (1036, \nsourceobs)  & float32  & Aperture-corrected, 1D PSF-weighted spectrum at \texttt{detectid} in units of \fluxden.  \\
    5:SPEC\_OBS\_ERR   & ImageHDU         &   (1036, \nsourceobs) &  float32 & Uncertainty in SPEC\_OBS in \fluxden.   \\
    6:APCOR         & ImageHDU         &   (1036, \nsourceobs)  & float32   & Aperture correction applied to spectrum and catalogue flux values\\
    7:WAVELENGTH & ImageHDU & (1036,) & float32 & Wavelength array from 3470\,\AA\ to 5540\,\AA\ in 2\,\AA\ bins \\
    \hline
    \end{tabular}
    \raggedright Note. This file contains Table 3 (also available in a simple .dat/.ecsv ASCII format) in HDU 1. The aperture correction value applied to the spectra. The aperture correction is the fractional fiber coverage of the $r=3\farcs5$ aperture centered on the detection identified in the column \texttt{detectid}. 
    \label{tab:spec_format}
\end{table*}

\section{Sample Properties}
\label{sec:sampleproperties}

In this section we discuss some basic properties of the HETDEX Catalog sample. More in depth studies will be reported in future HETDEX papers. A basic discussion of the source count and magnitude distribution for each source type is provided in Section\,\ref{sec:source_dist}. Section\,\ref{sec:line_properties} provides an overview of line fit parameters. We outline some comparisons to imaging counterparts where available in Section\,\ref{sec:imag_counterpart} and measure the imaging counterpart fraction for both the LAE and OII samples. Sections\,\ref{sec:lum} and \ref{sec:zdist} cover the line luminosity and redshift distributions for the high-$z$ and low-$z$ samples. Using a internally confirmed sample of emission-line sources, we provide an upper limit on the LAE false-positive rate in Section\,\ref{sec:sample_validation}.

\subsection{Source Distribution}
\label{sec:source_dist}

\begin{figure}[ht]
    \centering
    \includegraphics[trim=0.5cm 0.5cm 0.5cm 0cm,clip,width=3.2in]{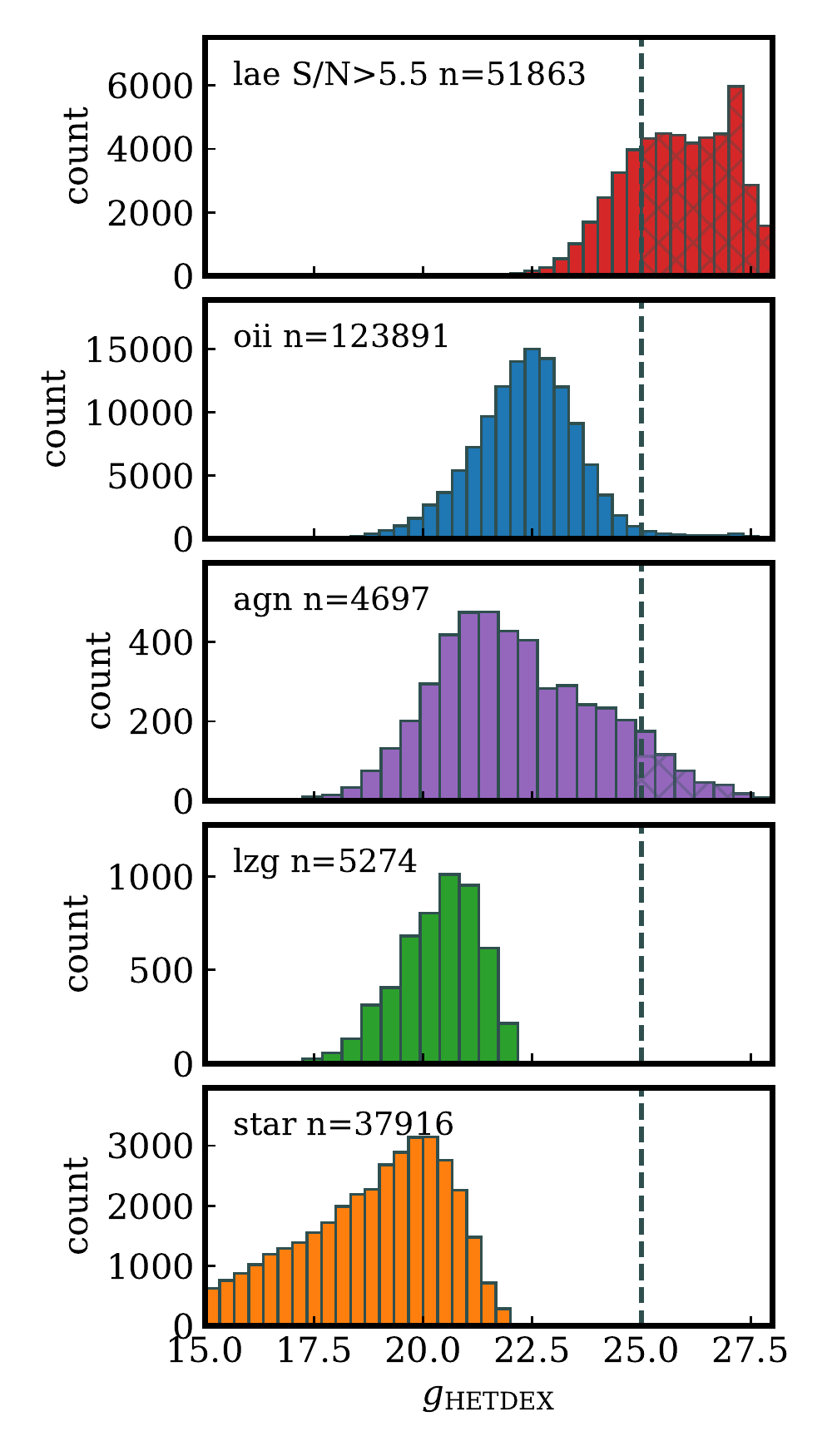}
    \caption{Histogram of \hetg\ magnitudes for each source type as measured by summing the 1D extracted spectra, weighted by the \sdss\ $g$-band response curve. If multiple detections comprise the source, the brightest detection is used. The vertical dashed line at \hetg=25 represents the HETDEX average sensitivity limit}
    \label{fig:g_hist}
\end{figure}

The unique count of astronomical sources is \nlae\ LAEs, \noii\ \OII\ emitting galaxies, \nagn\ AGN, \nlzg\ LZGs, and \nstar\ stars. Since the HETDEX observations contain several fields with repeated observations, a number of the sources have more than one entry.  Thus the total number of source detections is greater than the number of unique sources, and consists of \nstarobs\ stars, \noiiobs\ \OII\  emitters, \nagnobs\ AGN, \nlzgobs\ LZGs, and \nlaeobs\ LAEs. In the catalog, each source is allocated a common integer \texttt{source\_id} and \texttt{source\_name} where the latter descriptor follows the IAU standard, i.e., of the form \texttt{HETDEX~J123449.19+511733.7}. To search for repeated observations of the same object, the unique integer observation identifier \texttt{shotid} can be used. Since multiple detections can comprise a single source, we have provided the additional column \texttt{selected\_det==True} in the \textit{Detection Info Table} to indicate which detection best represents the source. If a source is observed multiple times, multiple \texttt{selected\_det==True} entries are in the catalog.

By field, the breakdown of sources for dex-spring:~\nspring, dex-fall:~\nfall, cosmos:~\ncosmos, and goods-n:~\ngoods, which is presented relative to field size and IFU count in Table\,\ref{tab:n_ifu}. In Figure~\ref{fig:g_hist}, the \hetg\ magnitude distribution is shown. The vertical dashed line shows the approximate limiting magnitude of observations; while the precise value varies with the  observing conditions, below \hetg$\approx25$, the ``magnitude'' is mostly the result of summed noise.  Note that the median continuum magnitude for $S/N>5.5$ LAEs is \hetg\ $\sim 25.9$, far below this threshold. In this sample, 84.7\% of the LAEs have \hetg$>24.5$.

The LAE sample used to produce the bright-end \lya\ luminosity function presented by \citet{Zhang2021} overlaps with the sample in this catalog. Their sample is derived from an earlier version of the catalog presented here. It also consists of an augmented LAE sample found by force extracting HETDEX spectra on known imaging sources in Hyper Suprime-Cam $r$-band \citep[HSC-$r$;][]{hsc} data and performing an independent line emission search. As they required HSC-$r$ band coverage for their work, their coverage consists of $\sim$45\% (11.4\,deg$^2$) of the coverage presented in this catalog (25.0\,deg$^2$). Cross-matching between both catalogs shows that 91\% of LAE candidates in \citet{Zhang2021} are recovered in our final source catalog and 95\% in the raw line emission-line database. The primary reason LAE candidates (about 1K LAE candidates) are not in the catalog presented here is they fail from stricter observational quality criteria cuts, updated masking and different line parameter criteria. Visual inspection suggests about three quarters of the missing sample are false positives due to noise and artifacts while the rest are confident LAE candidates that are culled due to quality cuts. We do note that a substantial fraction (about 10\%) of their LAE candidates are mis-classified. Many of these LAEs are actually low-redshift emission-line sources (primarily due to \OII\ emission, but some are due to \hbeta, \OIII\ and other line emission) in the \textit{HETDEX Source Catalog}.

Low-$z$, \OII\ emitting galaxies are most numerous in our catalog. These objects span a wide range of magnitudes ranging from \hetg$\sim 17.5$ to below the catalog's sensitivity limit (based on a threshold continuum requirement for object detection not the sensitivity limit of HETDEX observations), with a median value of  \hetg$\sim 22.4$. The faint end of this distribution overlaps that of the LAEs and illustrates the need for additional observational criteria in determining whether a line emitter is from a low-$z$ galaxy or from \lya.

The third panel in Figure~\ref{fig:g_hist} shows the AGN \hetg\ distribution. This distribution has a median magnitude that is slightly brighter than that for the \OII\ galaxies, \hetg$\sim 22.0$, and contains both bright continuum sources with broad line emission and a fainter population in which either multiple emission lines (i.e., \lya, \CIV, \HeII) or a single broad emission feature is found. In some cases, broad extended line emission can have very little continuum associated with it and it is often spatially extended suggesting the possibility that the heating mechanism is perhaps not due to an AGN.

The LZG population is mostly detected in the continuum detection search, although a detection of line emission can also result if there is a continuum peak within two absorption troughs or if there is an abrupt change in an object's continuum level. Not surprisingly, LZG galaxies are generally brighter than \OII\ emitters, with a median \hetg\ of 20.4 mag, and they are much less numerous.  But there is considerable overlap in the populations, and both object classes are present in our continuum catalog. Initially many \OII\ galaxies were missed by our line-emission search, and were only discovered by measuring spatially resolved \OII\ aperture fluxes (see Section~\ref{sec:aperflux}). If we consider just continuum detected sources, which can loosely be considered analogous to a magnitude limited survey, then 16.3\% of the low-$z$ galaxies have no measured line emission, while 82.9\% have measured continuum-subtracted \OII\ emission.  Alternatively, if we consider all HETDEX low-$z$ detections (both emission-line and continuum objects) then just 2.7\% of the catalog consists of LZGs and 97.2\% are \OII\ systems. 

Both the HETDEX line and continuum emission catalogs contain entries from stellar emission. As with LZGs, discontinuity jumps in the continuum can be mistaken for emission features; our spatial clustering ensures that duel detections are properly merged into a single source.  The stars in the HETDEX catalog can be as bright as \hetg$\sim12.3$ with a median value of \hetg$\approx19.0$. The number of stars in this catalog is considerably smaller than that reported in the HETDEX-\textit{Gaia} catalog \citep{hawkins2021}, due to the 
very different selection criteria and methods of detection applied. In \citet{hawkins2021}, spectral extractions were performed at the known locations of $10<G<22$ \textit{Gaia} DR2 stars \citep{Gaia2018} using the same data release (HDR2) presented here. Their catalog consisted of 98,736 unique stellar candidates; our catalog contains \nstar\ of these objects. 

The major difference between the two catalogs lies in the hetereogeneous sensitivity of the HETDEX continuum detection level. At magnitudes of \hetg$<15$, bright stars and galaxies saturate the detectors, making it impossible to properly perform a flux calibration.  When this happens, the detector may be useless for HETDEX science, but may still produce detectable and classifiable stellar spectra.  Moreover, as discussed in Section~\ref{sec:quality}, frames with bright stars or additional detector issues will fail our observation quality criteria and are removed from the survey; the forced extraction methods performed by \citet{hawkins2021} did not apply these additional criteria, and so will include more objects.  In addition, our continuum detection search was essentially counts limited, rather than magnitude limited. Depending on observing conditions, this restricted our continuum-selected sample to objects brighter than $g\sim21-22.5$. Finally, we note that it appears the HETDEX-\textit{Gaia} catalog has some likely galaxy contaminants: out of the \nstar\ overlapping sources, 87.3\% were classified in the HETDEX Source Catalog as stars, 3.9\% as AGN, 6.6\% as \OII\ galaxies and 2.0\% as LZGs. Indeed when measuring radial velocities, \citet{hawkins2021} report low level ($\sim2\%$) contamination by galaxies.

\subsection{Line Parameter Properties}
\label{sec:line_properties}

\begin{figure}[t]
    \centering
    \includegraphics[width=3.in]{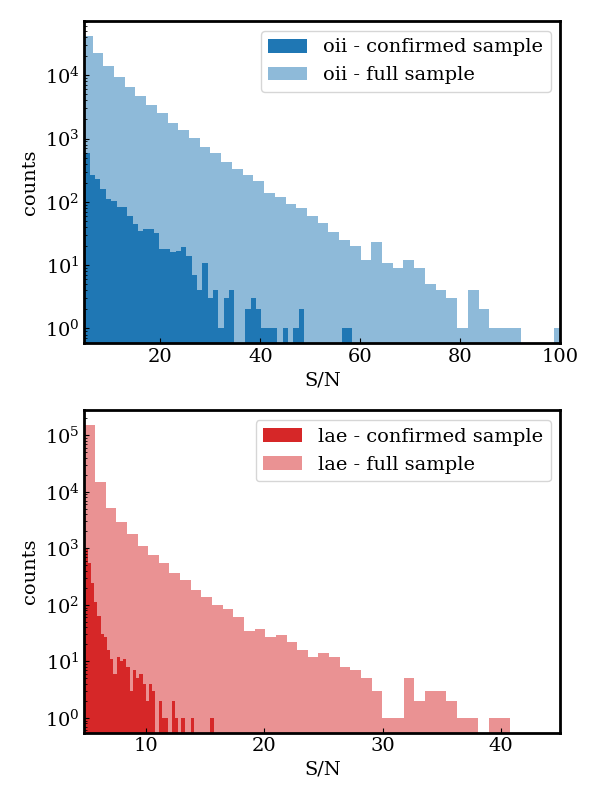}
    \caption{Signal-to-noise ratio distributions. The dark colors show the confirmed sample of emission-line sources detected multiple times in independent HETDEX observations; the lighter colors display the full catalog distribution.  LAEs are generally fainter than \OII\ galaxies and have lower values of $S/N$. Note the difference in the $x$-axis ranges in the panels.}
    \label{fig:sn_dist}
\end{figure}

As described in Section~\ref{sec:linecriteria}, the Gaussian amplitude (which provides the line flux), line width ($\sigma$) and continuum level are the free parameters of the emission-line fit. The central location of the line detection is determined by rastering the line-fit at high spatial resolution to maximize on $S/N$ of the fit. The \textit{Detection Info Table} reports theses values and the quality of fit, $\chi^2$. We employ variable cuts on the line parameters when we create the curated line emission catalog (as discussed in Section~\ref{sec:linecriteria} and described in Equation\,\ref{eq:criterion-1} and \ref{eq:criterion-2}). The internal HETDEX Source Catalog is limited to line emission detections with $S/N>4.8$, while the public version excludes all LAEs with $S/N<5.5$. At the lower value, confidence in the line detection goes down and the visual line-fits are of poor quality.

In Figure~\ref{fig:sn_dist}, we plot the $S/N$ distribution for the LAEs and \OII\ galaxies in the top and bottom panels, respectively. The lighter colors are for the full sample included in the catalog; the solid region is for an isolated ``confirmed'' sample with at least three independent observations. Many of these objects are from the science verification fields, which were subjected to multiple visits (see Section~\ref{sec:counterpartfraction} for a complete discussion).  The sample of sources confirmed by multiple observations have a similar $S/N$ distribution as the objects in the full catalog although the number counts go significantly down at higher $S/N$ as sample variance plagues the ``confirmed'' sample. For both samples, as expected, the number of sources increases as $S/N$ decreases. Unsurprisingly, the \OII\ galaxies extend to much higher $S/N$ than the LAEs.

\begin{figure*}[t]
    \centering
    \includegraphics[width=7in]{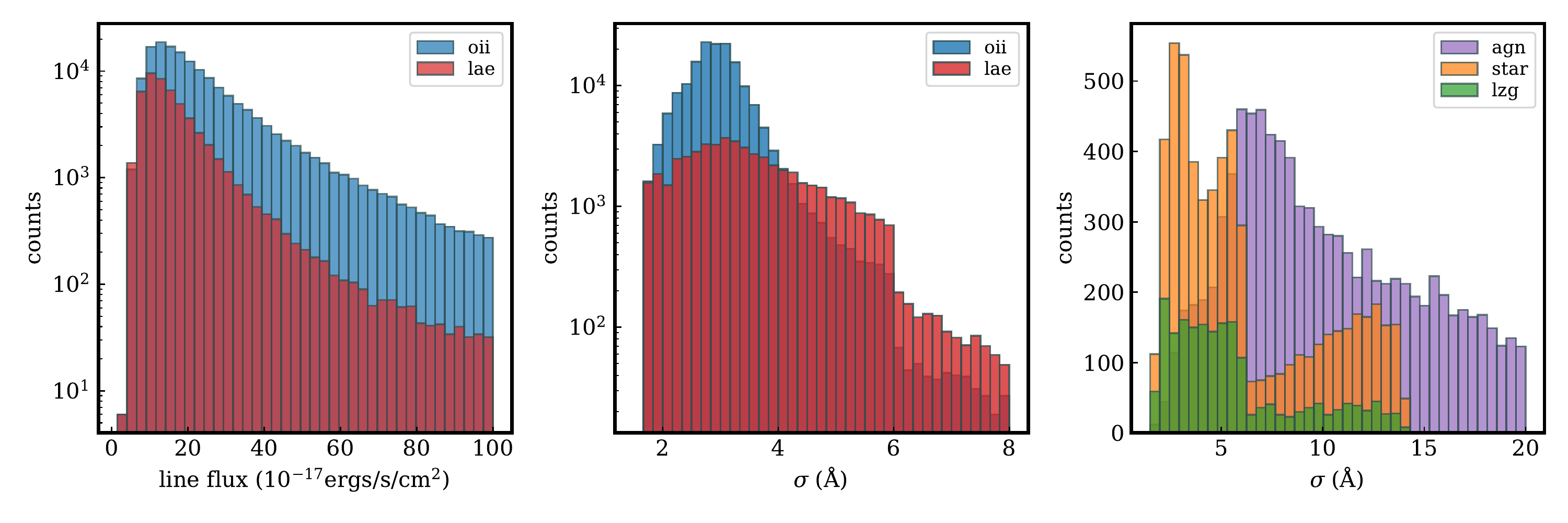}
    \caption{The left panel shows the observed line fluxes for the for LAEs (in red) and \OII\ galaxies (in blue). The middle panel shows the fitted Gaussian line width distribution in \AA\ for the LAEs (in red) and \OII\ galaxies (in blue).  The right panel gives the same information for AGN, LZGs and stars.  These classes are less numerous and exhibit a broader line distribution that for emission-line galaxies.}
    \label{fig:sigma_dist}
\end{figure*}

The left panel of Figure~\ref{fig:sigma_dist} shows the distribution of \lya\ and \OII\ line fluxes for the two main line emission samples. \lya\ line flux values range from 3.4 to 1030~$\times$~\fluxden\ where the 10th, 50th, 90th percentiles are 8.3, 14.5, 32.8~$\times$~\fluxden\ respectively. The \OII\ line fluxes range from 3.7 to 2960~$\times$~\fluxden\ where the 10th, 50th, 90th percentiles are 10.2, 20.0, 53.2~$\times$~\fluxden\ respectively. Distributions for the AGN sample can be found in \citet{liu2022}. Some line emission sources were observed multiple times in this release, and have multiple measurements in the catalog. These can be isolated by searching for a common \texttt{source\_id} value but a different observation id (\texttt{shotid}) value. The rms scatter in repeated line measurements is 13.1\% if comparison sources are required to be within 0.5\arcsec. This value increases to 13.8\% without any sky separation requirement suggesting source position plays a role in line flux accuracy as discussed earlier in Section\,\ref{sec:sdss_comp}. Simulations described in \citet{Gebhardt2021} suggest even higher statistical uncertainty of 25-30\% that is signal-to-noise dependent. A better flux accuracy is measured in the presented catalog likely due to a stricter $S/N$ requirement of 5.5.

\begin{figure}[t]
    \centering
    \includegraphics[width=3.25in]{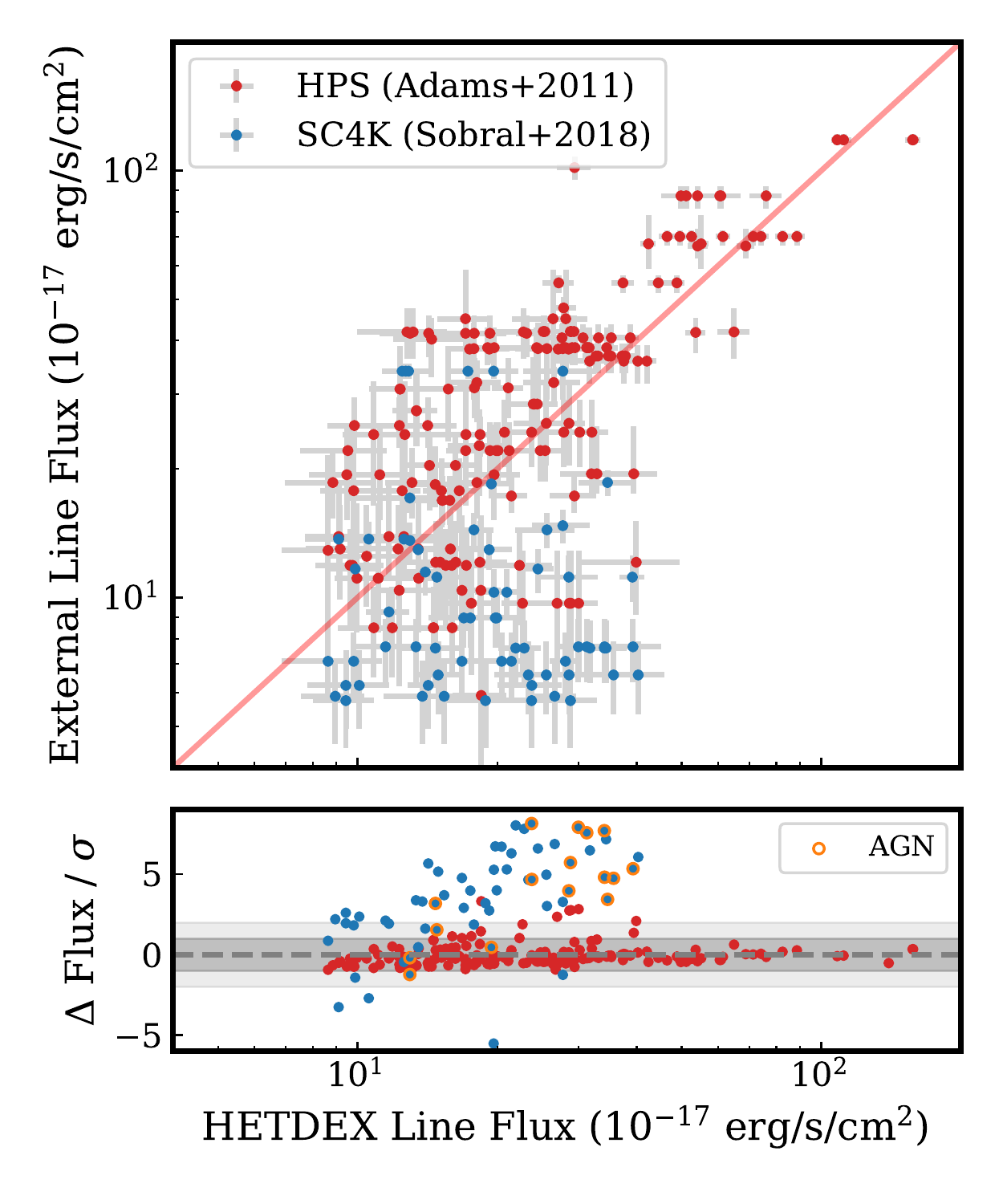}
    \caption{Top panel is a comparison of point-source line fluxes measured from the HETDEX line detection pipeline to two external surveys, SC4K \citep{Sobral2018} in blue, and HPS \citep{Adams2011a} in red. The comparison is comprised of LAEs, AGN and OIIs although the SC4K sample is limited to just LAEs and AGNs at high-$z$. The bottom panel compares the line flux difference (HETDEX minus the external values) in the samples relative to the combined uncertainty. The dark-grey shaded region indicates 1$\sigma$ agreement and the light-grey shaded region indicates 2$\sigma$ agreement. For the HPS sample, they agree within 1$\sigma$ 92\% of the time. HETDEX line fluxes are overestimated relative to the SC4K sample. AGNs from the HETDEX-SC4K overlap sample are indicated in orange and represent 1/3rd of the comparison sample whose fluxes disagree.}
    \label{fig:flux_comp}
\end{figure}

A comparison to two external samples of published line flux values is shown in Figure\,\ref{fig:flux_comp}. Here we compare a strictly LAE sample from the SC4K survey \citep{Sobral2018} shown in blue, as well as a mix of \OII\ and \lya$-$emitting galaxies from the HETDEX Pilot Survey \citep[HPS;][]{Adams2011a}. SC4K uses 16 different narrowband and medium-band filters over the COSMOS field to select a large sample of LAEs. For these emitters, we require a match to within 1\arcsec\ spatially and within 300\,\AA\ of the HETDEX emission-line wavelength. We find 50 LAEs and 17 AGNs overlap with our catalog. We find 30\% of the HETDEX line fluxes are more than three times the reported SC4K fluxes with 50\% of these being AGN in our catalog. SC4K measures their fluxes in 2\arcsec\ apertures and continuum is measured in multiple narrowband filters potentially leading to inconsistencies in the measurement which is most significant for the AGN sources. Designed as a HETDEX validation survey, HPS is similar in design to HETDEX. HPS used the VIRUS prototype IFU \citep[VIRUS-P;][]{hill2008} on the 2.7\,m Harlan J. Smith telescope at the McDonald. Three-dithered exposures provide similar coverage at similar resolving power, with the main difference being VIRUS-P's larger fiber size (4\farcs2 compared to VIRUS's 1\farcs5 diameter fibers). Data is reduced in a completely independent pipeline from the current HETDEX pipeline and line fluxes are measured differently. In total, 179 detections overlap within 8\,\AA\ spectrally and 1\,\arcsec\ spatially; 128 of these are OIIs, 37 are LAEs and 12 are AGN. The bottom panel shows the line flux differences (HETDEX - HPS) are consistent to within 1$\sigma$ of the combined lines flux uncertainties 92\% of the time.

The fitted Gaussian line width distributions for the \OII\ and LAE samples have roughly the same range.   As shown in the middle panel of Figure~\ref{fig:sigma_dist}, Ly$\alpha$ emitters show a broader distribution and have a higher frequency of broad line emitters relative to the \OII\ sample, as might be expected \citep{kulas2012,chonis2013}. The figure also shows a
substantial decrease in the distribution at $\sigma=6$: our fitting criteria removes all lower signal-to-noise ($S/N<6.5$) lines with $\sigma>6$ (see Equation\,\ref{eq:criterion-1}). The drop off suggests we are losing real sources with this criteria, especially in the LAE sample at a 0.001\% level. Fortunately, the HETDEX AGN Catalog selection does explore this parameter space. In the higher line width regime, visual inspection confirms the existence of many artifacts due to issues involving both the detector and the calibration.  

\subsection{Imaging Counterpart}
\label{sec:imag_counterpart}

We run each line and continuum detection through our source classifying software \elixer\ \citep{Davis2022} to obtain photometric magnitudes at the direct location of the HETDEX detection. We use the multiwavelength coverage provided by the Hyper Suprime-Cam through the Subaru Strategic Program \citep[HSC-SSP;][]{HSC-SSP,HSC-SSP-PDR3} as well as internally obtained HSC-$r$-band imaging as described in \citet{Davis2022}. Data reduction and source detections were performed with version 6.7 of the HSC pipeline, hscPipe \citep{Bosch2018a}, and produced $r$-band images with a 10$\sigma$ limit of $r = 25.1$ mag in a $2\arcsec$ diameter circular aperture. 

\begin{figure}[t]
    \centering
    \includegraphics[width=3.5in]{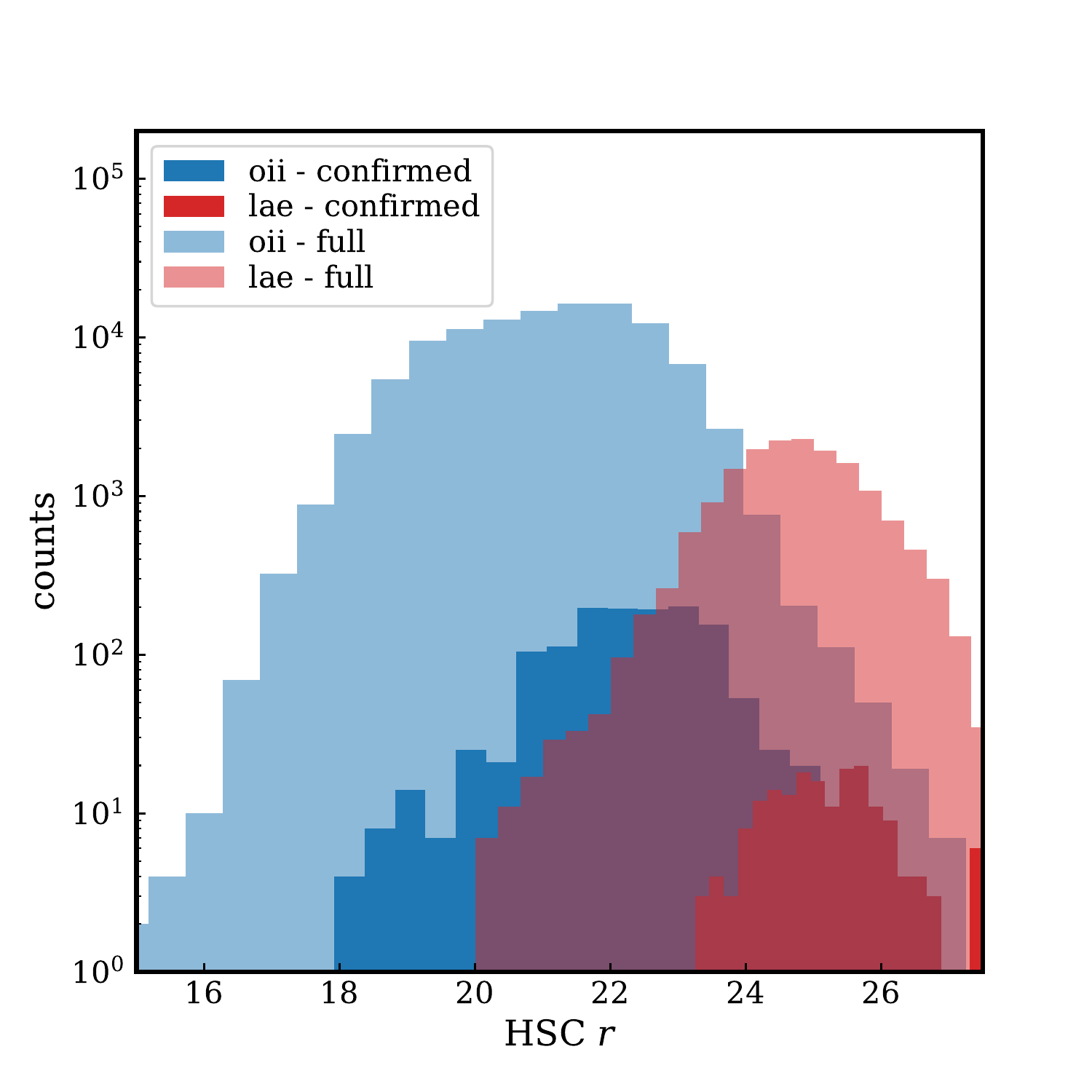}
    \caption{Distribution of HSC $r$-band image counterparts for the main catalog LAE and OII samples. Also shown are for a ``confirmed'' sample of sources which are high-confidence HETDEX detections, which are detected on at least three independent nights. These values come from the \elixer\ catalog SEP measurements performed on the image at the location of the HETDEX sources as described in \citet{Davis2022}. The comparison between the samples is to show that the confirmed sample is a good representation of the full line emission.}
    \label{fig:counter_rhist}
\end{figure}

The magnitude distribution of imaging counterpart to HETDEX sources, as measured in the HSC $r$-band, are shown in Figure~\ref{fig:counter_rhist} for the LAE and OII catalog sources. We show the full sample in shaded opacity while highlighting the confirmed samples in solid colors to show the overall consistencies in the two populations. \OII\ galaxies are brighter than the LAEs with a median value of $r=21.6$. In contrast, the LAE sample has a median magnitude of $r=24.7$ but we note that this distribution is biased due to the sensitivity limit of the $r$-band imaging data. 

\subsubsection{$g$-band Magnitude Comparison}

\begin{figure*}[t]
    \centering
    \includegraphics[width=6.5in]{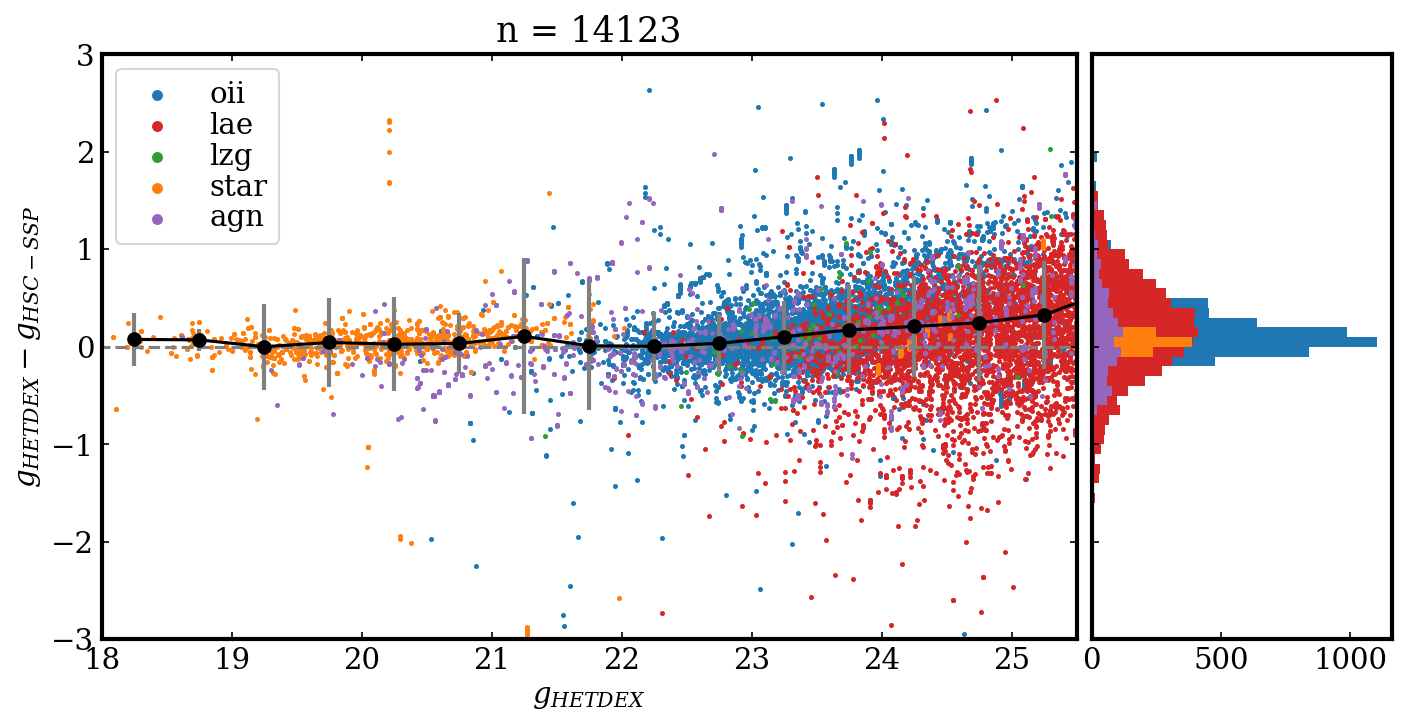}
    \caption{We compare the $g$-band magnitude to all imaging counterparts in deep $g$-band data from HSC-SSP. Depending on conditions, HETDEX can be sensitive to $g\sim25$\,mag.}
    \label{fig:gmag_comps} 
\end{figure*}

In Section~\ref{sec:source_dist}, we described the \hetg\ magnitude distributions for the catalog. Continuum sensitivity from a single extracted HETDEX spectrum varies based on observing conditions, but is generally reliable to \hetg$\sim 25$, although the uncertainties can be large. We compare these magnitudes with $g$-band measurements calculated with our \elixer\ software on $g$-band data from the HSC-SSP program which reach a sensitivity of $g$~=~26.5\,mag.

In Figure~\ref{fig:gmag_comps}, the difference in magnitude is plotted as a function of \hetg. The stellar sample shows an offset of 0.05 mag with a scatter of 0.51 mag. The \OII\ sample has an offset of 0.14 mag with a scatter of 0.41 mag. The LZG has the largest offset of -0.34 mag with a scatter of 0.57 mag. The AGN sample has an offset of 0.13 mag with a scatter of 0.54 mag. For both the \OII\ and LZG samples, the aperture effects are important, as HETDEX magnitudes are for point-source measurements and the HSC-SSP measurements are for varying apertures. We intentionally exclude any sources with a major-axis greater than 3\arcsec\ to mitigate aperture affects. Unsurprisingly, the faint LAE sample which pushes the sensitivity limits of HETDEX shows the largest scatter of 0.59 mag, but a modest offset of 0.17 mag. An overall trend line shows that the offset is largest at faint magnitudes where lower a $S/N$ in the HETDEX spectra leads to a lower integrated flux.

\subsubsection{Imaging Counterpart Fraction}
\label{sec:counterpartfraction}

In HSC $r$-band, our preferred photometric data set, many LAEs in our sample have no apparent counterpart. This is not surprising given the LAE sample exhibits stellar masses ranging from $\sim10^{8}-10^{10}\,M_\odot$ \citep{Hagen2016, Oyarzun2017, McCarron2022}. However, we also recognize that a lack of imaging counterparts can be an indication of potential false-positive contamination. This is specific to the high-$z$ sample since the low-$z$ identifications require that a continuum be detected in either the spectral data or the accompanying imaging.   Thus by definition the \OII\ sample has less contamination from noise and artifacts.

One way to investigate the amount of contamination present in our sample is to compare the fraction of confirmed LAEs with imaging counterparts to that of the full catalog sample. There are a number of ways to confirm the presence of line-emission for a HETDEX detection. For example, sources confirmed by other instruments  can confirm an object as real (as well as provide confirmation of the redshift and classification). Alternatively, we can use the HETDEX data themselves: if line-emission is detected in three or more independent observations, we confirm the source to be real.  Although HETDEX tiling is designed to visit the sky just once, there are a number of science verification fields (in legacy regions such as COSMOS and GOODS-N) that HETDEX has visited on numerous occasions.  In a few cases, observations were redone, due to sub-standard observing conditions. Although the unacceptable observations are not in our final catalog, the emission-line detections remain in the raw line database and are used for object verification. Finally, in a few cases, the corners of IFUs overlap (see for example the overlap in the right panel of Figure~\ref{fig:coverage2}), due to tiling changes associated with the increasing number of active IFUs over time. We call the set of objects identified in three or more independent observations our ``confirmed'' sample. We show the $S/N$ and HSC-$r$-band imaging counterpart magnitudes for the confirmed sample relative to the full catalog sample in Figures\,\ref{fig:sn_dist} and \ref{fig:counter_rhist}. The similar distributions indicate the sample is representative of the full catalog.

We consider two different catalog samples, all of which require accompanying HSC-$r$ imaging coverage, for this test: the publicly provided catalog LAE sample (n=39,083) and OII sample (n=86,357). We compare these to their counterparts in the confirmed dataset: n=422 for the LAE sample, n=1529 for the OII sample. To be in these sub-samples, we require that each source be contained on an HSC $r$-band image, which we assume have a depth better than $r=26.2$\,mag ($5\sigma$ limit). For each sample, we calculate the number of HETDEX detections with an $r$-band counterpart as measured in either of two ways: (1) through on-demand source extraction applied to the HSC image using the \elixer\ software tool or (2) a forced extraction at the exact position of the HETDEX detection and measured within an $r=1\farcs5$ circular aperture. 

In Table~\ref{tab:counterpart}, we summarize the net fraction of LAEs and OII galaxies with HSC $r$-band counterparts. The first thing to note is that the fractions for the confirmed sample and the main sample are consistent to within a few percentage points. This implies that the catalog is relatively free of contamination by false positives. In a catalog with many false detections, the fraction of LAEs with counterparts would be much greater in the confirmed sample than in the main sample.  

\begin{figure}[t]
    \centering
    \includegraphics[width=3.7in,trim=0cm 1cm 0cm 2cm,clip]{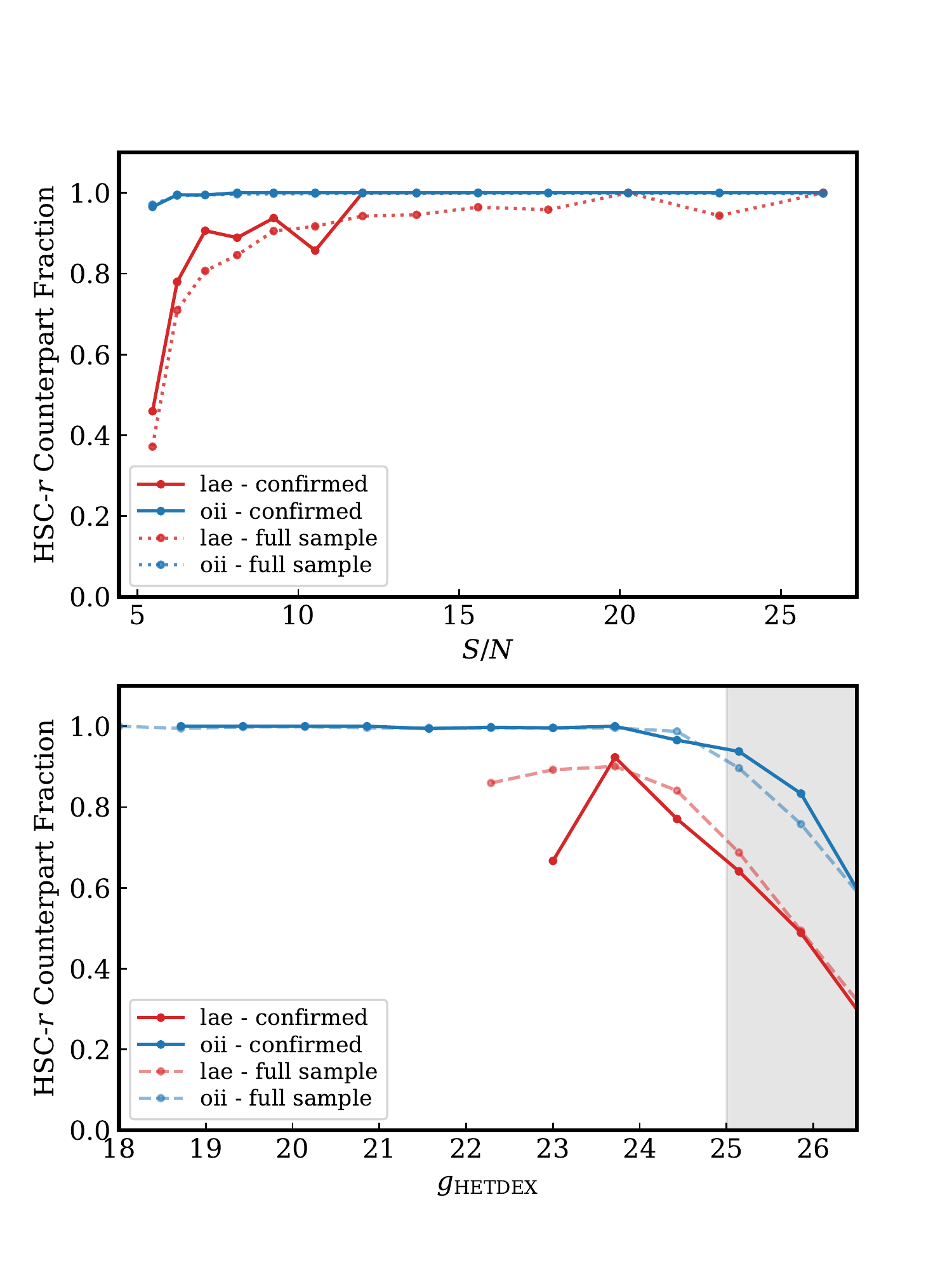}
    \caption{The fraction of OII and LAE sources with $r<26.2$-band HSC counterparts, as a function of S/N (top panel) and \hetg\ (bottom panel). The solid line is for a subset of sources that are confirmed by three independent HETDEX observations of the source.}
    \label{fig:counter_fraction}
\end{figure}

In Figure~\ref{fig:counter_fraction}, we consider how the fraction of objects with $r$-band counterparts depends on the $S/N$ of the line (top) and \hetg\ (bottom). The counterpart fraction increases with the $S/N$ on the LAE emission line.  This implies the brighter emitters tend to also have higher line flux in our catalog as seen in previous LAE studies where equivalent width has a minimal dependence on line luminosity (e.g. \citealt{Gronwall2007a, Ciardullo2013b}).  By design, nearly all OII sources are present on the $r$-band images, as a continuum detection (either through imaging or spectroscopy) is needed for a detection to be classified as OII. There are some OII  detections that have faint \hetg\ values and are consequently less likely to be observable in the $r$. These are sources that have been classified as OII by \elixer\ even though they have a very weak continuum. Upon inspection, these often tend to be false positives, due either to calibration issues or satellites.

The measured imaging counterpart fraction demonstrates the strength of an IFU-based LAE survey. Just over half of the LAE sample have image counterparts brighter than $r=26.2$. A survey based on imaging preselection at this sensitivity would miss half of the objects that HETDEX is using to trace the large-scale structure of the $1.9 < z < 3.5$ universe.

\begin{table}[t]
    \centering
    \caption{HSC $r$-band Imaging Counterpart Fraction} 
    \begin{tabular}{ccc}
    \hline
    \hline
            & Full Catalog & Confirmed  \\
    \hline
    LAE &	55.8\% (21,824/39,083) &	52.4\% (221/422) \\
    OII &	99.3\% (85,754/86,357) & 99.2\% (1517/1529) \\
    \hline
    \end{tabular}
    \raggedright The ``Confirmed'' sample of line-emitters consists of objects that have been independently detected in three different observations. The detected objects have $r$-band counterparts brighter than $r=26.2$.
    \label{tab:counterpart}
\end{table}

\subsubsection{Source positioning}

\begin{figure}[t]
    \centering
    \includegraphics[width=3.25in]{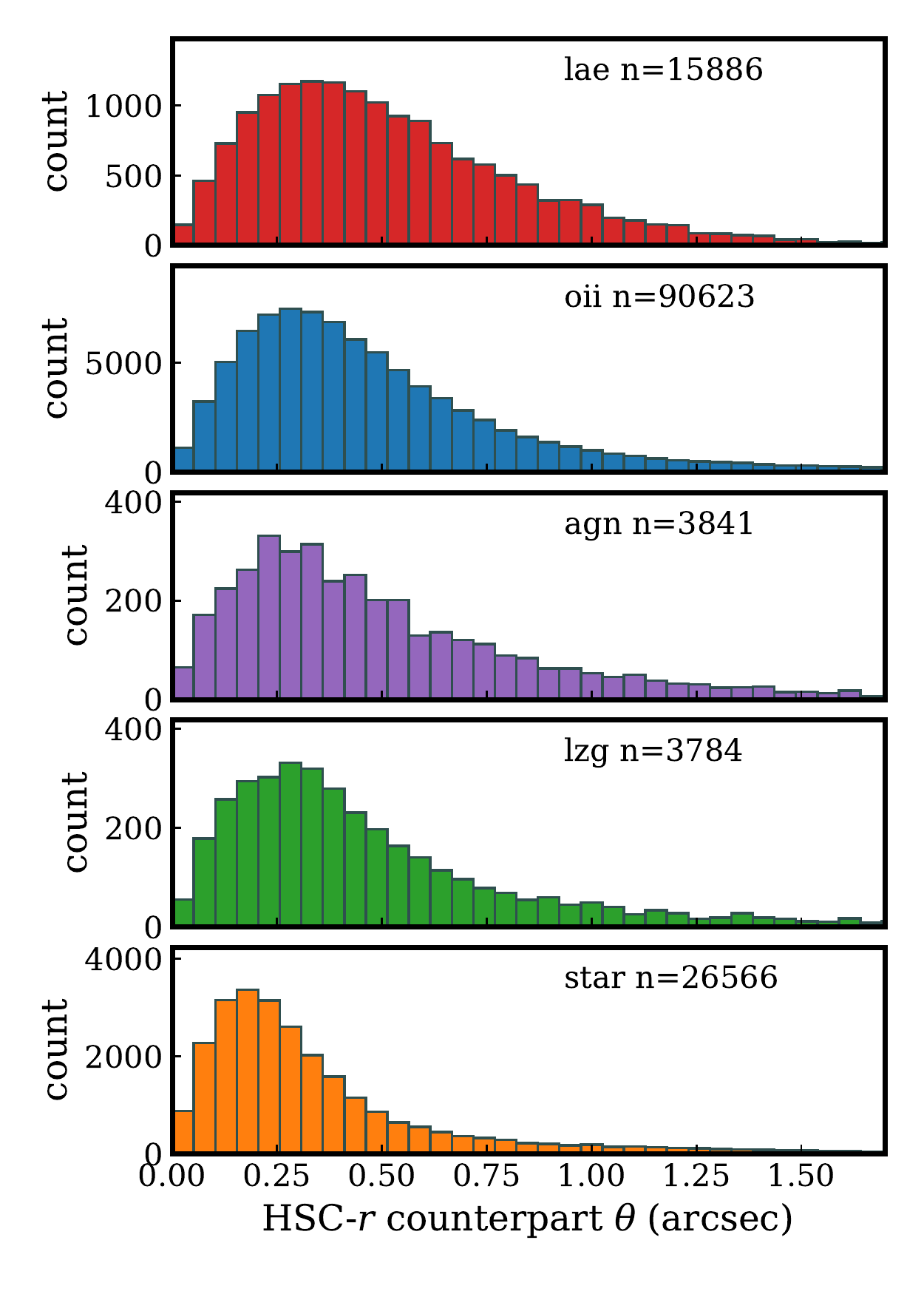}
    \caption{Distributions of sky separations between HETDEX source detections and their imaging counterparts. The LAE sample shows the widest spatial distribution primarily due to poorer centroid positioning of low $S/N$ HETDEX detections. The star sample in orange at the bottom shows the tightest distribution with a median sky separation of 0\farcs27.}
    \label{fig:counterpart_separation}
\end{figure}

The \textit{Detection Info Table} contains selected output from the \elixer\ catalog about the imaging counterparts of HETDEX detections. Included in these data are the separations between the HETDEX sources and the position of the most likely imaging counterparts (labeled as \texttt{counterpart\_dist}). Figure~\ref{fig:counterpart_separation} shows distributions of these separations for each source type. The LAE sample shows the widest distribution. This is primarily attributed to a poorer ability to center low $S/N$ emission lines within the PSF-weighted VIRUS spectral extraction. \citet{McCarron2022} find in deep GOODS-N HST imaging, HETDEX imaging counterparts can be up to 1\arcsec\ in separation from the HETDEX detection center.

The star sample in orange at the bottom shows the tightest distribution with a median sky separation of 0\farcs27. This is comparable to 
our astrometric uncertainties ($\sim 0\farcs 2$).  OII galaxies, AGNs and LZGs, have a wider distribution as there can be differences in where the source center lies.

\subsection{Luminosities}
\label{sec:lum}

\begin{figure}[t]
    \centering
    \includegraphics[width=3in]{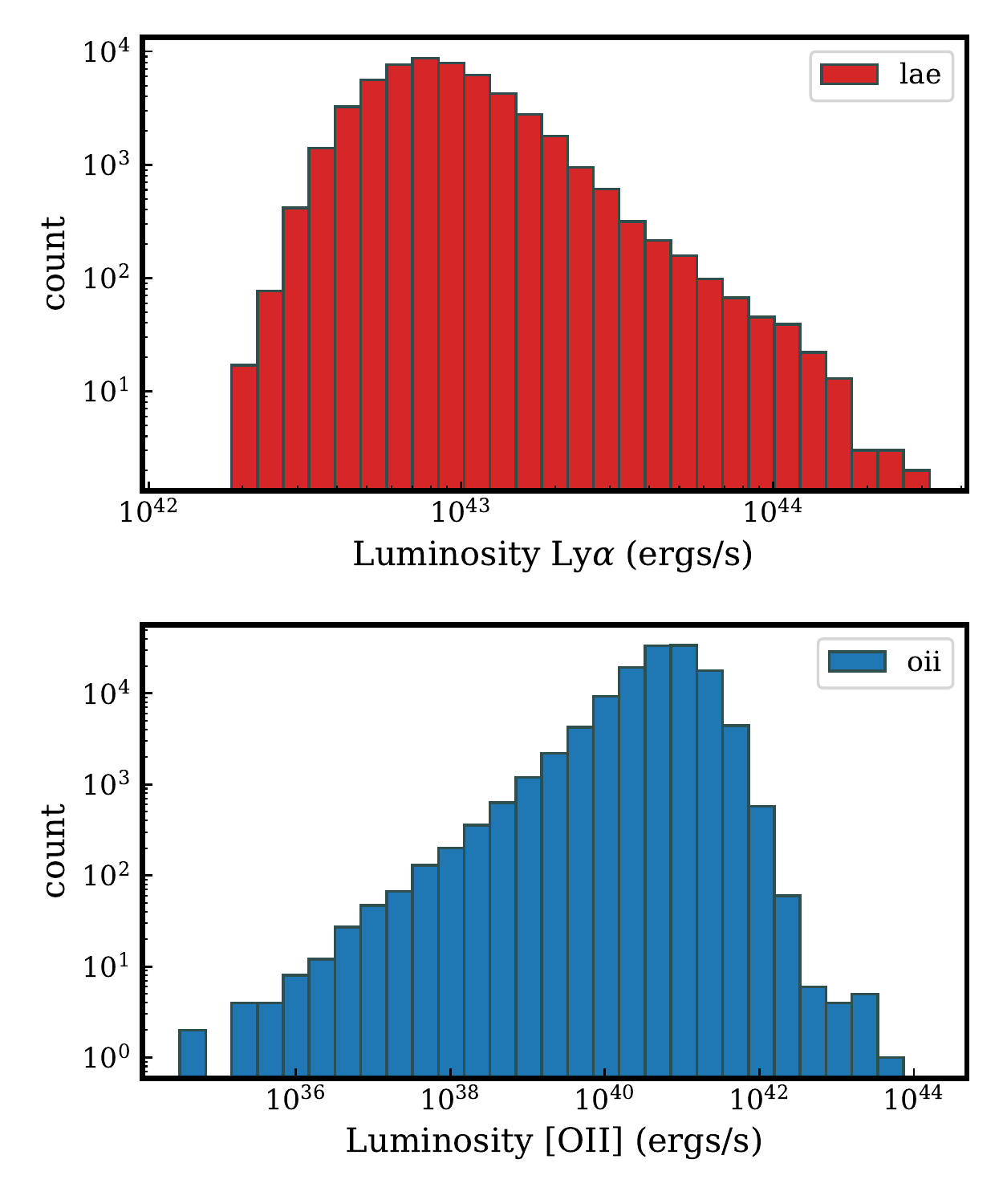}
    \caption{Top: \lya\ Luminosity distribution for \nlae\ LAEs (solid red). The lower $S/N$ dataset has a larger fraction of lower luminosity sources but both samples span a similar luminosity range. Bottom: \OII\ luminosities for the low-$z$ line emission galaxies. Here 88\% of the values come from resolved aperture line fluxes (\texttt{flux\_aper}). The rest are assumed to be point-source like and are the HETDEX pipeline fluxes (\texttt{flux}). }
    \label{fig:lum_hist}
\end{figure}

In Figure~\ref{fig:lum_hist}, we show the emission-line luminosities for LAEs and the OII galaxies. The \lya\ line luminosities range from $1.84\times10^{42}$~ergs~s$^{-1}$ to $2.85\times10^{44}$~ergs~s$^{-1}$ with a median value of
$8.31\times10^{42}$~ergs~s$^{-1}$. For each LAE source, the corresponding \lya\ flux and luminosity are found in the columns \texttt{flux\_lya} and \texttt{lum\_lya}.  For the OII galaxies, 87.9\% of the values (as indicated by \texttt{flag\_aper==1}) are from resolved aperture line fluxes (\texttt{flux\_aper});  the rest are assumed to be point-like sources and come from HETDEX pipeline (\texttt{flux}). The selected OII flux values can be found in the \textit{Source Observation Table} (see Table~\ref{tab:column_info}) in the column \texttt{flux\_oii} and the luminosities are given in column \texttt{lum\_oii}. The \OII\ line luminosities of our sample range from $6.13\times10^{32}$~ergs~s$^{-1}$ to $1.76\times10^{44}$~ergs~s$^{-1}$ with a median value of
$1.96\times10^{40}$~ergs~s$^{-1}$. 

\subsection{Redshift Distribution}
\label{sec:zdist}

\begin{figure}[t]
    \centering
    \includegraphics[width=3.25in]{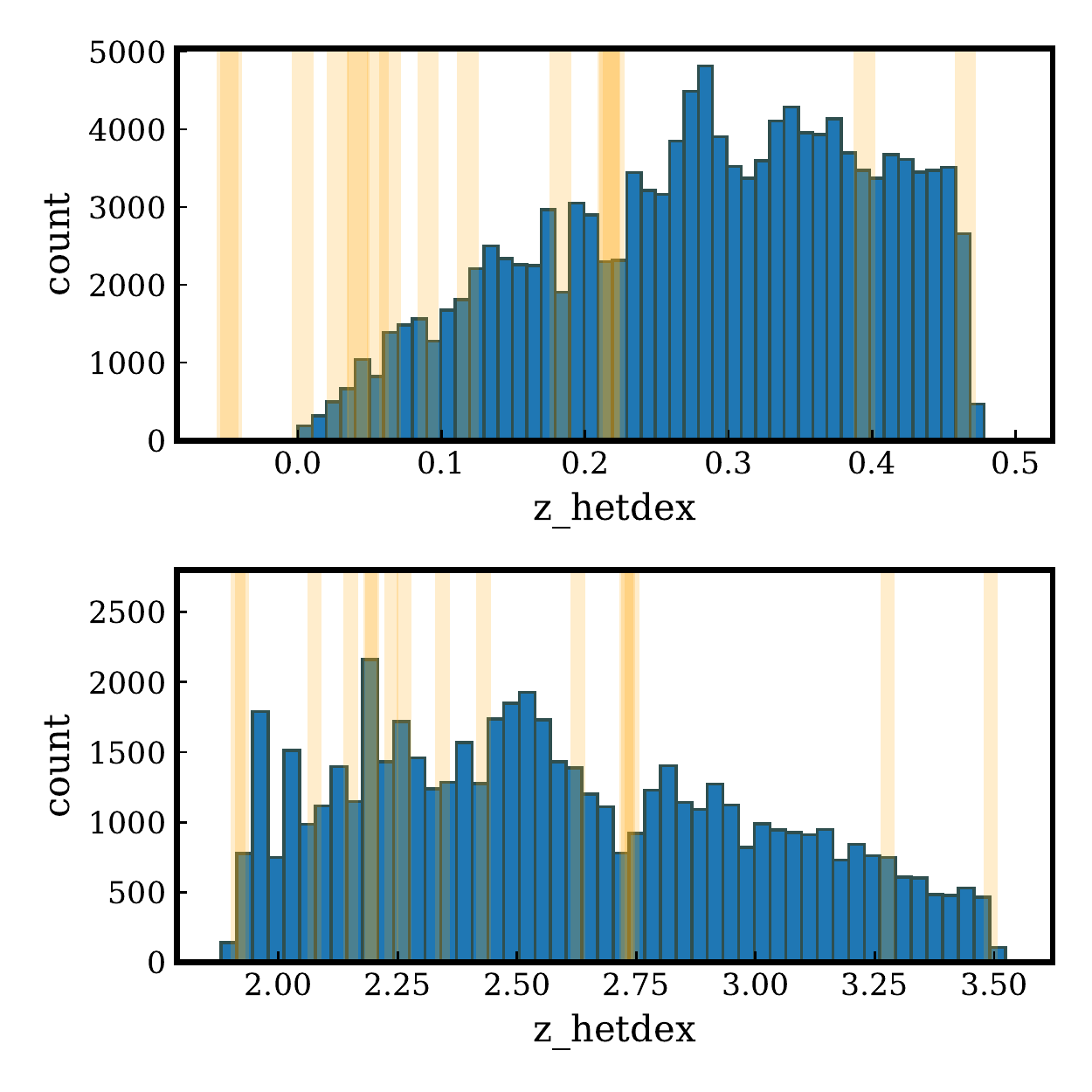}
    \caption{The redshift distribution of the low-$z$ and high-$z$ galaxy samples in the top and bottom panels respectively. The low-$z$ sample is a combination of OII emitters and LZGs; the high-$z$ dataset is limited to LAEs with $S/N>5.5$. The brightest sky lines are marked by light yellow bars, which cause a suppression in number counts in the high-$z$ distribution.}
    \label{fig:z_hist}
\end{figure}

In Figure~\ref{fig:z_hist}, we show the redshift distribution of the low-$z$ and high-$z$ galaxy samples.  For the low-$z$ sample, galaxies with (\texttt{source\_type==`oii'}) and without (\texttt{source\_type==`lzg'}) \OII\ line emission  are shown together; their counts increase with $z$, as the greater volume at higher $z$ is more important than the accompanying decrease in survey depth.  The dip at \zhet$\sim0.21$ in the low-$z$ sample (and at \zhet$\sim2.7$ in the high-$z$ dataset) is due to a mask that is applied at the center of 50\% of the detectors as well as an increase in night sky emission. Night sky emission, particularly in the blue, causes marked decreases in number counts in the lower redshift regions of both datasets but is most notable at high-$z$. The brightest sky lines are marked by light yellow bars on Figure~\ref{fig:z_hist}. At these epochs, the loss in depth due to increased distance outweights the volume effect.  
Sample variance can play a small role in the variation in counts, but given the size of the survey this should largely be mitigated. The remaining variability in counts as a function of wavelength is due to the complex sensitivity variations caused by variable observing conditions, detector spectral response, and fiber-to-fiber (and amplifier-to-amplifier) variations. Details concerning HETDEX's complex selection will be described in \citet{Farrow2023}.

\subsection{Overall Sample Validation}
\label{sec:sample_validation}

HETDEX is designed to search for faint, low $S/N$ emission lines in a large amount of data. HDR2 consists of 208 million fiber spectra, each with 1036 spectral resolution elements. This means examining over 210 billion resolution elements. Noise is ultimately the biggest contaminator in our catalog and the vast majority of our spectra observe the blank sky.  Attempts to quantify the HETDEX false-positive rate are ongoing but here we briefly summarize our efforts to confirm sources.

Multiple methods are used to measure the confirmation rate of HETDEX line emission sources. The method that provides the highest number of confirmed sources involves using the HETDEX data themselves. As described in Section~\ref{sec:counterpartfraction}, we assume that any emission-line source found in three independent observations is real. We create a validation sample by considering every OII and LAE in the catalog and checking to see if the location of the source has been targeted multiple times. This is done by cross-matching the catalog with the fiber database. If a source has fiber coverage from at least six observations, we put it in the validation sample. We note that just because a sky position has fiber coverage, that does not mean the observation is useful, as varying observing conditions may prevent a real source from being observed. Ultimately the confirmation rate provides only an upper limit on our false-positive rate. Consequently, we also use spectroscopic redshifts from the literature to validate  HETDEX detections if the redshift matches the redshift in the literature to within $\Delta z<0.02$

For the $S/N>5.5$ validation sample, 91.0\% of LAEs are confirmed; for the OII sample, the fraction is 99.3\%. The combined fraction is 98.1\% because of the high fraction of \OII-emitting galaxies in the catalog (\noii) compared to the number of $S/N>5.5$ LAEs (\nlae).

\section{Summary}

HETDEX is a medium-wide area, IFU spectroscopic survey that covers the wavelength range 3500\,\AA $\lesssim \lambda \lesssim$ 5500\,\AA\ at a resolving power of $750<R<950$. The survey will ultimately cover $\sim540$\,deg$^2$ with noncontiguous tiling leading to 94\,deg$^2$ of complete sky coverage. With the clustering of over one million high-$z$ \lya-emitting galaxies, HETDEX aims to measure the Hubble expansion rate, $H(z)$, and the angular diameter distance, $D_{A}(z)$, to better than 1\% accuracy.

This paper describes the first publicly released version of the HETDEX Source Catalog. The catalog is generated by combining raw HETDEX line emission and continuum emission detections which are performed in a grid search under point-source assumptions.  While there is some overlap between the two samples, the line emission search offers the unique capability of detecting very distant galaxies with relatively modest continuum emission and stellar mass through their bright \lya-emission. The catalog contains \nlae\ LAEs, \noii\ \OII-emitting galaxies, \nstar\ stars, \nlzg\ low-$z$ non-line- emitting galaxies, and \nagn\ AGN. By utilizing a three-prong classification approach, we provide robust spectroscopic redshifts and classifications for the entire catalog. When compared to external catalog spectroscopic redshifts, 96.1\% of the sources are within $\Delta z < 0.02$.

Using a sample of repeat source observations, we create a `confirmed' sample of confident sources. This allows us to validate line emitters that have not been observed in any other data set. In this `confirmed' sample, we find that we can confirm 91.0\% of the LAE sample and 99.3\% of the OII sample through evidence in repeat observations, suggesting an upper limit of 9\% for the false-positive rate in the LAE sample. 

Without any imaging preselection, HETDEX offers a blind search for LAEs. A search for imaging counterparts to the LAE sample in deep ancillary HSC $r$-band imaging shows that 45\% of the LAE sample has no detected imaging counterpart down to a limiting magnitude of $r=26.2$. A sample with imaging preselection at this sensitivity would miss half the HETDEX LAE sample presented in this paper.

Data access and details about the catalog can be found online at \url{http://hetdex.org}. A copy of the HETDEX Public Source Catalog (Version 3.2) is available on Zenodo \dataset[doi:10.5281/zenodo.7448504]{https://doi.org/10.5281/zenodo.7448504}. This Zenodo deposit includes the \textit{Source Observation Table} (columns described in Table~\ref{tab:column_info}) and the \textit{Detection Info Table} (columns described in Table~\ref{tab:det_col_info}) in multiple formats as well as a Jupyter notebook with access examples.

\acknowledgements

HETDEX is led by the University of Texas at Austin McDonald Observatory and Department of Astronomy with participation from the Ludwig-Maximilians-Universit\"at M\"unchen, Max-Planck-Institut f\"ur Extraterrestrische Physik (MPE), Leibniz-Institut f\"ur Astrophysik Potsdam (AIP), Texas A\&M University, The Pennsylvania State University, Institut f\"ur Astrophysik G\"ottingen, The University of Oxford, Max-Planck-Institut f\"ur Astrophysik (MPA), The University of Tokyo, and Missouri University of Science and Technology. In addition to Institutional support, HETDEX is funded by the National Science Foundation (grant AST-0926815), the State of Texas, the US Air Force (AFRL FA9451-04-2-0355), and generous support from private individuals and foundations.

The Hobby-Eberly Telescope (HET) is a joint project of the University of Texas at Austin, the Pennsylvania State University, Ludwig-Maximilians-Universit\"at M\"unchen, and Georg-August-Universit\"at G\"ottingen. The HET is named in honor of its principal benefactors, William P. Hobby and Robert E. Eberly.

The authors acknowledge the Texas Advanced Computing Center (TACC) at The University of Texas at Austin for providing high performance computing, visualization, and storage resources that have contributed to the research results reported within this paper. URL: http://www.tacc.utexas.edu

The authors are thankful to the Dark Energy Spectroscopic Instrument Survey team for providing invaluable early Survey Validation observations of a subset of the HETDEX emission-line sample.

The Institute for Gravitation and the Cosmos is supported by the Eberly College of Science and the Office of the Senior Vice President for Research at the Pennsylvania State University. The Kavli IPMU is supported by World Premier International Research Center Initiative (WPI), MEXT, Japan. 

This work makes use of the Sloan Digital Sky Survey IV, with funding provided by the Alfred P. Sloan Foundation, the U.S. Department of Energy Office of Science, and the Participating Institutions. \sdss-IV acknowledges support and resources from the Center for High-Performance Computing at the University of Utah. The \sdss\ web site is www.sdss.org.

This work makes use of the Pan-STARRS1 Surveys (PS1) and the PS1 public science archive, which have been made possible through contributions by the Institute for Astronomy, the University of Hawaii, the Pan-STARRS Project Office, the Max-Planck Society and its participating institutes.

This work makes use of data from the European Space Agency (ESA) mission {\it Gaia} (\url{https://www.cosmos.esa.int/gaia}), processed by the {\it Gaia} Data Processing and Analysis Consortium (DPAC, \url{https://www.cosmos.esa.int/web/gaia/dpac/consortium}). Funding for the DPAC has been provided by national institutions, in particular the institutions participating in the {\it Gaia} Multilateral Agreement.

This work makes use of observations made with the NASA/ESA Hubble Space Telescope obtained from the Space Telescope Science Institute, which is operated by the Association of Universities for Research in Astronomy, Inc., under NASA contract NAS 5–26555.

KG acknowledges support from NSF-2008793.

Software: This research was made possible by the open-source projects astropy \citep{astropy:2018}, python \citep{pythonref}, numpy \citep{harris2020array}, Scipy \citep{scipy}, hetdex-api (\url{https://github.com/HETDEX/hetdex_api}), elixer \citep[\url{https://github.com/HETDEX/elixer};][]{Davis2021}, diagnose \citep[\url{https://github.com/grzeimann/Diagnose};][]{Zeimann2022}, photutils \citep{photutils_1.3.0}, dustmaps \citep{dustmaps}, extinction (\url{https://github.com/kbarbary/extinction}) 

\bibliography{hetdex}
\bibliographystyle{aasjournal}

\newpage
\appendix
\section{Detection Info Table}
\label{appendix:1}

This appendix describes the \textit{Detection Info Table} which contains information for every line and continuum detection from our object detection search method (as described in Section~\ref{sec:detection}). As described in Section~\ref{sec:det_group}, a HETDEX source can be composed of a collection of line emission and continuum emission detections. The \textit{Source Observation Table}, outlined in Table\,\ref{tab:column_info}, provides a simplified version of the \textit{Detection Info Table} with one row per source observation, providing basic information about a source such as coordinates, redshift, \hetg\ magnitude, and the \OII\ and Ly$\alpha$ line flux and luminosity where applicable. The \textit{Detection Info Table} presented in this Appendix is expanded to provide additional information on every detection in a source. While many columns are the same to those in the \textit{Source Observation Table}, such as \texttt{source\_id}, \texttt{source\_name}, \texttt{RA}, \texttt{DEC}, \zhet. Additional information is provided regarding line fit parameter information. This includes the specific position of the detection (\texttt{RA\_det}, \texttt{Dec\_det}) and wavelength (\texttt{wave}) for the detection, the detection's line width, ($\sigma$: \texttt{sigma}), continuum-subtracted line flux and the local continuum measurement. Each observed wavelength is checked to see if it is a restframe match to a common line species at \zhet. Specifically, we consider \CIII, \CIV, \hbeta, \hdelta, \hgamma, \HeII, \lya, \OII and \OIII\ \footnote{\url{http://classic.sdss.org/dr6/algorithms/linestable.html}}. If a match is found it is listed in \texttt{line\_id}. Not all detections have a \texttt{line\_id} as some HETDEX line emission detections can result from jumps in a spectrum or calibration issues. We attempt to mitigate these by excluding high line width sources that are not selected as the main detection (ie. \texttt{selected\_det}==True) of a source. Other information as described in the text is also provided. This includes detection group information from 3D and 2D FOF detection grouping and \elixer\ imaging counterpart information. Also included are specific observation parameters such as the image quality of the observation, \texttt{fwhm}, and its observation ID information (e.g. \texttt{shotid}, \texttt{date}, \texttt{obsid}, \texttt{field} and specific information related to the highest weight fiber in the spectral extraction of the detection (such as \texttt{multiframe}, \texttt{fiber\_id}, \texttt{weight} and others). The detection whose spectrum is included in the \textit{Source Observation Table} that is the best representative of a source (typically the brightest magnitude detection) is identified by \texttt{selected\_det==True}. The description of all the parameters is provided in Table~\ref{tab:det_col_info}.

\begin{longtable}{ll}
\caption{\textit{Detection Info Table} Column Descriptions}\\ 
\toprule
\hline 
Name & Description \\
\hline
source\_id & HETDEX Source Identifier \\
source\_name & HETDEX IAU designation \\
RA & source\_id R.A. (ICRS deg) \\
DEC & source\_id decl. (ICRS deg) \\
z\_hetdex & HETDEX spectroscopic redshift \\
z\_hetdex\_src & HETDEX spectroscopic redshift source \\
z\_hetdex\_conf & 0 to 1 confidence HETDEX spectroscopic redshift source \\
source\_type & options are \texttt{star}, \texttt{lae}, \texttt{agn}, \texttt{lzg}, \texttt{oii}, and \texttt{none} \\
detectid & emission line or detection ID \\
selected\_det & best detectid for Ly$\alpha$ flux or \OII\ line flux \\
det\_type & detection type: `line' or `continuum' \\
line\_id & line identification at observed wavelength (wave) assuming redshift of z\_hetdex \\
RA\_det & detectid R.A. (ICRS deg) \\
DEC\_det & detectid decl. (ICRS deg) \\
src\_separation & separation in degrees between the detectid (RA\_det, DEC\_det) and the source\_id center (RA, DEC) \\
n\_members & number of detections in the source group \\
gmag\_err & MCMC uncertainty in gmag \\
gmag & \sdss-g magnitude measured in HETDEX spectrum \\
Av & applied dust correction in the $V$-band \\
ebv & applied selective extinction \\
wave & central wavelength of line emission (\AA) \\
wave\_err & MCMC error in wave (\AA) \\
flux & dust corrected line flux $10^{-17}\mathrm{erg/s/cm^2}$ \\
flux\_err & MCMC error in dust corrected line flux \\
flux\_obs & observed line flux $10^{-17}\mathrm{erg/s/cm^2}$ \\
flux\_obs\_err & MCMC error in observed line flux \\
flux\_aper & dust corrected, OII line flux measured in elliptical galaxy aperture in $10^{-17}\mathrm{erg/s/cm^2}$  \\
flux\_aper\_err & error in flux\_aper \\
flux\_aper\_obs & OII line flux measured in elliptical galaxy aperture in $10^{-17}\mathrm{erg/s/cm^2}$  \\
flux\_aper\_obs\_err & error in flag\_aper\_obs \\
flag\_aper & 1 = aperture line flux used for lum\_oii, -1= PSF-line flux used from "flux" column \\
sigma & sigma linewidth in gaussian line fit (\AA) \\
sigma\_err & MCMC error in sigma linewidth (\AA) \\
continuum & local fitted observed continuum in \fluxden  \\
continuum\_err & MCMC error in continuum in \fluxden\ \\
continuum\_obs & local fitted observed continuum in \fluxden\ \\
continuum\_obs\_err & MCMC error in continuum in \fluxden \\
sn & signal-to-noise for line emission \\
sn\_err & MCMC error in signal-to-noise \\
chi2 & reduced $\chi^2$ quality of line fit \\
chi2\_err & MCMC uncertainty in reduced $\chi^2$ \\
flux\_noise\_1sigma\_obs & observed 1 sigma flux sensitivity in $10^{-17}\mathrm{erg/s/cm^2}$ \\
flux\_noise\_1sigma & dust corrected 1 sigma flux sensitivity in $10^{-17}\mathrm{erg/s/cm^2}$ \\
apcor & aperture correction applied to spectrum at 4500\AA \\
counterpart\_mag & selected closest counterpart mag from source extracting on image data \\
counterpart\_mag\_err & uncertainty in counterpart\_mag \\
counterpart\_dist & distance to closest counterpart \\
counterpart\_catalog & image catalog source of counterpart \\
counterpart\_filter\ & image filter of counterpart \\
plya\_classification & \elixer\ likelihood line is Ly$\alpha$ ranges 0 to 1 (1=high probility line is \lya) \\
best\_z & \elixer\ best redshift \\
best\_pz & confidence in best\_z \\
z\_diagnose & best fit redshift from \texttt{Diagnose} \\
cls\_diagnose & best classification from \texttt{Diagnose}. Options are `STAR', `GALAXY', `QSO', `UNKNOWN' \\
stellartype & \texttt{Diagnose} spectral type classification for stars \\
agn\_flag & -1 not an AGN, 0 broad line source but not confirmed AGN, 1 confident AGN and z\_hetdex \\
wave\_group\_id & id for 3D Friend-of-Friends (FOF) clustering at common ra, dec, wave \\
wave\_group\_a & semi-major axis from 3D FOF clustering \\
wave\_group\_b & semi-minor axis from 3D FOF clustering \\
wave\_group\_pa & positional angle from 3D FOF clustering \\
wave\_group\_ra & mean ra from 3D FOF clustering \\
wave\_group\_dec & mean dec from 3D FOF clustering \\
wave\_group\_wave & mean wavelength from 3D FOF clustering \\
fwhm & measured seeing of the observation in arcsec \\
throughput & relative spectral response at 4540 assuming a 360 s nominal exposure \\
shotid & integer represent observation ID: int( date+obsid) \\
field & field ID: cosmos, goods-n, dex-fall, dex-spring \\
date & date \\
obsid & observation number \\
multiframe & string identifier for the ifuslot/specid/ifuid/amp combination \\
fiber\_id & string identifier for the highest weight fiber \\
weight & flux weight of the highest weight fiber \\
x\_raw & x value on the CCD of the detection (ds9 x value) \\
y\_raw & y value on the CCD of the detection (ds9 y value) \\
x\_ifu & x position in the ifu in arcsec \\
y\_ifu & y position in the ifu in arcsec \\
ra\_aper & Right Ascension of aperture center of imaging counterpart in degrees \\
dec\_aper & Declination of aperture center of imaging counterpart in degrees \\
catalog\_name\_aper & imaging source for measuring OII resolved apertures \\
filter\_name\_aper & filter of imaging used for measuring OII resolved apertures \\
dist\_aper & distance between aperture center and detectid position in arcsec \\
mag\_aper & photometric magnitude in aperture in imaging source \\
mag\_aper\_err & photometric magnitude error in aperture in imaging source \\
major & major axis of aperture ellipse of resolved OII galaxy defined by imaging \\
minor & minor axis of aperture ellipse of resolved OII galaxy defined by imaging \\
theta & angle in aperture ellipse \\
 &  \\
\hline
\end{longtable}
\label{tab:det_col_info}

\facilities{The Hobby-Eberly Telescope (McDonald Observatory)}

\end{document}